\documentclass[12pt]{article}
\usepackage[latin9]{inputenc}
\usepackage{geometry}
\usepackage{amsmath}
\usepackage{amssymb}
\usepackage{graphicx}
\usepackage{mathrsfs}
\usepackage{stmaryrd}
\usepackage{setspace}
\usepackage[authoryear]{natbib}
\usepackage[unicode=true,
 bookmarks=false,
 breaklinks=false,pdfborder={0 0 1},backref=section,colorlinks=false]
 {hyperref}
\hypersetup{
 colorlinks,citecolor=blue,pdftex}

\makeatletter

\usepackage{amsfonts}
\usepackage{amsthm}
\usepackage{lscape}
\usepackage{appendix}
\usepackage{dcolumn}
\usepackage{rotating}
\usepackage{comment}
\usepackage{color}

\usepackage{url}
\usepackage{todonotes}
\usepackage{verbatim}
\usepackage{titletoc}
\usepackage{pdflscape}

\usepackage[subtle]{savetrees}
\usepackage[small,compact]{titlesec}

\usepackage{diagbox}
\usepackage[flushleft]{threeparttable}
\usepackage{subfigure}
\usepackage{pgf}
\usepackage{tikz}\usepackage{caption}
\usepackage{tkz-tab}
\usepackage{tikz-3dplot}
\usetikzlibrary{calc}
\usetikzlibrary{arrows, automata}

\usepackage{changepage}
\usepackage{booktabs}

\allowdisplaybreaks

\newcolumntype{d}[1]{D{.}{.}{#1}}
\newcolumntype{t}[1]{D{,}{,}{#1}}
\newcolumntype{i}[1]{D{.}{}{#1}}
\newtheorem{theorem}{Theorem}[section]

\newtheorem{assumption}{Assumption}[section]

\newtheorem{definition}{Definition}[section]
\newtheorem{example}{Example}
\newtheorem{lemma}{Lemma}[section]

\newtheorem{proposition}{Proposition}[section]

\theoremstyle{plain}

\newcommand{\bd}{\boldsymbol{\delta}}
\newcommand{\hd}{\widehat{d}}
\newcommand{\hp}{\widehat{p}}

\newcommand{\cD}{\mathcal{D}}
\newcommand{\hcD}{\widehat{\mathcal{D}}}
\newcommand{\Rr}{\mathbb{R}}

\usepackage[font=small, labelfont=bf]{caption}

\DeclareMathOperator*{\argmax}{argmax\;}
\DeclareMathOperator*{\Var}{Var}
\DeclareMathOperator*{\sign}{sign}

\DeclareMathOperator{\argmin}{argmin}

\DeclareMathOperator{\diag}{Diag}

\numberwithin{equation}{section}

\usepackage{xcolor}

\makeatother

\begin{document}

\title{Post-Selection Inference for Partially Identified Parameters\footnote{We thank the editor, St\'ephane Bonhomme, three anonymous referees, Isaiah Andrews, Irene Botosaru, Patrik Guggenberger, Chris Muris and Jonathan Roth for helpful discussions and comments. We also acknowledge the use of ChatGPT and Claude for literature searches, proofreading and aiding in the development of some mathematical arguments. McCloskey acknowledges funding by the National Science Foundation (Grant SES-2341730).}}

\author{Sukjin Han\footnote{School of Economics, University of Bristol, vincent.han@bristol.ac.uk} \qquad Adam McCloskey\footnote{Department of Economics, University of Colorado, adam.mccloskey@colorado.edu}  
}

\date{\today}

\maketitle
\vspace{-0.2cm}

\begin{abstract}
Partial identification of parameters such as average treatment effects, average partial effects and welfare is particularly common when using observational data and experimental data with imperfect compliance due to the endogeneity of individuals' treatment uptake. In this setting, the researcher is typically interested in a treatment or policy that is either selected from the estimated set of best-performers or arises from a data-dependent selection rule.  In this paper, we develop new inference tools for partially-identified parameters chosen via these forms of selection.  We develop three types of confidence intervals for data-dependent and partially-identified parameters, discuss how they apply to several examples of interest and prove their uniform asymptotic validity under weak assumptions.

\vspace{0.2in}


\end{abstract}

\section{Introduction}\label{sec:intro}

Policymakers are often interested in choosing policies or treatments based on empirical evidence and understanding the welfare consequences of implementing their selections. An empirically-grounded policy is commonly chosen to optimize an estimated criterion (e.g., empirical welfare itself or regret), leading to a statistical inference problem for a data-dependent parameter.   It is now well-known that selecting an object of interest from data invalidates conventional inference tools (e.g.,~\citealp{AKM24}). In addition, optimal treatments and policies are often partially identified for various reasons, such as treatment endogeneity in observational data or experimental data with imperfect compliance (\citealp{stoye2012minimax,kallus2021minimax,pu2021estimating,d2021orthogonal,yata2021optimal}), imperfect external validity of experimental evidence (\citealp{ishihara2021evidence,AC22,MQS26}), nonlinearity of welfare (\citealp{cui2026policy}), and lack of overlap in algorithmic decisions (\citealp{BGIJ22}).  The failure of conventional inference tools after data-dependent selection is only compounded by the presence of such partial identification. At the same time, policymakers may hope to avoid imposing strong point-identifying assumptions on the underlying data-generating processes (e.g., parametric restrictions) lest policy decisions be inadvertently driven by such potentially false assumptions.

This paper aims to address the inference challenges simultaneously posed by partial identification and the data-driven selection of parameters.  We focus on a general class of data-dependent parameter selection, called \emph{polyhedral selection}, and develop new inference tools for parameters selected according to this class.  We show polyhedral selection captures a wide range of attitudes of a policymaker towards a policy choice, which includes two broad categories: (a) a policymaker, as a statistical decision maker under ambiguity \citep{man00}, choosing a single policy (e.g., via maximin, minimax) and conducting inference on welfare evaluated at that policy, and (b) one conducting inference on welfare for a policy in the estimated identified set for optimal policy. The confidence intervals (CIs) we propose here provide such a policymaker with the ability to accurately infer how well their selected treatments/policies can be expected to perform, for instance to gauge whether implementation of them is worthwhile from a welfare standpoint.

The current literature has not yet developed valid post-selection inference tools in partially-identified contexts, an important deficiency that the methods proposed in this paper aim to correct.  The methods we propose here build upon the ideas of conditional and hybrid inference employed in various point-identified contexts by, e.g.,~\cite{LSST16}, \cite{FST17}, \cite{TRTW18},  \cite{AKM24} and \cite{McC24} to produce CIs for partially-identified parameters such as welfare or ATEs chosen from an estimated set or via a data-dependent selection rule. Our framework formally nests the point-identified winners setting of \cite{AKM24}.\footnote{Under the embedding described in Appendix \ref{sec: AKM generalization}, their Assumptions 6--8 imply Assumptions \ref{ass: joint normality}--\ref{ass: variance restriction} of the present paper.  Our conditional CIs then coincide with the \cite{AKM24} conditional CIs.  The projection and hybrid CIs use the same principles as \cite{AKM24} but retain separate one-sided safeguards for the lower and upper endpoints of the identified set in order to cover the set in its entirety; see Appendix \ref{sec: AKM generalization} for a step-by-step comparison.}  Although \cite{ARP23} also propose conditional and hybrid inference methods in the partial identification context of moment inequality models, they do not allow for data-dependent selection of objects of interest, one of the main focuses of the present paper.  Finally, this paper directly relies upon results in the literature on partial identification of welfare, treatment effects and partial outcomes such as \cite{Man90}, \cite{BP97,BP13}, \cite{MP00}, \cite{MST18}, \cite{HY24} and \cite{Han24}.  We apply our inference methods to a general class of problems nesting these examples.

We start by motivating the problem of parameter selection and inference for a partially-identified parameter in a leading example of selecting dynamic treatment regimes (\citealp{Han24}) in Section \ref{sec:motivating_ex}. After sketching the ideas behind our inference methods in a simplified version of this example in Section \ref{sec:simple example}, we introduce in Section \ref{sec: general framework} the general inference framework to which our methods can be applied.  We formally introduce a (weak) polyhedral selection event and show that it incorporates selections based on maximin welfare, Hurwicz welfare, and minimax regret, and selection from the estimated identified set of optimal policies.  These are all leading cases of the types of selection in the literature on treatment choice under partial identification mentioned above. Our general framework also incorporates treatment choice based on Manski bounds for average counterfactual outcomes or ATEs and bounds on treatment parameters derived from linear programming (e.g.,~\citealp{BP97,BP13}; \citealp{MST18}; \citealp{Han24}; \citealp{HY24}) as well as parameters chosen via asymptotically optimal treatment choice rules (e.g.,~\citealp{CMS23}). These latter examples are covered in Section \ref{sec:examples} and Appendix \ref{sec: additional example}.

Within the general inference framework in Section \ref{sec: CIs}, we develop three types of CIs for data-dependent and partially-identified parameters whose coverage requirements can be broken down into two categories:~conditional and unconditional. It is natural for a researcher to require conditional validity when they want reliable inference on a particular policy \emph{only} when that policy is selected or retained for consideration from the data, rather than allowing poor coverage in those cases to be offset by better coverage in others.  On the other hand, unconditional validity is a natural requirement for a policymaker wishing to evaluate the overall procedure of selecting and assessing a policy according to a selection rule, requiring reliable coverage \emph{on average} across all repeated samples, no matter which particular policy was selected from the data.  As the name suggests, the \emph{conditional} CIs are asymptotically valid conditional on the parameter of interest corresponding to a treatment or policy chosen from an estimated set (i.e.,~case (b)).  The construction of this CI does not require a specific rule for choosing the parameter of interest from the estimated set.  Importantly, we show that these CIs are asymptotically valid \emph{uniformly} across a large class of data-generating processes (DGPs).  Uniform asymptotic validity is especially important for approximately correct finite-sample coverage in post-selection contexts like those in this paper (see, e.g., \citealp{AG09}).

The second and third types of CIs we develop in this paper are designed for inference on parameters chosen by data-dependent selection rules for which the object of interest is uniquely determined by the selection rule (i.e.,~case (a)).  The \emph{projection} CIs do not require knowledge of the selection rule to be asymptotically valid, whereas the \emph{hybrid} CI construction utilizes the particular form of a selection rule to improve upon the length properties of the projection CI.  Since they are valid conditionally, the law of total probability implies that the conditional CIs that condition on the realization of the selection rule are also valid unconditionally.  Although the conditional CIs are short for selections that occur with high probability, they can become exceptionally long when this probability is small (see, e.g.,~\citealp{KL21}).  Conversely, projection CIs are overly conservative when selection probabilities are high.  Our hybrid CIs interpolate the length properties of the conditional and projection CIs in order to attain good length properties regardless of the value selection probabilities take.  In analogy with the conditional CIs, we formally show that both projection and hybrid CIs are asymptotically valid in a uniform sense.

We analyze in Section \ref{sec: Monte Carlo} the finite-sample coverage and length properties of our proposed CIs in the context of a simple Manski bound example.  Since, to our knowledge, these are the first uniformly valid CIs for data-dependent selections of partially-identified parameters, there are no existing CIs to which we can directly compare.  Nevertheless, since our CIs can also be used for inference on a priori chosen partially-identified parameters, we conduct a power comparison with one of the leading methods for inference on a partially-identified parameter in Appendix \ref{sec: power comp}.  In particular, we compare the power of the test implied by our hybrid CIs to the power of the hybrid test of \cite{ARP23}, a test that applies to a general class of moment-inequality models that is also based upon a (different) hybrid between conditional and projection-based inference.  Encouragingly, the power of the test implied by our hybrid CI is quite competitive even in this environment for which it was not designed.  We also find that the finite-sample coverage of all of our CIs is approximately correct in a simple Manski bound example.  Finally, we analyze the length tradeoffs between the three different CIs across different DGPs, finding the hybrid CI to perform best overall. The final section, Section \ref{sec: application}, contains an empirical application corresponding to the leading example in Section \ref{sec:motivating_ex}, where we apply our procedures to dynamic policies of schooling and post-school training.

In the remaining appendices, Appendix \ref{sec: AKM generalization} provides a detailed discussion of precisely how the inference framework and methods of this paper relate to those of \cite{AKM24}, Appendix \ref{sec: additional simulation} contains additional simulation results corresponding to the dynamic treatment regime example detailed in Section \ref{sec: application},  Appendix \ref{sec: appendix} contains the mathematical proofs of the main results, while Appendix \ref{sec: degenerate affine endpoint components} describes an extension of our inference methods that permits affine endpoint components with zero asymptotic variance when they are uniformly separated from the corresponding population endpoint.


\section{Motivating Example:~Dynamic Treatment Regimes}\label{sec:motivating_ex}

Consider
schooling and post-school training as a sequence of treatments. A policymaker (PM) may be interested in understanding the welfare consequences of implementing a policy that comprises a particular (potentially adaptive) sequence of the treatments called a \emph{dynamic treatment regime} (\citealp{murphy2003optimal,robins2004optimal}). To be concrete, similar to \cite{Han24}, consider a dynamic treatment regime $\bd(\cdot)\equiv(\delta_{1},\delta_{2}(\cdot,\delta_1))\in \cD$, where $\delta_{1}$ indicates receipt of a high school (HS) diploma, $\delta_{2}(y_{1},\delta_{1})$ indicates receipt of the
job training program given pre-program employment status $y_{1}$ and first-stage policy $\delta_{1}$ and $\cD$ denotes the set of treatment regimes under consideration. By having
policy $\delta_{2}$ as a function of $y_{1}$, the treatment allocation policy adaptively
incorporates information about unobserved characteristics of the individuals
reflected in the response $Y_{1}$ to $\delta_{1}$. By implementing policy $\bd(\cdot)$, the policymaker aims to improve the employment status of individuals in the terminal stage. Let $Y_{1}(d_{1})$ and $Y_{2}(d_{1},d_{2})$ denote the counterfactual employment status each respective period for a fixed (i.e., non-adaptive) treatment (sequence) and define the counterfactual employment status given $\bd(\cdot)$ as $Y_{2}(\bd(\cdot))\equiv Y_{2}(\delta_{1},\delta_{2}(\cdot,\delta_1))$. Given this definition, the policymaker may be interested in maximizing the employment share $W(\bd)\equiv \mathbb{E}[Y_{2}(\bd(\cdot))]$
as a welfare objective over the set of possible policies. All possible adaptive policies in $\cD$ are listed in
Table \ref{tab:regimes}.

\begin{table}
\begin{centering}
\begin{tabular}{|c|c|c|c|}
\hline 
Regime \# & $\delta_{1}$ & $\delta_{2}(1,\delta_{1})$ & $\delta_{2}(0,\delta_{1})$\tabularnewline
\hline 
\hline 
1 & 0 & 0 & 0\tabularnewline
\hline 
2 & 1 & 0 & 0\tabularnewline
\hline 
3 & 0 & 1 & 0\tabularnewline
\hline 
4 & 1 & 1 & 0\tabularnewline
\hline 
5 & 0 & 0 & 1\tabularnewline
\hline 
6 & 1 & 0 & 1\tabularnewline
\hline 
7 & 0 & 1 & 1\tabularnewline
\hline 
8 & 1 & 1 & 1\tabularnewline
\hline 
\end{tabular}
\par\end{centering}
\caption{Possible Adaptive Policies $\bd(\cdot)$ When $T=2$ and $\delta_{1}$ Does Not Depend on Covariates}
\label{tab:regimes}
\end{table}

In practice, this policymaker would collect data to conduct inference on the welfare of the policy they select from the data in order to gauge how well the policy is expected to perform. This turns out to be a challenging problem due to modeling and sampling uncertainty. Suppose the policymaker observes $(D_1,D_2,Y_1,Y_2)$, where $D_{1}$ indicates whether an individual received a HS diploma, $D_{2}$ indicates whether the individual participated in the job training program, $Y_{1}$ indicates the employment status after $D_1$ (but before $D_2$) is realized, and $Y_{2}$ indicates the employment status after $D_2$ is realized. As $(D_1,D_2)$ can be endogenous, suppose the policymaker is equipped with IVs $(Z_1,Z_2)$ for them, respectively, where $Z_{1}$ is an indicator for the density of high schools and $Z_{2}$ is an indicator for whether the individual was randomly assigned to the job training program. Note that even with random assignment in the second stage ($Z_2$), noncompliance is a common source of endogeneity in $D_2$.

In general, the marginal and joint distributions of $(Y_{1}(d_{1}),Y_{2}(d_{1},d_{2}))$ are partially identified for reasons analogous to those for which average treatment effects are partially identified \citep{Man90}, as will become clear in later sections. This implies that the welfare $W(\bd)$ is also partially identified \emph{for any given} $\bd$, because
\begin{align}\label{eq:welfare}
W(\bd)\equiv \mathbb{P}[Y_{2}(\bd(\cdot))=1] & =\sum_{y_{1}\in\{0,1\}}\mathbb{P}[Y_{1}(\delta_{1})=y_{1},Y_{2}(\delta_{1},\delta_{2}(y_{1},\delta_{1}))=1]\nonumber\\
 & =\mathbb{P}[Y_{1}(\delta_{1})=1,Y_{2}(\delta_{1},\delta_{2}(1,\delta_{1}))=1]+\mathbb{P}[Y_{1}(\delta_{1})=0,Y_{2}(\delta_{1},\delta_{2}(0,\delta_{1}))=1].
\end{align}

Among the candidate parameters $\{W(\bd):\bd\in\cD\}$, for which parameter should the policymaker conduct inference (e.g., construct CIs)? Ideally, she may want to select $W(\bd^*)$ as a parameter of interest, where $\bd^*\in\cD$ is the
optimal policy that maximizes $W(\bd)$. This choice of parameter is infeasible for at least two reasons. First, $\bd^*$ is partially identified in general, as $W(\bd)$ is partially identified even with IVs.  Second, even if it were point-identified, say, from imposing strong parametric assumptions, the value of $\bd^*\in\cD$ is unknown to the PM in practice because it is a population object.  On the other hand, the PM would naturally be interested in inference on a value of $W(\bd)$ for a value of $\bd$ that she has selected to perform well from the data because this is the welfare objective for a policy the PM may implement in practice.

Given that the PM is naturally interested in inference on $W(\bd)$ for some data-dependent value of $\bd$, we consider the following two scenarios:

\noindent\emph{Case (a)}. When the PM is willing to choose the parameter from among $\{W(\bd):\bd\in\cD\}$, we assume that the PM is a statistical decision maker under ambiguity \citep{man00}: She would map the data to a policy in the form of $r:\{S_i\}\mapsto \widehat{\bd}$, where $S\equiv(D_1,D_2,Y_1,Y_2,Z_1,Z_2)$ is the vector of observables and $\{S_i\}$ denotes the sample of observables.  We consider a class of selection rules for $r$ that we call \emph{polyhedral selection rules}, for reasons that will become clear later. Examples are a selection rule that maximizes worst-case welfare and one that minimizes worst-case regret. Given the selection $\widehat{\bd}$, $W(\widehat{\bd})$ becomes the parameter for which the PM will form CIs.

\noindent\emph{Case (b)}. Alternatively, the PM may be interested in conducting inference on $W(\bd)$ for a value of $\bd$ in the estimated identified set for $\bd^*$ (or equivalently the set of undominated policies).  That is, the PM is interested in inference on $W(\bd)$ conditional on the \emph{polyhedral selection event} that $\bd\in \widehat{\cD}$.  Since $\bd^*$ is the ideal infeasible evaluation point for welfare, it is natural for the PM to be interested in conducting inference that is conditional on this feasible event.

When the parameter is selected in a data-dependent manner (case (a)) or inference is conditioned on a data-dependent event (case (b)), additional econometric challenges emerge; see the next section. We propose a computationally simple inference procedure that overcomes these challenges and yields CIs that are asymptotically valid uniformly across the DGPs that could have generated $S$.

For our application to the adaptive policies described above, we construct CIs for $W(\bd)$ evaluated at $\bd = \widehat{\bd}$ obtained from a maximin rule, which can be expressed as $\widehat\bd = \argmax_{\bd\in\cD} \widehat{L}(\bd)$, where $\widehat{L}(\bd)$ is the estimated lower bound on $W(\bd)$. When we apply our inference method to the dataset constructed for the application (details of this application are contained in Section \ref{sec: application}), we obtain the following 95\% CIs for the welfare (i.e., the probability of being employed in the terminal stage):\footnote{In the actual application in Section \ref{sec: application}, the welfare is the probability that 30-month earnings are above the median; the CIs below come from that application.}

\begin{center}
    \begin{tabular}{ccc}
    Conventional & (a) Hybrid & (b) Conditional \tabularnewline
    \hline
    (0.32, 0.78)	&(0.28, 0.73)	&(0.28, $\infty$)  \tabularnewline
    \hline
    \end{tabular}    
\end{center}
In this table, the conventional CI for $W(\bd)$ is invalid since $\bd$ was selected from the data and $W(\bd)$ is partially identified.  On the other hand, the hybrid CI is valid unconditionally while the conditional CI is valid conditional on the event that the regime $\bd = \argmax_{\bd\in\cD} \widehat{L}(\bd)$ is realized in the data.\footnote{See Section \ref{sec: application} for an in-depth discussion of when conditional vs unconditional validity would be desirable in the context of this application.}  The following are the coverage probabilities for the three CIs at the $0.95$ nominal level using simulated data based on the same setting as the above application (details are contained in Appendix \ref{sec: additional simulation}):

\begin{center}
    \begin{tabular}{lcccc}
     &  & Conventional & (a) Hybrid & (b) Conditional 
     \tabularnewline
    \hline
    DGP 1  &  &  0.66	&0.94	&0.94  \tabularnewline
    DGP 2  &  & 0.88	&0.95	&0.94 \tabularnewline
    \hline
    \end{tabular}
\end{center}

\section{Theoretical Challenges and How We Address Them: An Illustration}\label{sec:simple example}

We first illustrate our proposed methods using a simplified version of the motivating example from the previous section. Consider a binary outcome of interest $Y$, a binary treatment indicator $D$ and a binary treatment assignment $Z$.  Furthermore, let $Y(1)$ and $Y(0)$ denote counterfactual outcomes under treatment ($D=1$) and no treatment ($D=0$).  Assuming $\mathbb{E}[Y(d)|Z]=\mathbb{E}[Y(d)]$, \cite{Man90} shows that we can bound the average counterfactual outcomes $W(d)=\mathbb{E}[Y(d)]$ in the absence and presence of treatment as follows: for $d=0,1$,
\begin{equation}
L(d)\equiv \max\left\{p^{1d0},p^{1d1}\right\}\leq W(d)\leq \min\left\{1-p^{0d0},1-p^{0d1}\right\}\equiv U(d), \label{eq: Manski bounds}
\end{equation}
where $p^{ydz} \equiv Pr(Y=y,D=d|Z=z)$.  Given \eqref{eq: Manski bounds} it is natural to define the set of best-performing options $\cD^{*}\subset\cD\equiv\{0,1\}$, as a subset of the two options of treatment and no treatment, to be those that are \emph{undominated} options; from an observed dataset of outcomes, treatments and treatment assignments $\{(Y_i,D_i,Z_i)\}_{i=1}^n$, such a set can be estimated as:
\[
\widehat{\cD}=\left\{d\in\{0,1\}:\nexists d'\in\{0,1\} \text{ s.th. }\widehat U(d)<\widehat L(d')\right\}=\left\{d\in\{0,1\}:\widehat U(d)\geq \widehat L(d^ \prime)\text{ }\forall d^{ \prime}\in\{0,1\} \text{ s.th. }d^{ \prime}\neq d\right\},
\]
where $\widehat L(d)\equiv \max\{\widehat p^{1d0},\hp^{1d1}\}$ and $\widehat U(d)\equiv\min\{1-\widehat p^{0d0},1-\widehat p^{0d1}\}$ with $\widehat p^{ydz}$ being an empirical estimate of the fitted probability $p^{ydz}$.

We are interested in inference on the identified interval $[L(d),U(d)]$ for the average counterfactual outcome $W(d)$ of option $d\in\{0,1\}$, which will immediately provide inference on $W(d)$ itself, after the researcher selects this option from $\widehat{\cD}$ (in the sense of either (a) or (b) in the previous section).  In other words, we would like to provide statistically precise statements about the true average counterfactual outcome of an option selected from the data to give the researcher an idea of how well this selected option should be expected to perform in the population.  We first provide some intuition for why standard inference techniques based upon asymptotic normality fail and then sketch our proposals for valid inference in this context.

\subsection{Why Does Standard Inference Fail?}\label{sec: std inf fail}

To fix ideas, let us focus for now on inference for the lower bound $L(d)$ of a selected option rather than the entire identified set $[L(d),U(d)]$.  More specifically, since $L(d)$ is a lower bound for the average counterfactual outcome $W(d)$, we would like to obtain a lower confidence bound for $L(d)$. Under standard conditions, a central limit theorem implies $\widehat p=(\widehat p^{100},\widehat p^{010},\widehat p^{110},\widehat p^{101},\widehat p^{011},\widehat p^{111})^\prime$ is normally distributed in large samples.  So why can't a researcher form a conventional CI using $\widehat L(d)$ and quantiles from a normal distribution as the basis for inference?  There are two reasons such an approach is (asymptotically) invalid:
\begin{enumerate}
\item Even in the absence of data-dependent selection, $\widehat L(d)= \max\{\widehat p^{1d0},\widehat p^{1d1}\}$ is not normally distributed in large samples.

\item Data-dependent selection of $d$ further complicates the distribution of $\widehat L(d)$.
	\end{enumerate}

Reason 1.~is easy to see since $\widehat L(d)$ is the maximum of two normally distributed random variables in large samples when $d$ is chosen a priori.  To better understand reason 2., note that the distribution of $\widehat L(d)$ given $d\in \widehat{\cD}$ is the conditional distribution of the maximum of two normally distributed random variables given that the minimum of two other normally distributed random variables, $\widehat U(d)\equiv\min\{1-\widehat p^{0d0},1-\widehat p^{0d1}\}$, exceeds the maximum of yet another set of two normally distributed random variables, $\widehat L(d')=\max\{\widehat p^{1d'0},\widehat p^{1d'1}\}$ for $d'\neq d$.  Unconditionally, $\widehat L(\widehat d)$ for any data-dependent choice of $\widehat d$ is distributed as a mixture of the distributions of $\widehat L(0)$ and $\widehat L(1)$, neither of which are themselves normally distributed. Even this simple example, in which the set of candidate parameters contains only two elements (i.e., $\{W(0),W(1)\}$), poses interesting theoretical challenges: complex non-normal distributions arise due to the non-smoothness of the lower bound in the underlying parameters and data-dependent selection.

\subsection{Case (b): Conditional Confidence Intervals}\label{sec:simple example cond CIs}

Even though Case (a) is the leading case, we consider Case (b) first, because the former relies on the same conditional inference framework. Suppose that a researcher's interest in inference on $L(d)$ only arises when $d$ is estimated to be in the set of best-performing options, viz., $d\in \widehat{\cD}$.  In such a case, we are interested in a lower confidence bound for $L(d)$ that is approximately valid across repeated samples for which $d\in \widehat{\cD}$ because, as alluded to in the motivating example above, a PM would only be interested in inference on $W(d)$ precisely because $d\in \widehat{\cD}$.  Formally, we would like to form a \emph{conditionally} valid lower confidence bound $\widehat L(d)_\alpha^C$ such that
\begin{equation}
\mathbb{P}\left(\left.L(d)\geq \widehat L(d)_{\alpha}^C\right|d\in \widehat{\cD}\right)\geq 1-\alpha \label{eq: cond cov requirement}
\end{equation}
for some $\alpha\in(0,1)$ in large samples.  To do so, we characterize the conditional distribution of $\widehat L(d)$.  Specifically, let $\widehat j_L(d)\equiv \argmax_{j\in\{0,1\}}\widehat p^{1dj}$ be the value of $Z$ at which the maximum between the two estimated probabilities is achieved. Then $\widehat L(d)=\widehat p^{1d\widehat j_L(d)}$. The conditioning event $\{d\in \widehat{\cD},\widehat j_L(d)=j_L^*\}$, where $j_L^*$ is a potential value of $Z$, can be written as a polyhedron in $\widehat p$ and $\widehat L(d)$ is equal to a (random) element of $\widehat p$ so that Lemma 5.1 of \cite{LSST16} implies 
\begin{equation}
\widehat L(d)\left|\left\{d\in \widehat{\cD},\widehat j_L(d)=j_L^*\right\}\right.\sim \widehat p^{1dj_L^*}\left|\left\{\widehat{\mathcal{L}}_L(\mathcal{Z}_L)\leq \widehat p^{1dj_L^*}\leq \widehat{\mathcal{U}}_L(\mathcal{Z}_L)\right\}\right. \label{eq: cond dist simple example}
\end{equation}
for some known functions $\widehat{\mathcal{L}}_L(\cdot)$ and $\widehat{\mathcal{U}}_L(\cdot)$, where $\widehat p^{1dj_L^*}\sim\mathcal{N}(p^{1dj_L^*},\Var(\widehat p^{1dj_L^*}))$ in large samples and $\mathcal{Z}_L$ is a sufficient statistic for the nuisance parameter $p=( p^{100}, p^{010}, p^{110}, p^{101}, p^{011}, p^{111})^\prime$ that is asymptotically independent of $\widehat p^{1dj_L^*}$.  The characterization in \eqref{eq: cond dist simple example} permits the straightforward computation of a conditionally quantile-unbiased estimator for $p^{1d\widehat j_L(d)}=p^{1dj_L^*}$, since after further conditioning on the realized value of $\mathcal{Z}_L$ in sample, the latter is equal to the mean of the underlying normally distributed random variable $\widehat p^{1dj_L^*}\sim\mathcal{N}(p^{1dj_L^*},\Var(\widehat p^{1dj_L^*}))$ that is subject to truncation.\footnote{See, e.g.,~\cite{Pfa94} for quantile-unbiased estimation of the underlying mean of a truncated normal.}  Denoting this $1-\alpha$ quantile-unbiased estimator as $\widehat L(d)_{\alpha}^C$, we have
\begin{equation}
\mathbb{P}\left(\left.p^{1d\widehat j_L(d)}\geq \widehat L(d)_{\alpha}^C\right|d\in \widehat{\cD}\right)= 1-\alpha \label{eq: cond cov of pseudo-bound manski}
\end{equation}
in large samples. However, noting that $L(d)\geq p^{1d\widehat j_L(d)}$ with probability one, we can see that \eqref{eq: cond cov requirement} holds for this choice of $\widehat L(d)_{\alpha}^C$.

Although \eqref{eq: cond cov requirement} does not hold with exact equality, we note that the left-hand side cannot be much larger than the right-hand side.  In other words, although $\widehat L(d)_{\alpha}^C$ is a conservative lower confidence bound for $L(d)$, it is not very conservative.  This can be seen heuristically by working through the two possible values that $L(d)$ can take:
\begin{enumerate}
\item If $L(d)=p^{1d\widehat j_L(d)}$ (i.e., the correct probability is selected), then \eqref{eq: cond cov requirement} holds with equality in large samples by \eqref{eq: cond cov of pseudo-bound manski}.

\item If $L(d)\neq p^{1d\widehat j_L(d)}$ (i.e., an incorrect probability is selected), then $L(d)\approx p^{1d\widehat j_L(d)}$ (i.e., the correct and incorrect probabilities cannot differ by much) since $\widehat L(d)=\widehat p^{1d\widehat j_L(d)}$ so that the left-hand side of \eqref{eq: cond cov requirement} cannot be much larger than the right-hand side.
	\end{enumerate}

Finally, a construction analogous to that described above for producing a lower confidence bound for $L(d)$ produces a conditionally valid upper confidence bound $\widehat U(d)_{1-\alpha}^C$ for $U(d)$ that satisfies
\begin{equation}
\mathbb{P}\left(\left.U(d)\leq \widehat U(d)_{1-\alpha}^C\right|d\in \widehat{\cD}\right)\geq 1-\alpha \label{eq: cond cov requirement ub}
\end{equation}
for some $\alpha\in(0,1)$ in large samples.  The lower and upper confidence bounds can then be combined to form a CI, $[\widehat L(d)_{\alpha/2}^C,\widehat U(d)_{1-\alpha/2}^C]$, that is conditionally valid for $[L(d),U(d)]$ in large samples since
\begin{gather}
\mathbb{P}\left(\left.L(d)\geq \widehat L(d)_{\alpha/2}^C,U(d)\leq\widehat U(d)_{1-\alpha/2}^C\right|d\in \widehat{\cD}\right) \\ \notag
\geq 1-\mathbb{P}\left(\left.L(d)< \widehat L(d)_{\alpha/2}^C\right|d\in \widehat{\cD}\right)-\mathbb{P}\left(\left.U(d)>\widehat U(d)_{1-\alpha/2}^C\right|d\in \widehat{\cD}\right)\geq 1-\alpha. \label{eq: conditional CI validity example}
\end{gather}

\subsection{Case (a): Unconditional Confidence Intervals}\label{sec:simple example uncond CIs}

Given the ingredients from Case (b), now we are ready to illustrate Case (a). Suppose now that the researcher uses a data-dependent rule to select a \emph{unique} option of inferential interest. For example, suppose the researcher is interested in choosing the option with the highest counterfactual outcome in the worst case across its identified set (i.e., maximin welfare) so that she chooses $\widehat d=\argmax_{d\in\{0,1\}} \widehat L(d)$.  In such a case, it is natural to form a lower confidence bound for $L(\widehat d)$ that is valid across \emph{all} repeated samples when the PM is interested in the average performance of $W(\widehat d)$ overall, not just at the $\widehat d$ value realized in the data.  Formally, we would like to form an \emph{unconditionally} valid lower confidence bound $\widehat L(\widehat d)_{\alpha}^U$ such that
\begin{equation}
\mathbb{P}\left(L(\widehat d)\geq \widehat L(\widehat d)_{\alpha}^U\right)\geq 1-\alpha \label{eq: uncond cov requirement}
\end{equation}
for some $\alpha\in(0,1)$ in large samples.  Given its conditional validity \eqref{eq: cond cov requirement}, the conditional lower confidence bound $\widehat L(d)_{\alpha}^C$ also satisfies \eqref{eq: uncond cov requirement} upon changing the definition of $\widehat{\cD}$ to $\widehat{\cD}=\{\widehat d\}$ in its construction.  However, it is well known in the literature on selective inference that conditionally-valid confidence bounds can be very uninformative (i.e.,~far below the true value) when the probability of the conditioning event is small (e.g.,~\citealp{KL21}, \citealp{AKM24} and \citealp{McC24}).  See Section \ref{sec: unconditional CIs} for a detailed explanation. Here, we propose two additional forms of confidence bounds that are only unconditionally valid but do not suffer from this drawback.

First, we can form a lower confidence bound for $L(\widehat d)$ by projecting a one-sided rectangular simultaneous confidence lower bound for all possible values $L(\widehat d)$ can take: $\widehat L(\widehat d)_\alpha^P \equiv\widehat L(\widehat d)-\widehat c_{1-\alpha,L}\sqrt{\widehat\Sigma_{L,2\widehat d+1+\widehat j_L(\widehat d)}},$
where $\widehat c_{1-\alpha,L}$ is the $1-\alpha$ quantile of $\max_i\widehat\zeta_i/\sqrt{\widehat\Sigma_{L,i}}$ for $\widehat\zeta\sim\mathcal{N}(0,\widehat\Sigma_L)$, $\widehat\Sigma_L$ is a consistent estimator of $\Sigma_L\equiv\Var(\widehat p^{100},\widehat p^{101},\widehat p^{110},\widehat p^{111})$ and, for any square matrix $A$, $A_i$ denotes the $i^{th}$ element of the main diagonal of $A$. Here, the maximum is taken to guarantee simultaneous coverage of all possible values of $L(\widehat d)$. Since $p^{1\widehat d\widehat j_L(\widehat d)}\in\{\widehat p^{100},\widehat p^{101},\widehat p^{110},\widehat p^{111}\}$ with probability one,
\begin{gather*}
\mathbb{P}\left(p^{1\widehat d\widehat j_L(\widehat d)}\geq \widehat L(\widehat d)_{\alpha}^P\right)=\mathbb{P}\left(p^{1\widehat d\widehat j_L(\widehat d)}\geq \widehat p^{1\widehat d\widehat j_L(\widehat d)}-\widehat c_{1-\alpha,L}\sqrt{\widehat\Sigma_{L,2\widehat d+1+\widehat j_L(\widehat d)}}\right) \\
\geq\mathbb{P}\left(( p^{100}, p^{101}, p^{110}, p^{111})\geq (\widehat p^{100},\widehat p^{101},\widehat p^{110},\widehat p^{111})-\widehat c_{1-\alpha,L}\sqrt{\diag(\widehat\Sigma_L)}\right)= 1-\alpha
\end{gather*}
in large samples and \eqref{eq: uncond cov requirement} holds for $\widehat L(\widehat d)_\alpha^U=\widehat L(\widehat d)_\alpha^P$ because $L(\widehat d)\geq p^{1\widehat d\widehat j_L(\widehat d)}$.  However, $\widehat L(\widehat d)_\alpha^P$ suffers from a converse drawback to that of $\widehat L(\widehat d)_\alpha^C$:~it is unnecessarily conservative when $\widehat d=d$ is chosen with high probability (see e.g.,~\citealp{AKM24} and \citealp{McC24}).

We propose a favorable second lower confidence bound for $L(\widehat d)$ that combines the complementary strengths of $\widehat L(d)_\alpha^C$ and $\widehat L(\widehat d)_\alpha^P$.  Construction of this hybrid lower bound $\widehat L(\widehat d)_\alpha^H$ proceeds analogously to the construction of $\widehat L( d)_\alpha^C$ after adding the additional condition $\{p^{1\widehat d\widehat j_L(\widehat d)}\geq \widehat L(\widehat d)_\beta^P\}$ for $\beta<\alpha$ to the conditioning event and instead computing a conditionally quantile-unbiased estimator for $p^{1\widehat d\widehat j_L(\widehat d)}$, denoted as $\widehat L(\widehat d)_{\alpha}^H$, satisfying 
\begin{equation*}
\mathbb{P}\left(\left.p^{1\widehat d\widehat j_L(\widehat d)}\geq \widehat L(\widehat d)_{\alpha}^H\right|\widehat d=d^*,p^{1\widehat d\widehat j_L(\widehat d)}\geq \widehat L(\widehat d)_\beta^P\right)= \frac{1-\alpha}{1-\beta} 
\end{equation*}
in large samples, where $d^*$ is any realized value of the random variable $\widehat d$.  Imposing this additional condition in the formation of the hybrid bound ensures that $\widehat L(\widehat d)_{\alpha}^H$ is always greater than $\widehat L(\widehat d)_\beta^P$, limiting its worst-case performance relative to $\widehat L(\widehat d)_\beta^P$ when $\mathbb{P}(\widehat d=d^*)$ is small.  On the other hand, when $\mathbb{P}(\widehat d=d^*)$ is large, the additional condition $\{p^{1\widehat d\widehat j_L(\widehat d)}\geq \widehat L(\widehat d)_\beta^P\}$ is far from binding with high probability so that $\widehat L(\widehat d)_\alpha^H$ becomes very close to $\widehat L(d)_{(\alpha-\beta)/(1-\beta)}^C$.  In this case, $\widehat L(d)_{(\alpha-\beta)/(1-\beta)}^C$ is close to the conventional lower confidence bound based upon the normal distribution $\widehat L(\widehat d)-z_{(1-\alpha)/(1-\beta)}\sqrt{\Var(\widehat p^{1dj_L^*})}$ because the truncation bounds in \eqref{eq: cond dist simple example} are very wide (Proposition 3 in \citealp{AKM24}).

To see how \eqref{eq: uncond cov requirement} holds for $\widehat L(\widehat d)_\alpha^U=\widehat L(\widehat d)_\alpha^H$, note first that 
\begin{equation*}
\mathbb{P}\left(\left.L(\widehat d)\geq \widehat L(\widehat d)_{\alpha}^H\right|\widehat d=d^*,p^{1\widehat d\widehat j_L(\widehat d)}\geq \widehat L(\widehat d)_\beta^P\right)\geq\mathbb{P}\left(\left.p^{1\widehat d\widehat j_L(\widehat d)}\geq \widehat L(\widehat d)_{\alpha}^H\right|\widehat d=d^*,p^{1\widehat d\widehat j_L(\widehat d)}\geq \widehat L(\widehat d)_\beta^P\right)= \frac{1-\alpha}{1-\beta} 
\end{equation*}
for all $d^*\in\{0,1\}$.  Then, note that
\begin{gather*}
\mathbb{P}\left(L(\widehat d)\geq \widehat L(\widehat d)_{\alpha}^H\right)\geq \mathbb{P}\left(\left.L(\widehat d)\geq \widehat L(\widehat d)_{\alpha}^H\right|p^{1\widehat d\widehat j_L(\widehat d)}\geq \widehat L(\widehat d)_\beta^P\right)\cdot\mathbb{P}\left(p^{1\widehat d\widehat j_L(\widehat d)}\geq \widehat L(\widehat d)_\beta^P\right) \\
\geq \frac{1-\alpha}{1-\beta}(1-\beta)=1-\alpha
\end{gather*}
by the law of total probability.  

By similar reasoning to that used for the conditional CIs in Section \ref{sec:simple example cond CIs} above, $\widehat L(\widehat d)_{\alpha}^H$ is not very conservative as a lower confidence bound for $L(\widehat d)$. The researcher's choice of $\beta\in(0,\alpha)$ trades off the performance of $\widehat L(\widehat d)_{\alpha}^H$ across scenarios for which $\mathbb{P}(\widehat d=d^*)$ is large and small with a small $\beta$ corresponding to better performance when $\mathbb{P}(\widehat d=d^*)$ is large.  See \cite{McC24} for an in-depth discussion of these tradeoffs.  We recommend $\beta=\alpha/10$. 

Finally, analogous constructions to those above produce unconditional projection and hybrid confidence upper bounds $\widehat U(\widehat d)_{1-\alpha}^P$ and $\widehat U(\widehat d)_{1-\alpha}^H$ that can then be combined with the lower bounds to form CIs $[\widehat L(d)_{\alpha/2}^P,\widehat U(d)_{1-\alpha/2}^P]$ and $[\widehat L(d)_{\alpha/2}^H,\widehat U(d)_{1-\alpha/2}^H]$ for $[L(d),U(d)]$ that are unconditionally valid in large samples by the same arguments as those used in \eqref{eq: conditional CI validity example} above.

\section{General Inference Framework}\label{sec: general framework}

We now introduce the general inference framework that we propose, nesting the examples in the previous sections as special cases.  After introducing the general framework, we describe several additional example applications that fall within this framework.

In the setting of this paper, a researcher's interest in $W(d)$ arises when the policymaker selects $d\in \cD$ in a data-driven manner. Suppose the policymaker selects a rule given data summarized in the empirical distribution $\hp$. We first define the type of policymaker's selection rules we focus on.

\begin{definition}[Polyhedral Selection]\label{def:poly_sel}
    Let a random set $\cD(\hp) \subseteq \cD\equiv \{d^0,\ldots,d^K\}$ be a correspondence of $\hp$, an estimator of a finite-dimensional parameter $p$. For all $d\in \cD$, the event $\{d\in \cD(\hp)\}$ is a weak polyhedral selection in $\hp$ if
    \begin{equation}\label{eq:poly_sel}
        \{d\in \cD(\hp)\} = \bigcup_{j=1}^J\{A_j \hp \le c_j\}
    \end{equation}
    almost surely (a.s.) for some matrix $A_j$ and vector $c_j$ for $j\in \{1,...,J\}$; the event is a polyhedral selection in $\hp$ if \eqref{eq:poly_sel} holds with $J=1$.
\end{definition}

Weak polyhedral selection events are common in the selective inference literature.  See, e.g.,~\cite{LSST16} for an example in the context of LASSO-based model selection and \cite{AKM21} for an example in the context of inference after estimating a regression threshold or structural break.

The (estimated) policy set $\cD(\hp)$ could correspond to an estimated set of optimal treatments or policies or other data-dependent index sets of interest. Continuing with the motivating example in Section \ref{sec:motivating_ex}, Scenario (a) corresponds to a policymaker selecting $d \in \cD(\hp) =\{\hd\}$ (i.e., a singleton selection $d = \hd$) where $\hd$ can be the minimax or maximin rule for $W(d)$. Scenario (b) corresponds to $\cD(\hp) =\hcD$ where $\hcD$ is the estimated set for $\argmax_{d\in \cD} W(d)$. In these examples, $\cD(\hp)$ is typically constructed via the estimated (sharp) lower and upper bounds on welfare $W(d)$:\footnote{In the first two examples, we implicitly assume that some tie-breaking procedure is applied in the maximum and minimum (respectively) to produce the final rule.}

\begin{example}[Maximin Welfare Rule]\label{ex:maximin}
    Let $\cD(\hp) =\{\hd\}$ a.s. with
    \begin{equation*}
        \hd = \argmax_{d \in \cD}\widehat{L}(d),
    \end{equation*}
    where $\widehat{L}(d)$ is the estimated lower bound on $W(d)$.
\end{example}

\begin{example}[Hurwicz Welfare Rule]\label{ex:hurwicz}
    Let $\cD(\hp) =\{\hd\}$ a.s. with
    \begin{equation*}
        \hd = \argmax_{d \in \cD}\{w_L\widehat L(d)+w_U\widehat U(d)\},
    \end{equation*}
    for some fixed  known weights $w_L,w_U\geq 0$ where $\widehat{L}(d)$ and $\widehat{U}(d)$ are the estimated lower and upper bounds on $W(d)$. This includes Example \ref{ex:maximin} ($w_U=0$) and the maximax rule ($w_L=0$) as special cases.
\end{example}

\begin{example}[Minimax Regret Rule]\label{ex:minimax}
Let $\cD(\hp)=\{\hd\}$ a.s. with $\hd=\argmin_{d\in\cD}\widehat R(d)$ where $\widehat R(d)\equiv\max_{d'\in\cD}\widehat U(d',d)$ and $\widehat U(d',d)$ estimates the upper bound on
$W(d')-W(d)$, with $\widehat U(d,d)=0$. Suppose, for every $d'\neq d$, $U(d')$ and $L(d)$ are attainable in the
DGP.\footnote{Formally, for every $d',d\in\cD$ with $d'\neq d$, there exists a DGP consistent
with $p$ under which $W(d')=U(d')$ and $W(d)=L(d)$.}  Then $U(d',d)=U(d')-L(d)$, and so
\[
\hd=\argmin_{d\in\cD}
\max\left\{0,\max_{d'\neq d}
\bigl[\widehat U(d')-\widehat L(d)\bigr]\right\}.
\]
\end{example}

\begin{example}[Set of the Undominated]\label{ex:undominated}
    Let $\cD(\hp) =\hcD$ a.s. with\footnote{This estimated set and $\hd$ in Example \ref{ex:maximin} are related as follows: $\widehat{\cD}=\left\{d\in \cD:\widehat U(d)\geq\widehat L(\hd)\right\}$.}
    \begin{align*}
        \hcD =&\left\{d\in\{d^0,\ldots,d^K\}:\nexists d'\in\{d^0,\ldots,d^K\} \text{ s.th. }\widehat U(d)<\widehat L(d')\right\}\\
        =&\left\{d\in\{d^0,\ldots,d^K\}:\widehat U(d)\geq \max_{d'\in\{d^0,\ldots,d^K\}}\widehat L(d')\right\}.
    \end{align*}
\end{example}

\begin{example}[Predetermined Policy]\label{ex:predetermined}
    Let $\cD(\hp)=\{d^*\}$ where $d^* \in \cD$ is a policy chosen a priori without consulting data.
\end{example}

Examples \ref{ex:maximin}--\ref{ex:hurwicz} concern maximizing the weighted sum of (estimated) upper and lower bounds on the welfare (\citealp{hur51}), with maximizing the lower bound as the most conservative special case from the policymaker's perspective. Example \ref{ex:minimax} considers minimizing worst-case regret, which is considered to be less pessimistic than maximizing the worst-case welfare of Example \ref{ex:maximin} (\citealp{sav51}). The estimated set $\hcD$ in Example \ref{ex:undominated} can be viewed as the sample counterpart of the identified set $\cD^*$ of $d^* \in \argmax_{d\in \cD}W(d)$. Examples \ref{ex:maximin}--\ref{ex:minimax} are established rules in the statistical decision theory (\citealp{man04,Man21,Man23}) and Example \ref{ex:undominated} can be viewed as a reasonable set-valued target (\citealp{Han24}). The inherent feature of these examples is that the welfare $W(d)$ is partially identified. This \emph{model ambiguity} in choosing policies can originate from endogeneity (\citealp{kallus2021minimax,pu2021estimating,d2021orthogonal,yata2021optimal}), imperfect internal and external validity (\citealp{stoye2012minimax,ishihara2021evidence,MQS26}), or the welfare being a function of the joint distribution of potential outcomes (\citealp{cui2026policy}). Moreover, many decision problems involve a finite menu (\citealp{man04,MT16,CMS23,Han24}) and a finite menu is practically relevant even when a hypothetical infinite menu is available because a policymaker may only find it feasible to choose the policy within a finite set (e.g., finite set of programs, finite types of characteristics).

Under the construction of $\cD(\hp)$ reflected in Examples \ref{ex:maximin}--\ref{ex:undominated}, the following assumption guarantees that the selection events $d\in \cD(\hp)$ in such examples are (weak) polyhedral selections.
This assumption posits that our parameter of inferential interest $W(d)$ for some $d\in \cD$ belongs to an identified set $[L(d),U(d)]$ taking a particular interval form that is common to many applications of interest.

\begin{assumption} \label{ass: welfare bounds}
For an unknown finite-dimensional parameter $p$,
\begin{enumerate}
\item $L(d)\equiv \max_{j\in \{1,\ldots,J_L\}}\{\tilde\ell_{d,j}+\ell_{d,j}p\}\leq W(d)$ for some fixed and known $J_L$, $\tilde\ell_{d,1},\ldots,\tilde\ell_{d,J_L}$ and nonzero row vectors $\ell_{d,1},\ldots,\ell_{d,J_L}$ such that $\ell_{d,j}\neq\ell_{d,j'}$ for $j\neq j'$.
\item $U(d)\equiv \min_{j\in \{1,\ldots,J_U\}}\{\tilde u_{d,j}+u_{d,j}p\}\geq W(d)$ for some fixed and known $J_U$, $\tilde u_{d,1},\ldots,\tilde u_{d,J_U}$ and nonzero row vectors $u_{d,1},\ldots,u_{d,J_U}$ such that $u_{d,j}\neq u_{d,j'}$ for $j\neq j'$.
\end{enumerate}
\end{assumption}

The lower and upper endpoints of identified sets for the welfare, average counterfactual outcome, or ATE typically take the form of $L(d)$ and $U(d)$, especially when (sequences of) outcomes, treatments and instruments are discrete.\footnote{Our general framework applies to settings with dynamic outcomes, treatments and instruments observed over time.  This is what is meant here by ``sequences''.} See Section \ref{sec:examples} and Appendix \ref{sec: additional example} for such examples.

Assume the estimated lower and upper bounds on $W(d)$ are defined as the sample counterpart of Assumption \ref{ass: welfare bounds}:
\begin{align}
    \widehat L(d)&\equiv \max_{j\in \{1,\ldots,J_L\}}\{\tilde\ell_{d,j}+\ell_{d,j}\widehat p\}\label{eq:L-hat}\\
    \widehat U(d)&\equiv \min_{j\in \{1,\ldots,J_U\}}\{\tilde u_{d,j}+u_{d,j}\widehat p\}.\label{eq:U-hat}
\end{align} 

\begin{proposition}\label{prop: weak polyhedral examples}
    Given \eqref{eq:L-hat}--\eqref{eq:U-hat}, the selection events in Examples \ref{ex:maximin}--\ref{ex:minimax} are weak polyhedral selections and the selection event in Example \ref{ex:undominated} is a polyhedral selection. 
\end{proposition}

In developing our inference procedures, it is useful to consider an additional event that a particular affine component is active in \eqref{eq:L-hat}--\eqref{eq:U-hat}. Define
\begin{equation}\label{eq:j_hats}
\widehat j_L(d)\equiv\argmax_{j\in \{1,\ldots,J_L\}}\{\tilde\ell_{d,j}+\ell_{d,j}\widehat p\}, \quad \widehat j_U(d)\equiv\argmin_{j\in \{1,\ldots,J_U\}}\{\tilde u_{d,j}+u_{d,j}\widehat p\},
\end{equation}
which are the indices at which the estimated lower and upper bounds are realized. Then, the estimated bounds can be expressed as $\widehat{L}(d)=\tilde\ell_{d,\widehat j_L(d)}+\ell_{d,\widehat j_L(d)}\widehat p$ and $\widehat{U}(d)=\tilde u_{d,\widehat j_U(d)}+u_{d,\widehat j_U(d)}\widehat p$.  We work under the high-level assumption that the following joint event can be written as a polyhedron in $\widehat p$: (i) an option index $d$ is in the policy set of interest $\cD(\hp)$, (ii) the estimated bounds on $W(d)$ are realized at a given value (i.e., $\widehat j_L(d)=j_L^*$ and $\widehat j_U(d)=j_U^*$) and (iii) (optionally) an additional random vector is realized at any given value (i.e., $\widehat\gamma_L(d)=\gamma_L^*$ and $\widehat\gamma_U(d)=\gamma_U^*$). Although the event (i) is typically a weak polyhedral selection, the joint event (i)--(iii) is typically a polyhedral selection.\footnote{In conjunction with Assumption \ref{ass: welfare bounds}, the property (ii) is also useful for establishing coverage separately for the true bounds $L(d)$ and $U(d)$ via a dominance argument:~by forming a lower (upper) confidence bound for $\tilde\ell_{d,j_L^*}+\ell_{d,j_L^*} p$ ($\tilde u_{d,j_U^*}+u_{d,j_U^*}p$), we automatically obtain a lower (upper) confidence bound for $L(d)\geq \tilde\ell_{d,j_L^*}+\ell_{d,j_L^*} p$ ($U(d)\leq \tilde u_{d,j_U^*}+u_{d,j_U^*}p$).  See, e.g.,~the argument following \eqref{eq: cond cov of pseudo-bound manski}.}

Recall our parameter of interest $W(d)$ for some $d\in \cD$ and the policy set $\cD(\hp)$ in Definition \ref{def:poly_sel}.

\begin{assumption} \label{ass: estimated ID'd set}
\begin{enumerate}
\item The joint event $\{d\in\cD(\hp)$, $\widehat j_L(d)=j_L^*$ and $\widehat\gamma_L(d)=\gamma_L^*\}$ is equivalent to $\{A^L\hp\leq c^L\}$ a.s. for some matrix $A^L$ and vector $c^L$ that depend on $d\in\cD$, $j_L^*\in \{1,\ldots,J_L\}$, and $\gamma_L^*$ in the support of $\widehat\gamma_L(d)$.

\item The joint event $\{d\in\cD(\hp)$, $\widehat j_U(d)=j_U^*$ and $\widehat\gamma_U(d)=\gamma_U^*\}$ is equivalent to $\{A^U\hp\leq c^U\}$ a.s. for some matrix $A^U$ and vector $c^U$ that depend on $d\in \cD$, $j_U^*\in \{1,\ldots,J_U\}$, and $\gamma_U^*$ in the support of $\widehat\gamma_U(d)$.
\end{enumerate}
\end{assumption}

Although not immediately obvious, this assumption holds in a variety of settings; see the examples below. In particular, the joint event can be shown to be polyhedral in $\hp$ even if the marginal event $\{d\in\cD(\hp)\}$ is only weakly polyhedral; see Proposition \ref{prop: selection rule} below. The auxiliary vectors $\widehat\gamma_L(d)$ and $\widehat\gamma_U(d)$ are a mechanical bookkeeping device.  Conditional on $\widehat j_M(d)=j_M^*$ ($M=L,U$), they record any other random variables needed to express the event $\{d\in\cD(\hp)\}$ as a single polyhedron in $\widehat p$ after their realizations are conditioned upon.\footnote{Conditioning on the realizations of these random variables produces computational tractability by allowing us to express the asymptotic conditional distributions of $\widehat{L}(\widehat d)$ and $\widehat{U}(\widehat d)$ as normal distributions truncated to a single interval.  This computational tractability may be at the expense of some power loss relative to not conditioning on $\widehat j_M(d)$ and/or $\widehat\gamma_M(d)$, a common choice in the literature.  See, e.g.,~\cite{LSST16}, who analogously condition on the additional random variables that characterize the signs of estimated regression coefficients after selecting a model via LASSO.}

For example, selection rules based on estimated intersection bounds are piecewise affine (Examples \ref{ex:maximin}--\ref{ex:minimax}):~before the relevant active indices $\widehat j_M(d)$ and auxiliary vectors $\widehat \gamma_M(d)$ for $M\in\{L,U\}$ are fixed, the selection event can be a union of polyhedra; after they are fixed, it is described by linear inequalities.  When $\{d\in\cD(\hp),\widehat j_M(d)=j_M^*\}$ is already polyhedral, no additional label is needed and $\widehat\gamma_M(d)$ can be vacuously set to a constant.\footnote{Because additional conditioning may reduce power, $\widehat\gamma_M(d)$ should contain only the indices required for the polyhedral representation.}  A leading example of these cases is the selection of empirically non-dominated options (Example \ref{ex:undominated}) for which $\cD(\hp)=\hcD$.  See the proof of Proposition \ref{prop: selection rule} below for details.  
We also note that Assumption \ref{ass: estimated ID'd set} is compatible with Example \ref{ex:predetermined}, where $\cD(\hp)=\{d^*\}$: absent data-dependent selection, the researcher is interested in forming a CI for an identified interval $[L(d^*),U(d^*)]$ chosen by the researcher a priori. In this case, $\widehat\gamma_M(d^*)$ can be vacuously set to a fixed constant and $A^M$ and $c^M$ no longer depend on $\gamma$. Indeed, we examine an example of this special case when conducting a finite-sample power comparison in Section \ref{sec: power comp} below.

The conditional CIs below do not require $\cD(\hp)$ to be a singleton.  Under Assumptions \ref{ass: welfare bounds}--\ref{ass: estimated ID'd set} (and other regularity conditions introduced later), they provide correct (asymptotic) coverage for $W(d)$ for each $d$ conditional on the event $d\in\cD(\hp)$.  By contrast, unconditional inference on a data-dependent parameter requires the final target itself to be a well-defined random variable.  We therefore impose the next assumption only for the projection and hybrid procedures that target $W(\widehat d)$ unconditionally.  The assumption concerns the final sample recommendation and places no restriction on whether the population identified set of best-performing policies $\cD^*$ is a singleton.

\begin{assumption} \label{ass: selection rule}
For the unconditional inference problem, the researcher specifies a selection rule whose final output is a singleton $\cD(\hp)=\{\hd\}$ a.s., where $\hd$ takes values in $\cD \equiv \{d^0,\ldots,d^K\}$.
\end{assumption}

If an initial procedure produces an estimated set $\widehat{\cD}$ and a policymaker subsequently chooses one element using a rule $r$, the relevant final recommendation is $\hd=r(\widehat{\cD},\hp,\eta)$, where $\eta$ may be an independent randomization device.\footnote{A deterministic tie-breaking convention is enough when ties otherwise have probability zero.  A randomized policy rule can also be represented by augmenting the data with the realized randomization draw, provided that the resulting selection event satisfies the polyhedral structure needed for Assumption \ref{ass: estimated ID'd set}.} Assumption \ref{ass: selection rule} applies to this final rule, not to the intermediate set.

Under Assumptions \ref{ass: welfare bounds}--\ref{ass: selection rule} (and other regularity conditions), the unconditional CIs provide correct (asymptotic)  coverage for $W(\hd)$. Recall Examples \ref{ex:maximin}--\ref{ex:minimax} for the type of selection rules considered in Assumption \ref{ass: selection rule}.  On the other hand, only Assumptions \ref{ass: welfare bounds} and \ref{ass: estimated ID'd set} (and other regularity conditions) are required for the conditional CIs to provide correct coverage for $W(d)$ conditional on $d\in\mathcal{D}(\widehat p)$, such as $\cD(\widehat p)=\widehat{\cD}$ in Example \ref{ex:undominated}.

Taken together, Assumptions \ref{ass: estimated ID'd set} and \ref{ass: selection rule} hold naturally when $\hd$ is selected by uniquely maximizing a linear combination of $(L(d),U(d))$ (Examples \ref{ex:maximin}--\ref{ex:hurwicz}) or minimizing the regret (Example \ref{ex:minimax}) and the additional conditioning vectors $\widehat\gamma_L(d)$ and $\widehat\gamma_U(d)$ are defined appropriately. Assumption \ref{ass: estimated ID'd set} holds when the policy is selected from the estimated set $\hcD$ (Example \ref{ex:undominated}). The following proposition formalizes these statements.

\begin{proposition}\label{prop: selection rule}
Suppose \eqref{eq:L-hat}--\eqref{eq:U-hat} hold. When $\cD(\hp)=\{\hd\}$ a.s. (i.e., Assumption \ref{ass: selection rule}),
Assumption~\ref{ass: estimated ID'd set} holds in
\begin{enumerate}
\item Example \ref{ex:maximin} for $\widehat\gamma_L(d)$ constant and
$\widehat\gamma_U(d)=\widehat j_L(d)$;
\item Example \ref{ex:hurwicz} with $w_L=0$ for
$\widehat\gamma_L(d)=\widehat\gamma_U(d)
=(\widehat j_U(d^0),\ldots,\widehat j_U(d^K))'$;
\item Example \ref{ex:hurwicz} with $w_L,w_U>0$ for
$\widehat\gamma_L(d)=\widehat\gamma_U(d)
=(\widehat j_L(d^0),\ldots,\widehat j_L(d^K),
  \widehat j_U(d^0),\ldots,\widehat j_U(d^K))'$;
\item Example \ref{ex:minimax} for $\widehat\gamma_L(d)=\widehat\gamma_U(d)
=\bigl(\widehat j_L(a),\widehat j_U(a),\widehat b(a)\bigr)_{a\in\cD}$, 
where
\[
    \widehat b(a)\equiv
    \argmax_{b\in\cD}\widehat U(b,a),
    \qquad
    \widehat U(b,a)\equiv
    \begin{cases}
        0, & b=a,\\
        \widehat U(b)-\widehat L(a), & b\neq a.
    \end{cases}
\]
\end{enumerate}
When $\cD(\hp)=\hcD$ (Example \ref{ex:undominated}) or $\cD(\hp)=\{d^*\}$ (Example \ref{ex:predetermined}),
Assumption~\ref{ass: estimated ID'd set} holds with
$\widehat\gamma_L(d)$ and $\widehat\gamma_U(d)$ constant.
\end{proposition}

This proposition establishes that the joint event stated in Assumption \ref{ass: estimated ID'd set} is polyhedral in all the leading examples. Expressions for $A^M$ and $c^M$ ($M=L,U$), which are needed to construct the CIs we introduce below, in the settings of Proposition \ref{prop: selection rule} are available for reference in its proof in Appendix \ref{sec: appendix}. As this proposition makes clear, the additional conditioning vectors needed for Assumption \ref{ass: estimated ID'd set} to hold depend upon the particular form of selection rule used by the researcher. For example, if $\widehat d$ maximizes $\widehat L(d)$, then $\{\widehat d=d,\widehat j_L(d)=j\}$ is polyhedral and $\widehat\gamma_L(d)$ can be constant.  To conduct upper-endpoint inference for the same selected policy, however, conditioning on $\widehat j_U(d)$ does not identify the lower-bound component that generated selection, so we additionally set $\widehat\gamma_U(d)=\widehat j_L(d)$.  If $\widehat d$ instead maximizes $\widehat U(d)$, comparisons across policies require the active upper-bound index for each competitor, so the vector $(\widehat j_U(d^0),\ldots,\widehat j_U(d^K))'$ is included in the auxiliary conditioning information.  Weighted combinations of lower and upper bounds generally require the corresponding vectors of both types of active indices.

\section{Examples\label{sec:examples}}

In this section, we show that Assumptions \ref{ass: welfare bounds} and \ref{ass: estimated ID'd set} are satisfied in the motivating example in Section \ref{sec:motivating_ex}. We then briefly mention other examples, while deferring the verification of the assumptions for them to Appendix \ref{sec: additional example}.

\subsection{Dynamic Treatment Regime (Example in Section \ref{sec:motivating_ex})}\label{subsec:DTR example}

Continuing with the motivating example in Section \ref{sec:motivating_ex}, consider constructing a CI for the welfare $W(\bd)\equiv \mathbb{P}[Y_{2}(\bd(\cdot))=1]$ defined in \eqref{eq:welfare} that is evaluated at a policy chosen from $\cD$, which is a finite set (Table \ref{tab:regimes}). For the selection rule, we focus on implementing the maximin welfare rule (Example \ref{ex:maximin}) or the Hurwicz welfare rule (Example \ref{ex:hurwicz}). In both cases, we need to derive bounds on $W(\bd)$ based on the distribution of data, which is expressed as the vector $p$
of the form
\begin{align*}
p & \equiv \{\mathbb{P}[D_{1}=d_{1},Y_{1}=y_{1},D_{2}=d_{2},Y_{2}=y_{2}|Z_{1}=z_{1},Z_{2}=z_{2}]\}_{(d_{1},y_{1},d_{2},y_{2},z_{1},z_{2})}.
\end{align*} 
First, consider bounds on $W_{y}(d)\equiv \mathbb{P}[Y(d)=y]$ for $d\equiv(d_{1},d_{2})$,
which are $L_{y}(d)\equiv\max_{z}L_{y}(d;z)$ and $U_{y}(d)\equiv\min_{z}U_{y}(d;z)$
where
\begin{align}
L_{y}(d;z) & \equiv \mathbb{P}[Y=y,D=d|Z=z],\label{eq:L_y}\\
U_{y}(d;z) & \equiv \mathbb{P}[Y=y,D=d|Z=z]+\mathbb{P}[Y_{1}=y_{1},D_{1}=d_{1},D_{2}=1-d_{2}|Z=z]\nonumber \\
 & \quad+\mathbb{P}[D_{1}=1-d_{1},D_{2}=d_{2}|Z=z]+\mathbb{P}[D_{1}=1-d_{1},D_{2}=1-d_{2}|Z=z]\nonumber \\
 & =\mathbb{P}[Y=y,D=d|Z=z]+\mathbb{P}[Y_{1}=y_{1},D_{1}=d_{1},D_{2}=1-d_{2}|Z=z]+\mathbb{P}[D_{1}=1-d_{1}|Z=z].\label{eq:U_y}
\end{align}
Using these bounds, we can calculate bounds on $W(\bd)$, which are
\begin{align}
L(\bd) & \equiv\max_{z}L_{(1,1)}(\delta_{1},\delta_{2}(1,\delta_{1});z)+\max_{z}L_{(0,1)}(\delta_{1},\delta_{2}(0,\delta_{1});z),\label{eq:L_delta}\\
U(\bd) & \equiv\min_{z}U_{(1,1)}(\delta_{1},\delta_{2}(1,\delta_{1});z)+\min_{z}U_{(0,1)}(\delta_{1},\delta_{2}(0,\delta_{1});z).\label{eq:U_delta}
\end{align}
For example, for the fourth regime in Table \ref{tab:regimes},
\begin{align*}
\mathbb{P}[Y_{2}(\bd_{(4)}(\cdot))=1] & =\sum_{y_{1}\in\{0,1\}}\mathbb{P}[Y_{1}(\delta_{1})=y_{1},Y_{2}(\delta_{1},\delta_{2}(y_{1},\delta_{1}))=1]\\
 & =\mathbb{P}[Y_{1}(1)=1,Y_{2}(1,\delta_{2}(1,1))=1]+\mathbb{P}[Y_{1}(1)=0,Y_{2}(1,\delta_{2}(0,1))=1]\\
 & =\mathbb{P}[Y_{1}(1)=1,Y_{2}(1,1)=1]+\mathbb{P}[Y_{1}(1)=0,Y_{2}(1,0)=1]
\end{align*}
is bounded by
\begin{align*}
L(\bd_{(4)}) & \equiv\max_{z}L_{(1,1)}(1,1;z)+\max_{z}L_{(0,1)}(1,0;z),\\
U(\bd_{(4)}) & \equiv\min_{z}U_{(1,1)}(1,1;z)+\min_{z}U_{(0,1)}(1,0;z).
\end{align*}
Since $\max_{z}L(z)+\max_{z}\tilde{L}(z)=\max_{z,\tilde{z}}\left\{ L(z)+\tilde{L}(\tilde{z})\right\} $
for any functions $L$ and $\tilde{L}$, we can express \eqref{eq:L_delta} as
\begin{align*}
L(\bd) & =\max_{z,\tilde{z}}\left\{ L_{(1,1)}(\delta_{1},\delta_{2}(1,\delta_{1});z)+L_{(0,1)}(\delta_{1},\delta_{2}(0,\delta_{1});\tilde{z})\right\} \\
 & \equiv\max_{z,\tilde{z}}L(\bd;z,\tilde{z}),
\end{align*}
and, analogously, \eqref{eq:U_delta} as $U(\bd)\equiv\min_{z,\tilde{z}}U(\bd;z,\tilde{z})$. Therefore, given \eqref{eq:L_y}--\eqref{eq:U_y}, the bounds satisfy Assumption \ref{ass: welfare bounds}.  Now, consider choosing a single optimal regime that maximizes $L(\bd)$.
Then, for example, we can show that the following event can be characterized
as a polyhedron in the space of $p$
\begin{align*}
\{\bd_{(4)} & =\arg\max_{\bd}L(\bd)\text{ and }L(\bd_{(4)};z^{*},\tilde{z}^{*})\ge L(\bd_{(4)};z,\tilde{z})\text{ for all }z,\tilde{z}\},
\end{align*}
where $(z^{*},\tilde{z}^{*})=\arg\max_{z,\tilde{z}}L(\bd_{(4)};z,\tilde{z})$.
Note that this event is equivalent to $\{L(\bd_{(4)};z^{*},\tilde{z}^{*}) \ge L(\bd;z,\tilde{z})\text{ for all }\bd\text{ and }z,\tilde{z}\}$ or equivalently,
\begin{align}
\{L(\bd;z,\tilde{z})-L(\bd_{(4)};z^{*},\tilde{z}^{*}) & \le0\text{ for all }\bd\text{ and }z,\tilde{z}\}.\label{eq:conditioning_event}
\end{align}
Note that each $L(\bd;z,\tilde{z})$ is a linear combination
of the elements in $p$ as shown in \eqref{eq:L_y} and \eqref{eq:U_y}.
More formally, we can show that \eqref{eq:conditioning_event} is
equivalent to $A_{L}p \le0$ for some $A_{L}$. Note that
\begin{align*}
L(\bd;z,\tilde{z}) & =L_{(1,1)}(\delta_{1},\delta_{2}(1,\delta_{1});z)+L_{(0,1)}(\delta_{1},\delta_{2}(0,\delta_{1});\tilde{z})\\
 & =A^{1}(\bd;z)p+A^{0}(\bd;\tilde{z})p
\end{align*}
for some row vectors $A^{1}$ and $A^{0}$ because, for $\bd=\bd_{(4)}$,
\begin{align*}
L_{(1,1)}(\delta_{1},\delta_{2}(1,\delta_{1});z) & =L_{(1,1)}(1,1;z)=\mathbb{P}[Y=(1,1),D=(1,1)|Z=z]\\
 & =\mathbb{P}[D_{1}=1,Y_{1}=1,D_{2}=1,Y_{2}=1|Z=z],\\
L_{(0,1)}(\delta_{1},\delta_{2}(0,\delta_{1});\tilde{z}) & =L_{(0,1)}(1,0;\tilde{z})=\mathbb{P}[Y=(0,1),D=(1,0)|Z=\tilde{z}]\\
 & =\mathbb{P}[D_{1}=1,Y_{1}=0,D_{2}=0,Y_{2}=1|Z=\tilde{z}].
\end{align*}
An analogous argument can be applied to $U(\bd;z,\tilde{z})$. The sample counterpart of this characterization implies that Assumption \ref{ass: estimated ID'd set} holds in this setting. Also, this characterization facilitates the CI calculations in the following section.

\subsection{Other examples}

In Appendix \ref{sec: additional example}, we list other examples that fit into our inference framework. These examples can be broadly categorized as (i) Manski bounds with a continuous outcome; (ii) bounds for dynamic treatment effects; (iii) bounds derived from linear programming; (iv) optimal treatment assignment with partially identified ATE; and (v)
empirical welfare maximization with observational data. In these examples, parameters of interest are chosen in data-driven manners that correspond to weak polyhedral selection. In particular, we show in the appendix how Assumptions \ref{ass: welfare bounds} and \ref{ass: estimated ID'd set} hold in these examples.

\section{Confidence Interval Construction and Validity}\label{sec: CIs}

We now generalize the CI construction heuristically described in Section \ref{sec:simple example} to apply in the general framework of Section \ref{sec: general framework}, covering all example applications discussed above and in Appendix \ref{sec: additional example}. After introducing a few additional mild assumptions on the DGP, we construct conditional CIs and then move to unconditional CIs.

\subsection{Additional Assumptions on the DGP}

We suppose that the sample of data with $n$ observations is drawn from some unknown distribution $\mathbb{P}\in\mathcal{P}_n$.  As an estimator for $p$, we assume that $\widehat p$ is uniformly asymptotically normal under $\mathbb{P}\in\mathcal{P}_n$.

\begin{assumption} \label{ass: joint normality}
For the class of Lipschitz functions that are bounded in absolute value by one and have Lipschitz constant bounded by one, $BL_1$, there exist functions $p(\mathbb{P})$ and $\Sigma(\mathbb{P})$ such that for $\xi_{\mathbb{P}}\sim\mathcal{N}(0,\Sigma(\mathbb{P}))$ with
\[\lim_{n\rightarrow\infty}\sup_{\mathbb{P}\in \mathcal{P}_n}\sup_{f\in BL_1}\left|{E}_{\mathbb{P}}\left[f\left(\sqrt{n}(\widehat p-p(\mathbb{P}))\right)\right]-{E}_{\mathbb{P}}\left[f\left(
\xi_{\mathbb P}\right)\right]\right|=0.\]
\end{assumption} 

The notation of this assumption makes explicit that the parameter $p$ and the asymptotic variance $\Sigma$ depend upon the unknown distribution of the data $\mathbb{P}$.  It holds naturally for standard estimators $\widehat p$ under random sampling or weak dependence in the presence of bounds on the moments and dependence of the underlying data. This assumption holds in all the examples considered in the previous section and Appendix \ref{sec: additional example}: with appropriate random samples, $\hp$ are sample means. For example, Assumption \ref{ass: joint normality} is satisfied in the example of dynamic treatment regime when the data $\{Y_i,D_i,Z_i\}_{i=1}^n$ form a random sample, since the entries of $\widehat p$ are sample means.

Next, we assume that the asymptotic variance of $\widehat p$ can be uniformly consistently estimated by an estimator $\widehat\Sigma$.

\begin{assumption} \label{ass: variance estimation}
For all $\varepsilon>0$, the estimator $\widehat\Sigma$ satisfies
\[\lim_{n\rightarrow\infty}\sup_{\mathbb{P}\in\mathcal{P}_n} \mathbb{P}\left(\left\|\widehat\Sigma-\Sigma(\mathbb{P})\right\|>\varepsilon\right)=0.\]
\end{assumption}

This assumption is again naturally satisfied when using a standard sample analog estimator of $\Sigma$ under random sampling or weak dependence in the presence of moment and dependence bounds.  


When a raw covariance estimator is not positive semidefinite, we replace it by its projection onto the cone of positive-semidefinite matrices and continue to denote the resulting estimator by $\widehat\Sigma$.  This replacement does not affect Assumption \ref{ass: variance estimation}: projection onto the positive-semidefinite cone is nonexpansive under the Frobenius norm, and all matrix norms are equivalent in the fixed-dimensional setting considered here.

We make a final assumption imposing directional nondegeneracy on the covariance matrix $\Sigma$ uniformly across $\mathcal{P}_n$. 

\begin{assumption}\label{ass: variance restriction}
For some finite $\bar\lambda>0$ and every $\mathbb P\in\mathcal P_n$:
\begin{enumerate}
\item $\Sigma(\mathbb P)$ is positive semidefinite and $\lambda_{\max}\{\Sigma(\mathbb P)\}\leq\bar\lambda$.
\item For every $d\in\{d^0,\ldots,d^K\}$, $\ell_{d,j}\Sigma(\mathbb P)\ell_{d,j}'\geq \|\ell_{d,j}\|^2/\bar\lambda$ for $j=1,\ldots,J_L$ and $u_{d,j}\Sigma(\mathbb P)u_{d,j}'\geq \|u_{d,j}\|^2/\bar\lambda$ for $j=1,\ldots,J_U$.
\item For $M=L,U$, every $d\in\{d^0,\ldots,d^K\}$, every $j\in\{1,\ldots,J_M\}$, every $\gamma$ in the support of $\widehat\gamma_M(d)$, and every nonzero row $v$ of $A^M(d,j,\gamma)$, $v\Sigma(\mathbb P)v'\geq \|v\|^2/\bar\lambda$.
\end{enumerate}
\end{assumption}

Part 2 ensures that every scalar affine statistic used to estimate a lower or upper endpoint has variance uniformly bounded away from zero.  Part 3 rules out degeneracy of comparisons that determine the binding component or a potentially active polyhedral selection inequality.  Identically zero or redundant rows should be deleted from $A^M(d,j,\gamma)$ before applying the assumption.  No lower bound is imposed on the eigenvalues of $\Sigma(\mathbb P)$ in directions orthogonal to these collections.  The assumption therefore permits the full covariance matrix to be singular, including when some reduced-form probabilities are zero or when the vector $\widehat p$ contains exact linear redundancies, provided that the singular directions do not determine an endpoint or a potentially active selection boundary.  Moreover, this assumption immediately holds if the eigenvalues of $\Sigma(\mathbb P)$ are uniformly bounded above and away from zero.  For projection inference, only Parts 1 and 2 are imposed.  The conditional and hybrid procedures additionally use Part 3 to control the behavior of the truncation limits.

Part 2 of Assumption \ref{ass: variance restriction} can be relaxed when an affine component has zero asymptotic variance but is uniformly separated from the corresponding population endpoint.  Intuitively, such a component is asymptotically irrelevant for the maximum or minimum defining the endpoint because the probability that it is selected as binding converges uniformly to zero.  Appendix \ref{sec: degenerate affine endpoint components} formalizes this relaxation and modifies the CI constructions only by omitting components whose estimated variances fall below a vanishing threshold from the studentized maxima used for projection inference and, if such a component is nevertheless selected, setting the corresponding conditional, projection, and hybrid bound equal to its affine estimate.  This situation can arise when reduced-form probabilities are exactly zero or one, making some empirical probabilities and affine endpoint components degenerate, as in the Balke--Pearl calibration used in Section \ref{sec: Monte Carlo}.

\subsection{Conditional Confidence Intervals}\label{sec: conditional CIs}

We first generalize the conditional CI construction described in Section \ref{sec:simple example cond CIs}.  As in Section \ref{sec:simple example cond CIs}, we are interested in forming lower and upper confidence bounds $\widehat L(\widehat d)_{\alpha}^C$ and $\widehat U(\widehat d)_{1-\alpha}^C$ that satisfy \eqref{eq: cond cov requirement} and \eqref{eq: cond cov requirement ub} for all $d\in\cD\equiv\{d^0,\ldots,d^K\}$ as endpoints in the formation of a conditionally valid CI. This is because the researcher's interest in inference on option $d$ only arises when it is a member of the policy set $\cD(\hp)$.  

To begin, we characterize the conditional distributions of $\widehat{L}(d)$ given the event $\{d\in\cD(\hp)$, $\widehat j_L(d)=j_L^*$ and $\widehat\gamma_L(d)=\gamma_L^*\}$ characterized by Assumption \ref{ass: estimated ID'd set}.  These conditional distributions depend upon the nuisance parameter $p$.  As a first step, we form sufficient statistics for $p$ that are asymptotically independent of $\widehat{L}(d)$ given $\widehat j_L(d)=j_L^*$ and $\widehat{U}(d)$ given $\widehat j_U(d)=j_U^*$. 
 Since $\widehat{L}(d)=\tilde \ell_{d,j_L^*}+\ell_{d,j_L^*}\widehat p$ given $\widehat j_L(d)=j_L^*$ by Assumption \ref{ass: welfare bounds} and \eqref{eq:j_hats} (and similarly for $\widehat{U}(d)$), such sufficient statistics can be constructed as
\[\widehat{\mathcal{Z}}_{L}(d,j_L^*)\equiv \sqrt{n}\widehat p-\widehat b_{L}(d,j_L^*)\sqrt{n}(\tilde \ell_{d,j_L^*}+\ell_{d,j_L^*}\widehat p)\] 
and 
\[\widehat{\mathcal{Z}}_{U}(d,j_U^*)\equiv \sqrt{n}\widehat p-\widehat b_{U}(d,j_U^*)\sqrt{n}(\tilde u_{d,j_U^*}+u_{d,j_U^*}\widehat p)\]
with
\[\widehat b_{L}(d,j_L^*)\equiv \widehat\Sigma\ell_{d,j_L^*}^{\prime}\left(\ell_{d,j_L^*}\widehat\Sigma\ell_{d,j_L^*}^{\prime}\right)^{-1} \quad \text{and} \quad \widehat b_{U}(d,j_U^*)\equiv \widehat\Sigma u_{d,j_U^*}^{\prime}\left(u_{d,j_U^*}\widehat\Sigma u_{d,j_U^*}^{\prime}\right)^{-1},\]
by the asymptotic normality of $\widehat p$ and asymptotic absence of correlation between $\sqrt{n}\widehat p$ and $\widehat{\mathcal{Z}}_{L}(d,j_L^*)$ and $\widehat{\mathcal{Z}}_{U}(d,j_U^*)$. Next, Assumption \ref{ass: estimated ID'd set} characterizes the conditioning events $\{d\in\cD(\hp),\widehat j_L(d)=j_L^*,\widehat \gamma_L(d)=\gamma_L^*\}$ and $\{d\in\cD(\hp),\widehat j_U(d)=j_U^*,\widehat \gamma_U(d)=\gamma_U^*\}$ as polyhedra in $\widehat p$, which can in turn be expressed as intervals for $\tilde \ell_{d,j_L^*}+\ell_{d,j_L^*}\widehat p$ and $\tilde u_{d,j_U^*}+u_{d,j_U^*}\widehat p$. Then, again since $\widehat{L}(d)=\tilde \ell_{d,j_L^*}+\ell_{d,j_L^*}\widehat p$ given $\widehat j_L(d)=j_L^*$ by Assumption \ref{ass: welfare bounds}, Lemma 1 of \cite{McC24} implies
\begin{gather}
\left.\sqrt{n}\widehat{L}(d)\right|\left\{d\in\cD(\hp),\widehat j_L(d)=j_L^*,\widehat \gamma_L(d)=\gamma_L^*\right\} \notag \\
\sim \left.\sqrt{n}(\tilde \ell_{d,j_L^*}+\ell_{d,j_L^*}\widehat p)\right|\left\{\widehat{\mathcal{V}}_{L}^-\left(\widehat{\mathcal{Z}}_{L}(d,j_L^*),d,j_L^*,\gamma_L^*\right)\leq \sqrt{n}(\tilde \ell_{d,j_L^*}+\ell_{d,j_L^*}\widehat p)\leq \widehat{\mathcal{V}}_{L}^+\left(\widehat{\mathcal{Z}}_{L}(d,j_L^*),d,j_L^*,\gamma_L^*\right),\right. \notag \\
\left.\widehat{\mathcal{V}}_{L}^0\left(\widehat{\mathcal{Z}}_{L}(d,j_L^*),d,j_L^*,\gamma_L^*\right)\geq 0\right\} \label{eq: truncated lb}
\end{gather}
and
\begin{gather*}
\left.\sqrt{n}\widehat{U}(d)\right|\left\{d\in\cD(\hp),\widehat j_U(d)=j_U^*,\widehat \gamma_U(d)=\gamma_U^*\right\} \\
\sim \left.\sqrt{n}(\tilde u_{d,j_U^*}+u_{d,j_U^*}\widehat p)\right|\left\{\widehat{\mathcal{V}}_{U}^-\left(\widehat{\mathcal{Z}}_{U}(d,j_U^*),d,j_U^*,\gamma_U^*\right)\leq \sqrt{n}(\tilde u_{d,j_U^*}+u_{d,j_U^*}\widehat p)\leq \widehat{\mathcal{V}}_{U}^+\left(\widehat{\mathcal{Z}}_{U}(d,j_U^*),d,j_U^*,\gamma_U^*\right),\right. \\
\left.\widehat{\mathcal{V}}_{U}^0\left(\widehat{\mathcal{Z}}_{U}(d,j_U^*),d,j_U^*,\gamma_U^*\right)\geq 0\right\}
\end{gather*}
with
\begin{gather*}
\widehat{\mathcal{V}}_{M}^-(z,d,j,\gamma)\equiv\max_{k:(A^M(d,j,\gamma)\widehat b_{M}(d,j))_k<0}\frac{\sqrt{n}(c^M(d,j,\gamma))_k-(A^M(d,j,\gamma)z)_k}{(A^M(d,j,\gamma)\widehat b_{M}(d,j))_k}, \\
\widehat{\mathcal{V}}_{M}^+(z,d,j,\gamma)\equiv\min_{k:(A^M(d,j,\gamma)\widehat b_{M}(d,j))_k>0}\frac{\sqrt{n}(c^M(d,j,\gamma))_k-(A^M(d,j,\gamma)z)_k}{(A^M(d,j,\gamma)\widehat b_{M}(d,j))_k}, \\
\widehat{\mathcal{V}}_{M}^0(z,d,j,\gamma)\equiv\min_{k:(A^M(d,j,\gamma)\widehat b_{M}(d,j))_k=0}\sqrt{n}(c^M(d,j,\gamma))_k-(A^M(d,j,\gamma)z)_k
\end{gather*}
for $M=L,U$.

Now, under Assumptions \ref{ass: joint normality} and \ref{ass: variance estimation}, the distribution of $\sqrt{n}(\tilde \ell_{d,j_L^*}+\ell_{d,j_L^*}\widehat p)$ can be approximated by a $\mathcal{N}(\sqrt{n}(\tilde \ell_{d,j_L^*}+\ell_{d,j_L^*} p),\ell_{d,j_L^*}\widehat\Sigma \ell_{d,j_L^*}')$-distributed random variable that is asymptotically independent of $\widehat{\mathcal{Z}}_{L}(d,j_L^*)$.  Using the distributional characterization in \eqref{eq: truncated lb}, we can therefore use the corresponding truncated normal cumulative distribution function to produce quantile-unbiased estimators of the underlying mean $\sqrt{n}(\tilde \ell_{d,j_L^*}+\ell_{d,j_L^*} p)$.  Let $F_{TN}(\cdot;\mu,\sigma^2|\mathcal V^-,\mathcal V^+)$ denote the truncated normal cumulative distribution function for an underlying normally-distributed random variable with mean $\mu$ and variance $\sigma^2$ that is truncated to lie between $\mathcal V^-$ and $\mathcal V^+$. For $\alpha\in(0,1)$, define $\widehat L(d)_{\alpha}^C$ to solve
\begin{gather*}
F_{TN}\left(\sqrt{n}(\tilde \ell_{d,\widehat j_L(d)}+\ell_{d,\widehat j_L(d)}\widehat p);\mu,\ell_{d,\widehat j_L(d)}\widehat\Sigma\ell_{d,\widehat j_L(d)}^{\prime}\right|\widehat{\mathcal{V}}_{L}^-\left(\widehat{\mathcal{Z}}_{L}(d,\widehat j_L(d)),d,\widehat j_L(d),\widehat \gamma_L(d)\right), \\
\left.\widehat{\mathcal{V}}_{L}^+\left(\widehat{\mathcal{Z}}_{L}(d,\widehat j_L(d)),d,\widehat j_L(d),\widehat \gamma_L(d)\right)\right)=1-\alpha
\end{gather*}
in $\mu$.  Similarly, define $\widehat U(d)_{\alpha}^C$ to solve
\begin{gather*}
F_{TN}\left(\sqrt{n}(\tilde u_{d,\widehat j_U(d)}+u_{d,\widehat j_U(d)}\widehat p);\mu,u_{d,\widehat j_U(d)}\widehat\Sigma u_{d,\widehat j_U(d)}^{\prime}\right|\widehat{\mathcal{V}}_{U}^-\left(\widehat{\mathcal{Z}}_{U}(d,\widehat j_U(d)),d,\widehat j_U(d),\widehat \gamma_U(d)\right), \\
\left.\widehat{\mathcal{V}}_{U}^+\left(\widehat{\mathcal{Z}}_{U}(d,\widehat j_U(d)),d,\widehat j_U(d),\widehat \gamma_U(d)\right)\right)=1-\alpha
\end{gather*}
in $\mu$.  Then, results in \cite{Pfa94} imply that $\widehat L(d)_{\alpha}^C$ and $\widehat U(d)_{\alpha}^C$ are optimal $\alpha$ quantile-unbiased estimators of $\sqrt{n}(\tilde \ell_{d,j_L^*}+\ell_{d,j_L^*} p)$ and $\sqrt{n}(\tilde u_{d,j_U^*}+u_{d,j_U^*} p)$ asymptotically.

Finally, combine these quantile-unbiased estimators to form a conditional CI for the identified interval $[L(d),U(d)]$:~
\begin{equation}
(n^{-1/2}\widehat L(d)_{\alpha_1}^C,n^{-1/2}\widehat U(d)_{1-\alpha_2}^C). \label{eq: cond CI}
\end{equation}
We establish the conditional uniform asymptotic validity of this CI.

\begin{theorem} \label{thm: cond'l coverage for cond'l bounds}
Suppose Assumptions \ref{ass: welfare bounds}, \ref{ass: estimated ID'd set} and \ref{ass: joint normality}--\ref{ass: variance restriction} hold.  Then, for any $d\in\{d^0,\ldots,d^K\}$ and $0<\alpha_1,\alpha_2<1/2$,
\begin{gather*}
\liminf_{n\rightarrow\infty}\inf_{\mathbb{P}\in\mathcal{P}_n}\left\{\left[\mathbb{P}\left(\left.[L(d),U(d)]\subseteq \left(n^{-1/2}\widehat L(d)_{\alpha_1}^C,n^{-1/2}\widehat U(d)_{1-\alpha_2}^C\right)\right|d\in\cD(\hp)\right)-(1-\alpha_1-\alpha_2)\right]\cdot\mathbb{P}(d\in\cD(\hp))\right\}\geq 0.
\end{gather*}
\end{theorem}

\subsection{Unconditional Confidence Intervals}\label{sec: unconditional CIs}

In parallel with the previous subsection, we now generalize the unconditional CI constructions described in Section \ref{sec:simple example uncond CIs} to the general framework of Section \ref{sec: general framework}.  While conditional inference is well-defined for any fixed $d$ after conditioning on $d\in\cD(\hp)$ even when $\cD(\hp)$ contains multiple policies,  unconditional inference instead averages over the random identity of the selected object.  It therefore requires the researcher to specify which policy is ultimately selected.  Assumption \ref{ass: selection rule} formalizes this requirement by defining a unique final recommendation $\cD(\hp)=\{\widehat d\}$.

As mentioned above, if only unconditional coverage of $[L(\widehat d),U(\widehat d)]$ is desired, the conditional CI \eqref{eq: cond CI} with $d=\widehat d$ can be unnecessarily long.  This length problem can be understood directly from the truncated-normal distribution.  Conditional on the sufficient statistic and selection labels, the endpoint estimates are normal truncated to an interval $[a,b]$.  The truncated-normal distribution function divides the probability mass between $a$ and the observation by the probability of the full truncation event.  When selection is rare, the observed statistic often lies in a far tail and only slightly inside a truncation boundary.  Conditional on selection, crossing that boundary is expected and is therefore no longer evidence of a large underlying mean.  For example, if $T\sim N(\mu,\sigma^2)$ is observed conditional on $T\geq a$ and $T=a+\delta$, inversion of the conditional distribution function implies approximately $\mu=a+\sigma^2\log(1-c)/\delta+\delta/2$ at conditional quantile $c$.  This value diverges to $-\infty$ as $\delta\downarrow0$.  Hence, a winner that barely exceeds the runner-up, or a policy that barely satisfies its selection inequalities, can generate a very long conditional CI.  To avoid this potential extreme length problem, we instead describe two different methods---projection and hybrid methods---to form the unconditional confidence bounds that constitute the endpoints of these unconditional CIs in this general framework.

The general formation of the confidence bounds based upon projecting simultaneous confidence bounds for all possible values of $\sqrt{n}L(\widehat d)$ and $\sqrt{n}U(\widehat d)$ proceeds by computing $\widehat{c}_{1-\alpha,M}$, the $1-\alpha$ quantile of $\max_{i\in \{1,\ldots,(K+1)J_M\}}\widehat \zeta_{M,i}/\sqrt{\widehat{\Sigma}_{M,i}}$, where $\widehat \zeta_M\sim\mathcal{N}(0,\widehat{\Sigma}_M)$ for $M=L,U$ with $\widehat{\Sigma}_L=\ell^{mat}\widehat{\Sigma}\ell^{mat\prime}$, $\widehat{\Sigma}_U=u^{mat}\widehat{\Sigma}u^{mat\prime}$, $\ell^{mat}=(\ell_{0,1}',\hdots,\ell_{0,J_L}',\hdots,\ell_{K,1}',\hdots,\ell_{K,J_L}')'$ and $u^{mat}=(u_{0,1}',\hdots,u_{0,J_U}',\hdots,u_{K,1}',\hdots,u_{K,J_U}')'$, recalling that $\Upsilon_i$ denotes the $i^{th}$ element of the main diagonal of any square matrix $\Upsilon$.  Here, the maximum is taken to guarantee simultaneous coverage.  The lower level $1-\alpha$ projection confidence bound for $\sqrt{n}L(\widehat d)$ is $\widehat L(\widehat d)_\alpha^P=\sqrt{n}\widehat L(\widehat d)-\widehat{c}_{1-\alpha,L}\sqrt{\widehat{\Sigma}_{L,\widehat dJ_L+\widehat{j}_L(\widehat d)}}$ and the upper level $1-\alpha$ projection confidence bound for $\sqrt{n}U(\widehat d)$ is $\widehat{U}_{1-\alpha}^P(\widehat d)=\sqrt{n}\widehat U(\widehat d)+\widehat{c}_{1-\alpha,U}\sqrt{\widehat{\Sigma}_{U,\widehat dJ_U+\widehat{j}_U(\widehat d)}}$, because e.g.,~$\sqrt{n}L(\widehat d)$ can take value equal to any entry of the vector $\sqrt{n}(\tilde\ell_{0,1},\hdots,\tilde\ell_{0,J_L},\hdots,\tilde\ell_{K,1},\hdots,\tilde\ell_{K,J_L})'+\sqrt{n}\ell^{mat} p$, $\sqrt{n}(\ell^{mat}\widehat p-\ell^{mat}p)$ is asymptotically distributed $\mathcal{N}(0,\Sigma_L)$ for $\Sigma_L=\ell^{mat}\Sigma\ell^{mat\prime}$ by Assumption \ref{ass: joint normality} and $\widehat\Sigma_L$ is consistent for $\Sigma_L$ by Assumption \ref{ass: variance estimation}.

Combining these two confidence bounds at appropriate levels yields an unconditional CI for the identified interval $[L(\widehat d),U(\widehat d)]$ of the selected $\widehat d$,
\begin{equation}
(n^{-1/2}\widehat L(\widehat d)_{\alpha_1}^P,n^{-1/2}\widehat U(\widehat d)_{1-\alpha_2}^P), \label{eq: proj CI}
\end{equation}
with uniformly correct asymptotic coverage, regardless of how $\widehat d$ is selected from the data.

\begin{theorem} \label{thm: uncond'l coverage for proj bounds}
Suppose Assumptions \ref{ass: welfare bounds}, \ref{ass: joint normality}, \ref{ass: variance estimation} and Parts 1--2 of Assumption \ref{ass: variance restriction} hold.  Then, for any (random) $\widehat{d}\in\{d^0,\ldots,d^K\}$ and $0<\alpha_1,\alpha_2<1/2$,
\begin{equation*}
\liminf_{n\rightarrow\infty}\inf_{\mathbb{P}\in\mathcal{P}_n}\mathbb{P}\left([L(\widehat d),U(\widehat d)]\subseteq \left(n^{-1/2}\widehat L(\widehat d)_{\alpha_1}^P,n^{-1/2}\widehat U(\widehat d)_{1-\alpha_2}^P\right)\right)\geq 1-\alpha_1-\alpha_2.
\end{equation*}
\end{theorem}

Note that the projection CI \eqref{eq: proj CI} has the benefit of correct coverage regardless of how $\widehat d$ is chosen from the data.  In this sense, it is more robust than the other CIs we propose in this paper.  On the other hand, by using the common selection structure of Assumption \ref{ass: estimated ID'd set}, we are able to produce a hybrid CI that combines the strengths of the conditional CI \eqref{eq: cond CI} and the projection CI \eqref{eq: proj CI} which, as described in Section \ref{sec:simple example uncond CIs}, are shorter under complementary scenarios.

In analogy with the construction of the conditional CIs, to construct the hybrid CIs we begin by characterizing the conditional distributions of $\widehat L(\widehat d)$ and $\widehat U(\widehat d)$ but now adding an additional component to the conditioning events.  More specifically, under Assumptions \ref{ass: estimated ID'd set} and \ref{ass: selection rule}, by intersecting the events 
\begin{gather*}
\left\{\widehat d=d^*,\widehat j_L(\widehat d)=j_L^*,\widehat \gamma_L(\widehat d)=\gamma_L^*\right\}=\left\{d^*\in\cD(\hp),\widehat j_L(d^*)=j_L^*,\widehat \gamma_L(d^*)=\gamma_L^*\right\} \\
=\left\{\widehat{\mathcal{V}}_{L}^-\left(\widehat{\mathcal{Z}}_{L}(d^*,j_L^*),d^*,j_L^*,\gamma_L^*\right)\leq \sqrt{n}(\tilde \ell_{d^*,j_L^*}+\ell_{d^*,j_L^*}\widehat p)\leq \widehat{\mathcal{V}}_{L}^+\left(\widehat{\mathcal{Z}}_{L}(d^*,j_L^*),d^*,j_L^*,\gamma_L^*\right),\right. \\
\left.\widehat{\mathcal{V}}_{L}^0\left(\widehat{\mathcal{Z}}_{L}(d^*,j_L^*),d^*,j_L^*,\gamma_L^*\right)\geq 0\right\}
\end{gather*}
and 
\begin{gather*}
\left\{\sqrt{n}(\tilde \ell_{\widehat d,\widehat j_L(\widehat d)}+\ell_{\widehat d,\widehat j_L(\widehat d)} p)\geq \widehat{L}_\beta^P(\widehat d)\right\} \\
=\left\{\sqrt{n}(\tilde \ell_{\widehat d,\widehat j_L(\widehat d)}+\ell_{\widehat d,\widehat j_L(\widehat d)}\widehat p)\leq \sqrt{n}(\tilde \ell_{\widehat d,\widehat j_L(\widehat d)}+\ell_{\widehat d,\widehat j_L(\widehat d)} p)+\widehat{c}_{1-\beta,L}\sqrt{\widehat{\Sigma}_{L,\widehat dJ_L+\widehat{j}_L(\widehat d)}} \right\}
\end{gather*}
for some $0<\beta<\alpha<1$, we have
\begin{gather*}
\left.\sqrt{n}\widehat{L}(\widehat d)\right|\left\{\widehat d=d^*,\widehat j_L(\widehat d)=j_L^*,\widehat \gamma_L(\widehat d)=\gamma_L^*,\sqrt{n}(\tilde \ell_{\widehat d,\widehat j_L(\widehat d)}+\ell_{\widehat d,\widehat j_L(\widehat d)} p)\geq \widehat{L}_\beta^P(\widehat d)\right\} \\
\sim \left.\sqrt{n}(\tilde \ell_{d^*,j_L^*}+\ell_{d^*,j_L^*}\widehat p)\right|\left\{\widehat{\mathcal{V}}_{L}^-\left(\widehat{\mathcal{Z}}_{L}(d^*,j_L^*),d^*,j_L^*,\gamma_L^*\right)\leq \sqrt{n}(\tilde \ell_{d^*,j_L^*}+\ell_{d^*,j_L^*}\widehat p)\right. \\
\left.\leq \widehat{\mathcal{V}}_{L}^{+,H}\left(\widehat{\mathcal{Z}}_{L}(d^*,j_L^*),d^*,j_L^*,\gamma_L^*,\sqrt{n}(\tilde \ell_{d^*,j_L^*}+\ell_{d^*,j_L^*} p)\right),\widehat{\mathcal{V}}_{L}^0\left(\widehat{\mathcal{Z}}_{L}(d^*,j_L^*),d^*,j_L^*,\gamma_L^*\right)\geq 0\right\},
\end{gather*}
where
\[\widehat{\mathcal{V}}_{L}^{+,H}(z,d,j,\gamma,\mu)\equiv\min\left\{\widehat{\mathcal{V}}_{L}^{+}(z,d,j,\gamma),\mu+\widehat{c}_{1-\beta,L}\sqrt{\widehat{\Sigma}_{L,dJ_L+j}}\right\}.\]
Similarly, 
\begin{gather*}
\left.\sqrt{n}\widehat U(\widehat d)\right|\left\{\widehat d=d^*,\widehat j_U(\widehat d)=j_U^*,\widehat \gamma_U(\widehat d)=\gamma_U^*,\sqrt{n}(\tilde u_{\widehat d,\widehat j_U(\widehat d)}+u_{\widehat d,\widehat j_U(\widehat d)} p)\leq \widehat{U}_{1-\beta}^P(\widehat d)\right\} \\
\sim \left.\sqrt{n}(\tilde u_{d^*,j_U^*}+u_{d^*,j_U^*}\widehat p)\right|\left\{\widehat{\mathcal{V}}_{U}^{-,H}\left(\widehat{\mathcal{Z}}_{U}(d^*,j_U^*),d^*,j_U^*,\gamma_U^*,\sqrt{n}(\tilde u_{d^*,j_U^*}+u_{d^*,j_U^*} p)\right)\right. \\
\left.\leq \sqrt{n}(\tilde u_{d^*,j_U^*}+u_{d^*,j_U^*}\widehat p)\leq \widehat{\mathcal{V}}_{U}^{+}\left(\widehat{\mathcal{Z}}_{U}(d^*,j_U^*),d^*,j_U^*,\gamma_U^*\right),\widehat{\mathcal{V}}_{U}^0\left(\widehat{\mathcal{Z}}_{U}(d^*,j_U^*),d^*,j_U^*,\gamma_U^*\right)\geq 0\right\},
\end{gather*}
where
\[\widehat{\mathcal{V}}_{U}^{-,H}(z,d,j,\gamma,\mu)\equiv\max\left\{\widehat{\mathcal{V}}_{U}^{-}(z,d,j,\gamma),\mu-\widehat{c}_{1-\beta,U}\sqrt{\widehat{\Sigma}_{U,dJ_U+j}}\right\}.\]
Since the distribution of $\sqrt{n}(\tilde \ell_{d^*,j_L^*}+\ell_{d^*,j_L^*}\widehat p)$ can be approximated by $\mathcal{N}(\sqrt{n}(\tilde \ell_{d^*,j_L^*}+\ell_{d^*,j_L^*} p),\ell_{d^*,j_L^*}\Sigma \ell_{d^*,j_L^*}')$ asymptotically and $\widehat{\mathcal{Z}}_{L}(d^*,j_L^*)$ is asymptotically independent, we again work with the truncated normal distribution to compute a hybrid lower confidence bound for $\sqrt{n}L(\widehat d)$: for $0<\beta<\alpha<1$, define $\widehat L(d)_{\alpha}^H$ to solve
\begin{gather*}
F_{TN}\left(\sqrt{n}(\tilde \ell_{d,\widehat j_L(d)}+\ell_{d,\widehat j_L(d)}\widehat p);\mu,\ell_{d,\widehat j_L(d)}\widehat\Sigma\ell_{d,\widehat j_L(d)}^{\prime}\right|\widehat{\mathcal{V}}_{L}^-\left(\widehat{\mathcal{Z}}_{L}(d,\widehat j_L(d)),d,\widehat j_L(d),\widehat \gamma_L(d)\right), \\
\left.\widehat{\mathcal{V}}_{L}^{+,H}\left(\widehat{\mathcal{Z}}_{L}(d,\widehat j_L(d)),d,\widehat j_L(d),\widehat \gamma_L(d),\mu\right)\right)=\frac{1-\alpha}{1-\beta}
\end{gather*}
in $\mu$.  Similarly, define $\widehat U(d)_{\alpha}^H$ to solve
\begin{gather*}
F_{TN}\left(\sqrt{n}(\tilde u_{d,\widehat j_U(d)}+u_{d,\widehat j_U(d)}\widehat p);\mu,u_{d,\widehat j_U(d)}\widehat\Sigma u_{d,\widehat j_U(d)}^{\prime}\right|\widehat{\mathcal{V}}_{U}^{-,H}\left(\widehat{\mathcal{Z}}_{U}(d,\widehat j_U(d)),d,\widehat j_U(d),\widehat \gamma_U(d),\mu\right), \\
\left.\widehat{\mathcal{V}}_{U}^+\left(\widehat{\mathcal{Z}}_{U}(d,\widehat j_U(d)),d,\widehat j_U(d),\widehat \gamma_U(d)\right)\right)=\frac{1-\alpha}{1-\beta}
\end{gather*}
in $\mu$.  

Here, $\widehat L(\widehat d)_{\alpha}^H$ is an unconditionally valid lower confidence bound for $L(\widehat d)$ and $\widehat U(\widehat d)_{1-\alpha}^H$ is an unconditionally valid upper confidence bound for $U(\widehat d)$.  Combining these two confidence bounds at appropriate levels yields an unconditional CI for the identified interval $[L(\widehat d),U(\widehat d)]$ of the selected $\widehat d$,
\begin{equation*}
(n^{-1/2}\widehat L(\widehat d)_{\alpha_1}^H,n^{-1/2}\widehat U(\widehat d)_{1-\alpha_2}^H),
\end{equation*}
with uniformly correct asymptotic coverage when $\widehat d$ is selected from the data according to Assumptions \ref{ass: estimated ID'd set} and \ref{ass: selection rule}.

\begin{theorem} \label{thm: uncond'l coverage for hybrid bounds}
Suppose Assumptions \ref{ass: welfare bounds}--\ref{ass: selection rule} and \ref{ass: joint normality}--\ref{ass: variance restriction} hold.  Then, for any $0<\alpha_1,\alpha_2<1/2$,
\begin{equation*}
\liminf_{n\rightarrow\infty}\inf_{\mathbb{P}\in\mathcal{P}_n}\mathbb{P}\left([L(\widehat d),U(\widehat d)]\subseteq \left(n^{-1/2}\widehat L(\widehat d)_{\alpha_1}^H,n^{-1/2}\widehat U(\widehat d)_{1-\alpha_2}^H\right)\right)\geq 1-\alpha_1-\alpha_2.
\end{equation*}
\end{theorem}

\section{Finite Sample Performance of Confidence Intervals}\label{sec: Monte Carlo}

Moving now to a context in which the object of interest is selected from the data, we compare the finite sample performance of our conditional, projection and hybrid CIs again in the setting of the Manski bounds example of Section \ref{sec:simple example}.  In this case, we are interested in inference on the average counterfactual outcome $W(\widehat d)$ for $W(d)=\mathbb{E}[Y(d)]$, where interest arises either in the average counterfactual outcome for treatment ($\widehat d=1$) or control ($\widehat d=0$) depending upon which has the largest estimated lower bound:~$\widehat d=\argmax_{d\in\{0,1\}}\widehat L(d)$.  This form of $\widehat d$ corresponds to Example \ref{ex:maximin}, treated in case 1.~of Proposition \ref{prop: selection rule}, which we use to specify $\widehat\gamma_L(\widehat d)$ and $\widehat\gamma_U(\widehat d)$ in the construction of the conditional and hybrid CIs.  We report analogous simulation results for the dynamic treatment regime example in Appendix \ref{sec: additional simulation} with $\widehat p$ generated from a multinomial, rather than normal, distribution.  In addition, we conduct a power comparison of the hypothesis tests associated with our CIs to one of the leading methods for inference on a partially-identified parameter in Appendix \ref{sec: power comp} in the Example \ref{ex:predetermined} setting of inference on an a priori chosen policy.  Encouragingly, the power of the test implied by our hybrid CI is quite competitive in this environment.

We strengthen the mean independence assumption $\mathbb{E}[Y(d)|Z]=\mathbb{E}[Y(d)]$ to full statistical independence $(Y(1),Y(0))\perp Z$ and use the sharp bounds on the ATE $W(1)-W(0)=\mathbb{E}[Y(1)]-\mathbb{E}[Y(0)]$ derived by \cite{BP97,BP13}:
\begin{equation*}
L=\max\left\{\begin{array}{c}
p^{111}+p^{000}-1 \\
p^{110}+p^{001}-1 \\
p^{110}-p^{111}-p^{101}-p^{010}-p^{100} \\
p^{111}-p^{110}-p^{100}-p^{011}-p^{101} \\
-p^{011}-p^{101} \\
-p^{010}-p^{100} \\
p^{001}-p^{011}-p^{101}-p^{010}-p^{000} \\
p^{000}-p^{010}-p^{100}-p^{011}-p^{001} 
\end{array}\right\}
\end{equation*}
and
\begin{equation*}
U=\min\left\{\begin{array}{c}
1-p^{011}-p^{100} \\
1-p^{010}-p^{101}\\
-p^{010}+p^{011}+p^{001}+p^{110}+p^{000} \\
-p^{011}+p^{111}+p^{001}+p^{010}+p^{000} \\
p^{111}+p^{001} \\
p^{110}+p^{000} \\
-p^{101}+p^{111}+p^{001}+p^{110}+p^{100} \\
-p^{100}+p^{110}+p^{000}+p^{111}+p^{101}
\end{array}\right\}.
\end{equation*}
For a sample size of $n=100$, we generated $\widehat p$ from a $\mathcal{N}(p,\Sigma)$ distribution.\footnote{Note that in this problem, the value of $\Sigma$ is implied by the value of $p$.} We examine three different DGPs within the framework of this example.  The DGP corresponding to $p=(.081,0,0,.073,.139,.473)'$ is calibrated to the probabilities estimated by \cite{BP13} in the context of a treatment for high-cholesterol (specifically, by the drug cholestyramine).  The DGP corresponding to $p=(.25,.25,.25,.25,.25,.25)'$ generates completely uninformative bounds for the ATE.  And the DGP corresponding to $p=(.01,.44,.01,.01,.01,.54)'$ generates quite informative bounds.

For the Balke and Pearl calibration, the two zero values in $p$ make the second lower-endpoint component and the second upper-endpoint component used in this selected-welfare exercise have zero variance.  A mechanical implementation of the formulas in Section \ref{sec: CIs} would therefore involve division by zero when studentizing these components for projection inference or when constructing the sufficient statistic and truncation limits after a degenerate component is selected.  We instead use the degenerate-endpoint-robust versions of the conditional, projection, and hybrid procedures described in Appendix \ref{sec: degenerate affine endpoint components}: components classified as degenerate are omitted from the studentized maxima used to compute the projection critical values and, if a degenerate component is selected, the corresponding  bound is set equal to its affine estimate.  At this calibration, the degenerate components are uniformly separated from the corresponding population endpoints, so Theorem \ref{thm: degenerate endpoint extension} implies that the modified procedures retain the asymptotic coverage guarantees established in Section \ref{sec: CIs}.

We compute the unconditional coverage frequencies of the conditional, projection and hybrid CIs. For comparison, we also report the coverage of the conventional CI that applies
normal critical values to the estimated endpoints while treating
the selected policy $\widehat d$ and the estimated binding indices
$\widehat j_L(\widehat d)$ and $\widehat j_U(\widehat d)$ as if they were fixed
in advance.  Specifically, its nominal
$(1-\alpha)$ form is
\begin{equation}
    \bigg[
    \widehat L(\widehat d)
    -\frac{z_{1-\alpha/2}}{\sqrt n}
    \sqrt{
        \ell_{\widehat d,\widehat j_L(\widehat d)}
        \widehat\Sigma
        \ell_{\widehat d,\widehat j_L(\widehat d)}'
    },
    \widehat U(\widehat d)
    +\frac{z_{1-\alpha/2}}{\sqrt n}
    \sqrt{
        u_{\widehat d,\widehat j_U(\widehat d)}
        \widehat\Sigma
        u_{\widehat d,\widehat j_U(\widehat d)}'
    }
    \bigg],
\label{eq: conventional CI}
\end{equation}
where $z_q$ is the $q$ quantile of the standard
normal distribution.\footnote{When a selected affine component is classified as degenerate,
we set the corresponding endpoint of the conventional CI equal
to its affine estimate without a standard-error adjustment,
following the convention in Appendix
\ref{sec: degenerate affine endpoint components}.}  These coverage frequencies are reported in Table \ref{table:cov freqs}.  Consistent with the asymptotic results of Theorems \ref{thm: cond'l coverage for cond'l bounds}, \ref{thm: uncond'l coverage for proj bounds} and \ref{thm: uncond'l coverage for hybrid bounds}, the conditional, projection and hybrid CIs all have correct coverage for all DGPs and the modest sample size of $n=100$.  Also consistent with Theorem \ref{thm: uncond'l coverage for proj bounds}, we note that the projection CI tends to be conservative with true coverage above the nominal level of 95\%.  Finally, we note that the conventional CI can substantially under-cover, consistent with the discussion in Section \ref{sec: std inf fail}.  Such under-coverage by the conventional CI is even more pronounced in our simulations for dynamic treatment regimes in Appendix \ref{sec: additional simulation}.

\begin{table}[htbp]
\caption{Unconditional Coverage Frequencies}\label{table:cov freqs}
\centering{}%
\begin{tabular}{lccccc}
 & & \multicolumn{4}{c}{Confidence Interval}\tabularnewline
Data-Generating Process &  & Conv & Cond  & Proj & Hyb \tabularnewline
\hline
$p=(.081,0,0,.073,.139,.473)'$  &  &  0.95	&0.95	&0.98	&0.95  \tabularnewline
$p=(.25,.25,.25,.25,.25,.25)'$  &  & 0.86	&0.95	&0.97	&0.95	\tabularnewline
$p=(.01,.44,.01,.01,.01,.54)'$  &  &  0.95	&0.95	&0.99	&0.96  \tabularnewline
\hline
\end{tabular}
\vspace{0.2in}
\caption*{\footnotesize This table reports unconditional coverage frequencies for the counterfactual outcome selected by maximizing the estimated lower bound on the counterfactual outcomes of either treatment or control, all evaluated at the nominal coverage level of 95\%.  Coverage frequencies are reported for conventional (``Conv''), conditional (``Cond''), projection (``Proj'') and hybrid (``Hyb'') CIs for a sample size of $n=100$.}
\end{table}

\begin{figure}
\centering
\includegraphics[scale=0.29,trim={2.5cm 7cm 1cm 7cm}]{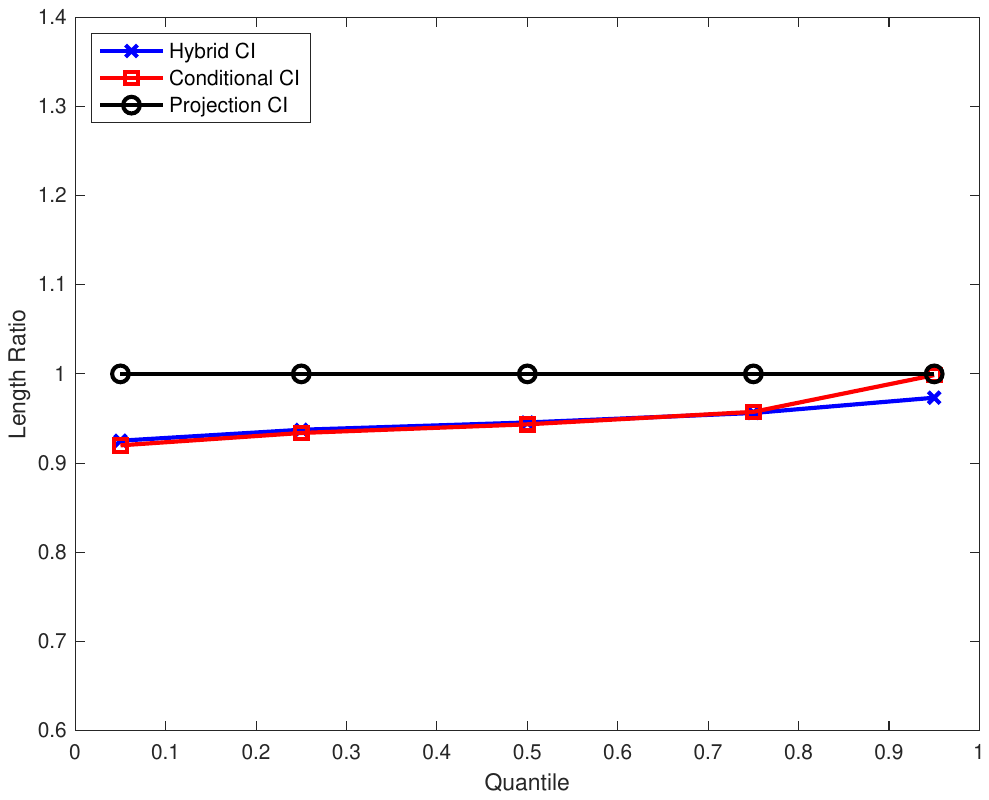}\includegraphics[scale=0.29,trim={1cm 7cm 1cm 7cm}]{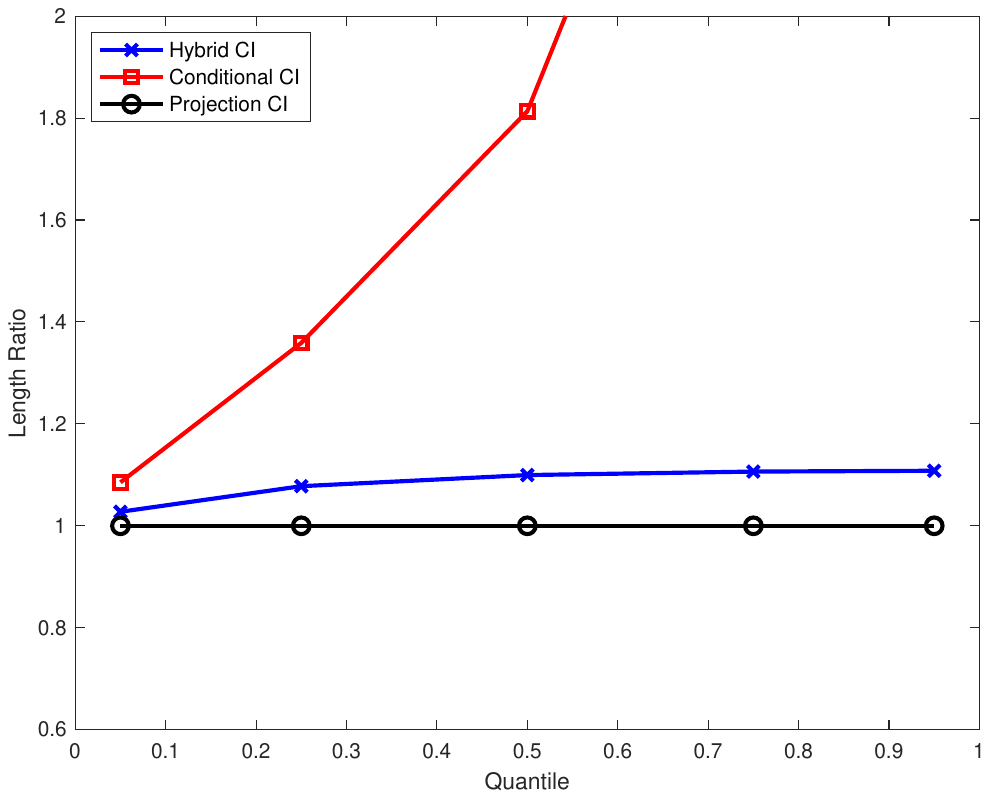}\includegraphics[scale=0.29,trim={1cm 7cm 1cm 7cm}]{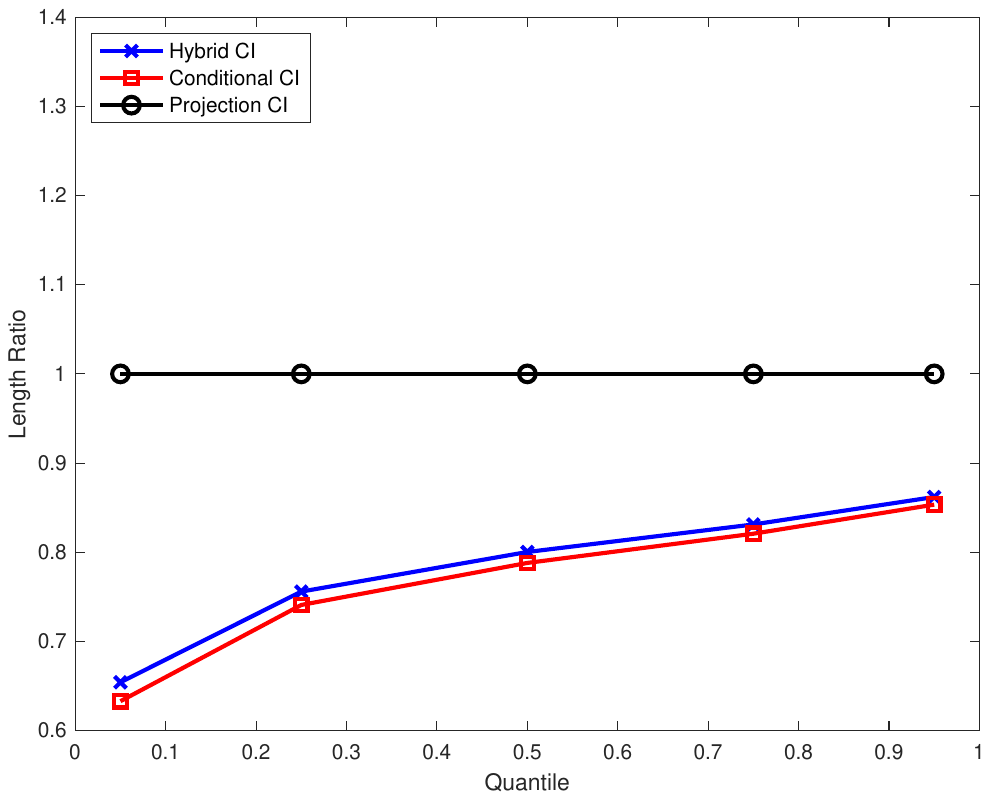}
\caption{Ratios of CI length quantiles relative to those of the projection CI for the conditional CI (red), projection CI (black) and hybrid CI (blue) of the average counterfactual outcome selected by maximizing the estimated lower bound when $p=(.081,0,0,.073,.139,.473)'$ (left), $p=(.25,.25,.25,.25,.25,.25)'$ (middle) and $p=(.01,.44,.01,.01,.01,.54)'$ (right) and $n=100$.}
\label{fig: CI length comp}
\end{figure}

Next, we compare the length quantiles of the CIs with correct coverage for these same DGPs.  Figure \ref{fig: CI length comp} plots the ratios of the $5^{th}$, $25^{th}$, $50^{th}$, $75^{th}$ and $95^{th}$ quantiles of the length of the conditional, projection and hybrid CIs relative to those same length quantiles of the projection CI.  As can be seen from the figure, the conditional CI has the tendency to become very long, especially at high quantile levels for certain DGPs, whereas the hybrid CI tends to perform the best overall by limiting the worst-case length performance of the conditional CI relative to the projection CI.  Relative to projection, the hybrid CI enjoys length reductions of 20-30\% for favorable DGPs while only showing length increases of 5-10\% for unfavorable DGPs.

\section{Empirical Application: Dynamic Treatment Regime\label{sec: application}}

Continuing the motivating example in Sections \ref{sec:motivating_ex} and \ref{sec:examples}, we revisit \citeauthor{Han24}'s (\citeyear{Han24}) application in which
schooling and post-school training are viewed as a sequence of treatments. \citet{Han24} considers the Job Training
Partnership Act (JTPA) program for post-school training and a
high school diploma (HS) (or its equivalents) for schooling. He considers
high school diplomas rather than college degrees because the former
are more relevant for the disadvantaged population of Title II of the
JTPA program. In this paper, we are interested in conducting inference
on welfare---earnings---evaluated at the regime chosen in a data-driven
manner. The dataset is constructed by combining the JTPA data with
the US Census and the National Center for Education Statistics (NCES).
The following is the set of variables:~$Y_{2}$ is an indicator for whether the individual is above or below the median
of 30-month earnings,\footnote{Note the difference from Section \ref{sec:motivating_ex}, which considers the employment status as $Y_2$ for illustration.} $D_{2}$ is an indicator for whether the individual participated in the job training program, $Z_{2}$ is an indicator for whether the individual was (randomly) assigned to the program, $Y_{1}$ is an indicator for whether the individual is above or below the 80th percentile
of pre-program earnings, $D_{1}$ is an indicator for whether the individual received a HS diploma or GED, and $Z_{1}$ is an indicator for the density of high schools \citep{Nea97}.\footnote{Specifically, $Z_{1}=1$ if
the number of high schools per square mile in each training site (e.g. a city) is above 35.} The number of individuals
in the sample is 9,223. We assume $Z\bot(Y(d),D(z))$.

Following \citet{Han24}, given the dynamic treatment regime $\bd(\cdot)$, the
counterfactual earning type in the terminal stage is defined as $Y_{2}(\bd(y_1))\equiv Y_{2}(\delta_{1},\delta_{2}(y_1,\delta_1))$,
where $y_1$ is the counterfactual earning type in the
first stage. Even though it is not a direct function of individual characteristics, the treatment regime $\bd(\cdot)$ is a function of $Y_1$, which is an important summary that reflects individual characteristics relevant to earnings. Recall all possible regimes in $\cD$ that are listed in
Table \ref{tab:regimes}. Suppose $\bd^{*}$ is the
optimal regime that maximizes the average terminal earning $W(\bd)\equiv \mathbb{E}[Y_{2}(\bd(\cdot))]$
as welfare. We are interested in constructing CIs for $W(\bd)$
evaluated at $\bd=\widehat{\bd}$ that is chosen by the maximin welfare rule (Example \ref{ex:maximin}): $\cD(\hp)=\{\widehat{\bd}\}$ where $\widehat{\bd} \in \argmax_{\bd} \widehat{L}(\bd)$ with $\widehat{L}(\bd)$ being the estimated version of the lower bound on $W(\bd)$. In Section \ref{subsec:DTR example}, we show how to derive the analytical lower bound $L(\bd)$ (i.e., \eqref{eq:L_delta}).

Recall that conditional inference is natural when the policymaker requires accountability for a particular program recommendation chosen from the data. In the context of the present example, suppose a funding body would consider supporting the schooling-and-targeted-training regime only when the empirical analysis recommends it. The body would want the reported CI for the regime's earnings prospects to be reliable specifically among samples in which the recommendation is made; good coverage in samples recommending no training or universal training would not compensate for systematically misleading assessments of the program it is being asked to fund.  On the other hand, unconditional inference is natural when the policymaker evaluates the reliability of an entire selection-and-reporting procedure.  In the context of the present example, suppose a planner is willing to implement any of the eight regimes and adopts the rule of choosing the regime according to $\widehat{\bd}$ above and would like a CI for the earnings prospects associated with this rule. The planner may require that this procedure reliably assess whichever regime it recommends across repeated samples, while accepting that coverage need not meet the nominal level separately for every possible recommendation; this is the criterion addressed by the projection and hybrid intervals.

Table \ref{table:regime_CI} reports 95\% CIs for the welfare selected by maximizing the estimated lower bound on the welfare. Recall that the welfare is the probability that the 30-month earnings are above the median.  It is notable that the hybrid CI is \emph{shorter} than the conventional CI, which is defined in analogy to \eqref{eq: conventional CI} above, even though the latter does not have (uniformly) correct coverage. We can also see that although the hybrid CI is not quite contained in the projection CI, it is somewhat shorter. Finally, the conditional CI has infinite length in this example, demonstrating an extreme example of how the conditional CIs can be uninformatively long (see \citealp{KL21}).

\begin{table}[htbp]
\caption{Confidence Intervals}\label{table:regime_CI}
\centering{}%
\begin{tabular}{cccc}
Conv & Cond  & Proj & Hyb \tabularnewline
\hline
(0.32, 0.78)	&(0.28, $\infty$)	&(0.29, 0.79)	&(0.28, 0.73)  \tabularnewline
\hline
\end{tabular}
\vspace{0.2in}
\caption*{\footnotesize This table reports 95\% CIs for the welfare selected by maximizing the estimated lower bound on the welfare: conventional (``Conv''), conditional (``Cond''), projection (``Proj'') and hybrid (``Hyb'') CIs.}
\end{table}

Appendix \ref{sec: additional simulation} contains Monte Carlo simulated coverage frequencies of the CIs with various DGPs in the setting of this application. Overall, the findings are consistent with the ones in Section \ref{sec: Monte Carlo}.  


\newpage

\begin{appendix}

\section{Additional Examples}\label{sec: additional example}

This appendix details examples in addition to those detailed in the main text, showing how they fall under the general framework of this paper.

\subsection{Revisiting Manski Bounds with a Continuous Outcome\label{subsec:Revisiting-Manski's-Bounds}}

We revisit the example with Manski bounds in Section \ref{sec:simple example}, now allowing for a continuous outcome with bounded support. Let $Y\in[y^{l},y^{u}]$
be continuously distributed and assume $\mathbb{E}[Y(d)|Z]=\mathbb{E}[Y(d)]$ for $d\in\{0,1\}$.  Note that the sharp bounds on $W(d)\equiv \mathbb{E}[Y(d)]$ are
\begin{align*}
L(d) & \equiv\max_{z\in\{0,1\}}\left\{\mathbb{E}[Y|D=d,Z=z]\mathbb{P}(D=d|Z=z)+(1-\mathbb{P}(D=d|Z=z))y^{l}\right\},\\
U(d) & \equiv\min_{z\in\{0,1\}}\left\{\mathbb{E}[Y|D=d,Z=z]\mathbb{P}(D=d|Z=z)+(1-\mathbb{P}(D=d|Z=z))y^{u}\right\},
\end{align*}
and the sharp bounds on the ATE are $L(1,0)$ and $U(1,0)$ for $L(\tilde{d},d) \equiv L(\tilde{d})-U(d)$ and $U(\tilde{d},d) \equiv U(\tilde{d})-L(d)$. We can define the identified set $\cD^* \subseteq \cD$ of optimal treatments as
\begin{align*}
\cD^{*} & \equiv\left\{d\in\{0,1\}:L(\tilde{d},d)\le0\text{ }\forall\tilde{d}\in\{0,1\}\right\}=\left\{d\in\{0,1\}:\max_{\tilde{d}\in\{0,1\}}L(\tilde{d})\le U(d)\right\}.
\end{align*}
Then, Assumption \ref{ass: welfare bounds} holds with
\begin{align*}
p & =\left(\begin{array}{c}
\mathbb{E}[Y|D=0,Z=1]\mathbb{P}(D=0|Z=1)\\
\mathbb{E}[Y|D=0,Z=0]\mathbb{P}(D=0|Z=0)\\
\mathbb{E}[Y|D=1,Z=1]\mathbb{P}(D=1|Z=1)\\
\mathbb{E}[Y|D=1,Z=0]\mathbb{P}(D=1|Z=0)\\
\mathbb{P}(D=0|Z=1)\\
\mathbb{P}(D=0|Z=0)
\end{array}\right)
\end{align*}
and $\tilde{\ell}_{0,j} =y^{l}$ and $\tilde{\ell}_{1,j} =0$ for $j\in\{0,1\}$ and
\begin{align*}
\ell_{0,0} & =(\begin{array}{cccccc}
0 & 1 & 0 & 0 & 0 & -y^{l}\end{array}), \quad \ell_{0,1} =(\begin{array}{cccccc}
1 & 0 & 0 & 0 & -y^{l} & 0\end{array}), \\
\ell_{1,0} & =(\begin{array}{cccccc}
0 & 0 & 0 & 1 & 0 & y^{l}\end{array}), \quad \ell_{1,1} =(\begin{array}{cccccc}
0 & 0 & 1 & 0 & y^{l} & 0\end{array})
\end{align*}
and symmetrically for $\tilde{u}_{d,j}$ and $u_{d,j}$. For $\widehat{L}(d)$ and $\widehat{U}(d)$ being the sample counterparts of $L(d)$ and $U(d)$ upon replacing $p$ with $\widehat p$, Assumption
\ref{ass: estimated ID'd set} holds for $\cD(\hp)=\widehat\cD$ by Proposition \ref{prop: selection rule}.

\subsection{Bounds for Dynamic Treatment Effects\label{subsec:Bounds-for-Dynamic}}

Consider binary $Y_{t}$ and $D_{t}$ for $t=1,...,T$. Let $Y\equiv(Y_{1},...,Y_{T})$
and $D\equiv(D_{1},...,D_{T})$. Suppose that we are equipped with
a sequence of binary instruments $Z\equiv(Z_{t_{1}},...,Z_{t_{K}})$,
which is a subvector of $(Z_{1},...,Z_{T})$. For $t=1,...,T$, let
$Y_{t}(d_{1},...,d_{t})$ be the counterfactual outcome at $t$ and $Y(d)\equiv(Y_{1}(d_{1}),...,Y_{T}(d_{1},...,d_{T}))$.
We assume that the instruments $Z$ are independent of $(Y(d),D(z))$.

When $T=2$, the setting is similar to that in the motivating example (Section \ref{sec:motivating_ex}). Let $Y\equiv(Y_{1},Y_{2})$ and $D\equiv(D_{1},D_{2})$. For given welfare $W(d)$ with $d\equiv (d_1,d_2)$, we are interested in the optimal policy $d^{*}$ that satisfies 
\begin{align*}
d^{*} & \in\arg\max_{d\in\cD}W(d),
\end{align*}
where $\cD\equiv\{(1,1),(1,0),(0,1),(0,0)\}$. Note that this is a non-adaptive version of the policy considered in Section \ref{sec:motivating_ex}. The sign of the welfare
difference, $W(d)-W(\tilde{d})$ for $d,\tilde{d}\in\cD$, is useful for establishing the ordering of $W(d)$ with respect to $d$
and thus to identifying $d^{*}$. However, without additional identifying
assumptions, we can only establish a partial ordering of $W(d)$ based
on the bounds on the welfare difference \citep{Han24}. This will produce the identified
set $\cD^{*}$ for $d^{*}$.

An example of the welfare is $W(d)\equiv \mathbb{E}[Y_{2}(d)]$, namely, the average
counterfactual terminal outcome. The bounds on welfare $W(d)$ are
\begin{align*}
L(d) & \equiv\max_{z}L(d;z),\qquad U(d) \equiv\min_{z}U(d;z), 
\end{align*}
where
\begin{align*}
L(d;z) & \equiv \mathbb{P}(Y_{2}=1,D=d|Z=z),\\
U(d;z) & \equiv \mathbb{P}(Y_{2}=1,D=d|Z=z)+\sum_{d'\neq d}\mathbb{P}(D=d'|Z=z), 
\end{align*}
which have forms analogous to those in the static case. Define the
dynamic ATE in the terminal period for a change in treatment from $d$ to $\tilde d$ as
\begin{align*}
W(\tilde d)-W({d})=\mathbb{E}[Y_{2}(\tilde d)-Y_{2}({d})] & =\mathbb{P}(Y_{2}(\tilde d)=1)-\mathbb{P}(Y_{2}({d})=1).
\end{align*}
Then the bounds on the dynamic ATE are as follows:
\begin{align*}
L(\tilde d,{d}) & \equiv L(\tilde d)-U({d}),\qquad U(\tilde d,{d}) \equiv U(\tilde d)-L({d}).
\end{align*}

Another example of welfare is the joint distribution $W(d)\equiv \mathbb{P}(Y(d)=(1,1))$
where $Y(d)\equiv(Y_{1}(d_{1}),Y_{2}(d))$. The bounds on $W(d)$
in this case are $L(d)\equiv\max_{z}L(d;z)$ and $U(d)\equiv\min_{z}U(d;z)$
where
\begin{align*}
L(d;z) & \equiv \mathbb{P}(Y_{2}=1,D=d|Z=z),\\
U(d;z) & \equiv \mathbb{P}(Y_{2}=1,D=d|Z=z)+\mathbb{P}(Y_{1}=1,D_{1}=d_{1},D_{2}=d_{2}'|Z=z)\\
 & \quad+\mathbb{P}(D_{1}=d_{1}',D_{2}=d_{2}|Z=z)+\mathbb{P}(D_{1}=d_{1}',D_{2}=d_{2}'|Z=z),
\end{align*}
with $d'_1\neq d_1$ and $d'_2\neq d_2$. Consider the effect of treatment on the joint distribution, for example,
\begin{align*}
W(1,1)-W(1,0) & =\mathbb{P}(Y(1,1)=(1,1))-\mathbb{P}(Y(1,0)=(1,1)).
\end{align*}
Then, with $\tilde d=(1,1)$ and ${d}=(1,0)$, the bounds on this parameter
are 
\begin{align*}
L(\tilde d,{d}) & \equiv\max_{z}\mathbb{P}(Y_{2}=1,D=(1,1)|Z=z)\\
 & \quad-\min_{z}\left\{\mathbb{P}(Y_{2}=1,D=(1,0)|Z=z)+\mathbb{P}(Y_{1}=1,D=(1,1)|Z=z)\right.\\
 & \left.\quad+\mathbb{P}(D=(0,1)|Z=z)+\mathbb{P}(D=(0,0)|Z=z)\right\},\\
U(\tilde d,{d}) & \equiv\min_{z}\left\{\mathbb{P}(Y_{2}=1,D=(1,1)|Z=z)+\mathbb{P}(Y_{1}=1,D=(1,0)|Z=z)\right.\\
 & \left.\quad+\mathbb{P}(D=(0,1)|Z=z)+\mathbb{P}(D=(0,0)|Z=z)\right\}\\
 & \quad-\max_{z}\mathbb{P}(Y_{2}=1,D=(1,0)|Z=z).
\end{align*}
In these examples, the identified set $\cD^{*}$ can be characterized
as a set of maximal elements: 
\begin{align*}
\cD^{*} & =\{d:\nexists\tilde{d}\neq d\text{ such that }L(\tilde{d},d)>0\} =\{d:L(\tilde{d},d)\le0\text{ }\forall\tilde{d}\neq d\}.
\end{align*}
These examples are special cases of the model in \citet{Han24}.\footnote{See also
\citet{HL24} for other examples of dynamic causal parameters that can be used to define optimal treatments.}

In both cases, it is easy to see that $L(d)$ and $U(d)$ satisfy
Assumption \ref{ass: welfare bounds} with $p$ being the vector of probabilities $p(y,d|z)\equiv \mathbb{P}(Y=y,D=d|Z=z)$.
To verify Assumption \ref{ass: estimated ID'd set} for $\cD(\hp)=\widehat\cD$, let $\widehat{L}(\tilde{d},d)$ be the
estimator of $L(\tilde{d},d)$ where the sample frequency replaces
the population probability. Then $\widehat{\cD}=\{d:\widehat{L}(\tilde{d},d)\le0\text{ }\forall\tilde{d}\neq d\}$.

Continuing with the first example, let $Z=Z_{1}\in\{0,1\}$, that is,
the researcher is only equipped with a binary instrument in the first
period and no instrument in the second period. We focus on inference
for $L(d)$ for $d\in\widehat{\cD}$. Let $\widehat{z}(d)\in\arg\max_{z\in\{0,1\}}\widehat{L}(d;z)$
so that $\widehat{L}(d)=\widehat{L}(d;\widehat{z}(d))$. Then we can
write the data-dependent event that $d$ is an element of $\widehat{\cD}$,
$\widehat{L}(d)=\widehat{L}(d;z^{*})$ as a polyhedron 
\begin{align*}
\{d\in\widehat{\cD},~\widehat{z}(d)=z^{*}\}=\{\widehat{L}(\tilde{d},d)\le0\text{ }\forall\tilde{d}\neq d,~\widehat{z}(d)=z^{*}\} & =\{A^L\widehat{p}\leq0\}
\end{align*}
for some matrix $A^L$, where $\widehat{p}$ is the vector of probabilities $\widehat{p}(y,d|z)$, so that Assumption \ref{ass: estimated ID'd set} holds.
This is due to the forms of $L(d;z)$ and $U(d;z)$ above and
$\widehat{L}(\tilde{d},d)\le0$ $\forall\tilde{d}\neq d$ if and only
if
\begin{align*}
\widehat{L}(\tilde{d};z) & \le\widehat{U}(d;z')\quad\forall\tilde{d}\neq d,\text{ }\forall z,z'
\end{align*}
and, for example, $\widehat{z}(d)=1$ if and only if $\widehat{L}(d;0)-\widehat{L}(d;1) \le0$.
A similar formulation follows for the second example. In fact, this
approach applies to a general parameter $W(d)$ with bounds that are
 minima and maxima of linear combinations of $p(y,d|z)$'s, such as parameters that have the following form:
\begin{align*}
W(d) & \equiv f(q_{d})
\end{align*}
for some linear functional $f$, where $q_{d}(y)\equiv \mathbb{P}(Y(d)=y)$.

\subsection{Bounds Derived from Linear Programming\label{subsec:Bounds-Derived-from}}

In more complex settings, calculating analytical bounds on $W(d)$ or $W(d)-W(\tilde{d})$ may be cumbersome. This is especially true when the researcher wants to incorporate additional identifying assumptions. In this situation,
the computational approach using linear programming can be useful
(\citealp{MST18,HY24}).

To incorporate many complicated settings, suppose that $W(d)=A_{d}q$
and $p=Bq$ for some known row vector $A_{d}$ and matrix $B$, an
unknown vector $q$ in a simplex $\mathcal{Q}$, and a vector $p$
that is estimable from data.  Typically $q$ is a vector of probabilities of a latent variable that
governs the DGP; see \citet{BP97,BP13},
\citet{Han24} and \citet{HY24}. The linearity in this assumption
is usually implied by the nature of a particular problem (e.g.,
discreteness). Then we have
\begin{align}
\begin{array}{c}
L(d)=\min_{q\in\mathcal{Q}}A_{d}q,\\
U(d)=\max_{q\in\mathcal{Q}}A_{d}q,
\end{array} & \quad s.t.\quad Bq=p\label{eq:LP0}
\end{align}
and ATE bounds for a change from treatment $d$ to treatment $\tilde d$
\begin{align}
\begin{array}{c}
L(\tilde d,{d})=\min_{q\in\mathcal{Q}}(A_{\tilde d}-A_{{d}})q,\\
U(\tilde d,{d})=\max_{q\in\mathcal{Q}}(A_{ \tilde d}-A_{{d}})q,
\end{array} & \quad s.t.\quad Bq=p.\label{eq:LP}
\end{align}
Note that $L(\tilde d,{d})\neq L(\tilde d)-U({d})$ in general because the
$q$ that solves \eqref{eq:LP0} for $L(\tilde d)$ and $U({d})$ may be different (and
similarly for $U(\tilde d,{d})$). As before, the identified set of optimal treatments here is
characterized as $\cD^{*}\equiv \{d:L(\tilde{d},d)\le0\text{ }\forall\tilde{d}\neq d\}$.

An example of this setting can be found in \citet{HY24}. Let $(Y,D,Z)$
be a vector of a binary outcome, treatment and instrument and let
$p$ be a vector with entries $p(y,d|z)\equiv \mathbb{P}(Y=y,D=d|Z=z)$ across $(y,d,z)\in\{0,1\}^{3}$.\footnote{See \citet{HY24} for the use of linear programming with continuous
$Y$.} Suppose $W(d)=\mathbb{E}[Y(d)]$ for $d\in\{0,1\}$. Then, we can define the
response type $\varepsilon\equiv(Y(1),Y(0),D(1),D(0))$ with a realized
value $e\equiv(y(1),y(0),d(1),d(0))$, where $Y(d)$ denotes the counterfactual outcome under treatment $d$ and $D(z)$ denotes the counterfactual treatment under instrument value $z$. Let $q(e)\equiv \mathbb{P}(\varepsilon=e)$
be the latent distribution. Then
\begin{align*}
W(d) & =\mathbb{P}[Y(d)=1]=\sum_{e:y(d)=1}q(e)\equiv A_{d}q,
\end{align*}
where $q$ is the vector of $q(e)$'s and $A_{d}$ is an appropriate
selector (a row vector).

Assume that $(Y(d),D(z))$ is independent of $Z$ for $d,z\in\{0,1\}$.  The data distribution $p$ is related to the latent distribution by
\begin{align*}
\mathbb{P}[Y=1,D=d|Z=z] & =\mathbb{P}[Y(d)=1,D(z)=d]=\sum_{e:y(d)=1,d(z)=d}q(e)\equiv B_{d,z}q,
\end{align*}
where the first equality follows by the independence assumption, $q$ is a vector of $q(e)$'s and $B_{d,z}$
is an appropriate selector (a row vector). Now define 
\begin{align*}
B & \equiv\left[\begin{array}{c}
B_{1,1}\\
B_{0,1}\\
B_{1,0}\\
\vdots
\end{array}\right],\qquad p\equiv\left[\begin{array}{c}
p(1,1|1)\\
p(1,0|1)\\
p(1,1|0)\\
p(1,0|0)\\
\vdots
\end{array}\right]
\end{align*}
so that all of the constraints relating the data distribution to the latent distribution can be expressed as $Bq=p$.

To verify Assumption \ref{ass: welfare bounds}, it is helpful to invoke strong duality for
the primal problems \eqref{eq:LP0} (under regularity conditions)
and write the following dual problems:
\begin{align*}
L(d)= & \max_{\lambda}-\tilde{p}'\lambda,\quad s.t.\quad\tilde{B}'\lambda\ge-A_{d}',\\
U(d)= & \min_{\lambda}\tilde{p}'\lambda,\quad s.t.\quad\tilde{B}'\lambda\ge A_d',
\end{align*}
where $\tilde{B}\equiv\left[\begin{array}{c}
B\\
\boldsymbol{1}'
\end{array}\right]$ is a $(d_{p}+1)\times d_{q}$ matrix with $\boldsymbol{1}$ being
a $d_{q}\times1$ vector of ones, and $\tilde{p}\equiv\left[\begin{array}{c}
p\\
1
\end{array}\right]$ is a $(d_{p}+1)\times1$ vector. By using a vertex enumeration algorithm (e.g., \citet{AF91}), one can find all (or a relevant subset)
of vertices of the polyhedra $\{\lambda:\tilde{B}'\lambda\ge-A_{d}'\}$
and $\{\lambda:\tilde{B}'\lambda\ge A_{d}'\}$. Let $\Lambda_{L,d}\equiv\{\lambda_{1},...,\lambda_{J_{L,d}}\}$
and $\Lambda_{U,d}\equiv\{\lambda_{1},...,\lambda_{J_{U,d}}\}$ be
the sets that collect such vertices, respectively. Then, it is easy
to see that $L(d)=\max_{\lambda\in\Lambda_{L,d}}-\tilde{p}'\lambda$
and $U(d)=\min_{\lambda\in\Lambda_{U,d}}\tilde{p}'\lambda$, and thus
Assumption \ref{ass: welfare bounds} holds.

To verify Assumption \ref{ass: estimated ID'd set}, we use the dual problems to \eqref{eq:LP}:
\begin{align*}
L(\tilde d,{d})= & \max_{\lambda}-\tilde{p}'\lambda,\quad s.t.\quad\tilde{B}'\lambda\ge-\Delta_{\tilde d,{d}}',\\
U(\tilde d,{d})= & \min_{\lambda}\tilde{p}'\lambda,\quad s.t.\quad\tilde{B}'\lambda\ge\Delta_{\tilde d,{d}}',
\end{align*}
where $\Delta_{\tilde d,{d}}\equiv A_{\tilde d}-A_{{d}}$. Analogous
to the vertex enumeration argument above, let $\Lambda_{L,\tilde d,{d}}\equiv\{\lambda_{1},...,\lambda_{J_{L,\tilde d,{d}}}\}$
and $\Lambda_{U,\tilde d,{d}}\equiv\{\lambda_{1},...,\lambda_{J_{U,\tilde d,{d}}}\}$
be the sets that collect all (or a relevant subset) of vertices of
the polyhedra $\{\lambda:\tilde{B}'\lambda\ge-\Delta_{\tilde d,{d}}'\}$
and $\{\lambda:\tilde{B}'\lambda\ge\Delta_{\tilde d,{d}}'\}$, respectively.
Then, $L(\tilde d,{d})=\max_{\lambda\in\Lambda_{L,\tilde d,{d}}}-\tilde{p}'\lambda$
and $U(\tilde d,{d})=\min_{\lambda\in\Lambda_{U,\tilde d,{d}}}\tilde{p}'\lambda$. Let $\widehat{\cD}=\{d:\widehat{L}(\tilde d,{d})\le0\text{ }\forall\tilde{d}\neq d\}$,
where $\widehat{L}(\tilde d,{d})$ is the sample counterpart of $L(\tilde d,{d})$
with $\widehat{\tilde{p}}\equiv\left[\begin{array}{c}
\widehat{p}\\
1
\end{array}\right]$ replacing $\tilde{p}\equiv\left[\begin{array}{c}
p\\
1
\end{array}\right]$. Partition $\lambda$ as $\lambda=(\lambda^{1\prime},\lambda^{0})'$
where $\lambda^{0}$ is the last element of $\lambda$. Note that
$d\in\widehat{\cD}$ if and only if
\begin{align*}
\max_{\lambda\in\tilde{\Lambda}_{L,d}}-(\widehat{p}'\lambda^{1}+\lambda^{0}) & \le0,
\end{align*}
where $\tilde{\Lambda}_{L,d}=\bigcup_{\tilde{d}\neq d}\Lambda_{L,\tilde d,{d}}$.
Also let $\widehat{\lambda}$ be such that $-\widehat{\tilde{p}}'\widehat{\lambda}=\max_{\lambda\in\tilde{\Lambda}_{L,d}}-\widehat{\tilde{p}}'\lambda$.
Then, $\widehat{\lambda}=\lambda_{j_{L}^{*}}$ if and only if 
\begin{align*}
\widehat{p}'\lambda_{j_{L}^{*}}^{1}+\lambda_{j_{L}^{*}}^{0}-(\widehat{p}'\lambda^{1}+\lambda^{0}) & \le0\quad\forall\lambda\in\tilde{\Lambda}_{L,d}\backslash\{\lambda_{j_{L}^{*}}\}
\end{align*}
so that Assumption \ref{ass: estimated ID'd set} holds for $\cD(\hp)=\widehat\cD$.

\subsection{Optimal Treatment Assignment with Partially-Identified ATE} \label{sec: optimal selection}

In recent work, \cite{CMS23} note that ``plug-in'' rules for determining treatment choice can be sub-optimal when ATEs are not point-identified since the bounds on the ATE are not smooth functions of the reduced-form parameter $p$.  Using an optimality criterion that minimizes maximum regret over the identified set for the ATE, conditional on $p$, they advocate bootstrap and quasi-Bayesian methods for optimal treatment choice.  More specifically, they consider settings for which the ATE of a treatment is identified via intersection bounds: 
\[b_L(p)\equiv \max_{j\in\{1,\ldots,J_L\}}\{\tilde{\ell}_j+\ell_jp\}\leq ATE\leq \min_{k\in\{1,\ldots,J_U\}}\{\tilde{u}_k+u_kp\}\equiv b_U(p)\]
for some fixed and known $J_L$, $J_U$, $\tilde{\ell}_1,\ldots,\tilde{\ell}_{J_L}$, $\tilde{u}_1,\ldots,\tilde{u}_{J_U}$, $\ell_1,\ldots,\ell_{J_L}$ and $u_1,\ldots,u_{J_U}$.\footnote{Although \cite{CMS23} do not write the form of their bounds as they are written here, the representation here is equivalent to the one in that paper upon proper definition of $p$ since the elements in the intersection bounds are smooth functions of a reduced-form parameter.} Therefore, Assumption \ref{ass: welfare bounds} trivially holds for the ATE.

Now, let $\psi(q)\equiv\max\{b_U(q),0\}+\min\{b_L(q),0\}$.  In the notation of Assumptions \ref{ass: joint normality} and
\ref{ass: variance estimation}, the quasi-Bayesian treatment rule of
\citet{CMS23} is
\[
    \widehat d_n^\infty
    =
    \mathbf 1\left\{
    E^*\left[
    \psi\left(
    \widehat p+\varepsilon
    \right)
    \right]\geq0
    \right\},
    \qquad
    \varepsilon\sim\mathcal N\left(0,\frac{\widehat\Sigma}{n}\right),
\]
where $E^*(\cdot)$ denotes expectation with respect to the
quasi-posterior draw conditional on the data.  In practice, this
expectation can be approximated using
\begin{equation*}
    \widehat d
    \equiv
    \widehat d_{n,m}
    =
    \mathbf 1\left\{
    \frac1m\sum_{i=1}^m
    \psi\left(
    \widehat p+\varepsilon_i
    \right)\geq0
    \right\},
    \qquad
    \varepsilon_1,\ldots,\varepsilon_m
    \overset{i.i.d.}{\sim}
    \mathcal N\left(0,\frac{\widehat\Sigma}{n}\right).
\end{equation*}
For each fixed $m$, this form of $\widehat d$ satisfies Assumption \ref{ass: estimated ID'd set}  for $\cD(\hp)=\{\widehat d\}$ when $\widehat\gamma_L(d)=\widehat\gamma_U(d)$ are specified properly.

To see this, let \[\widehat\gamma_L(d)=\widehat\gamma_U(d)=(\varepsilon_1^\prime,\ldots,\varepsilon_m^\prime,\underbar k_1,\ldots,\underbar k_m,\bar k_1,\ldots,\bar k_m,s_1^{\ell},\ldots,s_m^{\ell},s_1^{u},\ldots,s_m^{u})',\]
where $\underbar k_i\equiv\argmin_{k\in\{1,\ldots,J_U\}}\{\tilde{u}_k+u_k(\widehat p+\varepsilon_i)\}$, $\bar k_i\equiv\argmax_{k\in\{1,\ldots,J_L\}}\{\tilde{\ell}_k+\ell_k(\widehat p+\varepsilon_i)\}$, $s_i^{\ell}\equiv\sign(\tilde{\ell}_{\bar k_i}+\ell_{\bar k_i}(\widehat p+\varepsilon_i))$ and $s_i^{u}\equiv\sign(\tilde{u}_{\underbar k_i}+u_{\underbar k_i}(\widehat p+\varepsilon_i))$ for $i=1,\ldots,m$.  Now note that 
\begin{enumerate}
\item \emph{$\underbar k_i=\underbar k_i^*$} and $\bar k_i=\bar k_i^*$ if and only if \emph{$(u_{\underbar k_i^*}-u_k)\widehat p\leq \tilde{u}_k-\tilde{u}_{\underbar k_i^*}+({u}_k-{u}_{\underbar k_i^*})\varepsilon_i$} for all $k=1,\ldots,J_U$ and $(\ell_k-\ell_{\bar k_i^*})\widehat p\leq \tilde{\ell}_{\bar k_i^*}-\tilde{\ell}_k+({\ell}_{\bar k_i^*}-{\ell}_k)\varepsilon_i$ for all $k=1,\ldots,J_L$;
\item $s_i^{\ell}=-$ and $s_i^{u}=-$ if and only if $\ell_{\bar k_i}\widehat p\leq -\tilde{\ell}_{\bar k_i}-{\ell}_{\bar k_i}\varepsilon_i$ and \emph{$u_{\underbar k_i}\widehat p\leq -\tilde{u}_{\underbar k_i}-{u}_{\underbar k_i}\varepsilon_i$};
\item $\widehat d=1$ if and only if \emph{$\sum_{i\in \underbar{m}}(-{u}_{\underbar k_i}\widehat p)+\sum_{i\in \bar{m}}(-{\ell}_{\bar k_i}\widehat p)\leq \sum_{i\in \underbar{m}}(\tilde{u}_{\underbar k_i}+{u}_{\underbar k_i}\varepsilon_i)+\sum_{i\in \bar{m}}(\tilde{\ell}_{\bar k_i}+{\ell}_{\bar k_i}\varepsilon_i)$}, where \emph{$\underbar{m}=\{i\in\{1,\ldots,m\}:s_i^{u}=+\}$} and $\bar{m}=\{i\in\{1,\ldots,m\}:s_i^{\ell}=+\}.$
\end{enumerate}

Since $\psi$ is Lipschitz,
\[
    E^*\left[
    n\left(
    \frac1m\sum_{i=1}^m
    \psi\left(
    \widehat p+\varepsilon_i
    \right)
    -
    E^*\left[
    \psi\left(
    \widehat p+\varepsilon
    \right)
    \right]
    \right)^2
    \right]
    \leq
    \frac{C_\psi^2\operatorname{tr}(\widehat\Sigma)}{m}
\]
for a finite constant $C_\psi$.  Thus, any sequence $m=m_n\to
\infty$ makes the Monte Carlo approximation error $o_p(n^{-1/2})$.  Under a
standard no-tie condition for the limiting treatment-choice
statistic, $\Pr(\widehat d_{n,m_n}\neq\widehat d_n^\infty)\to0$.  Allowing $m=m_n\to\infty$ in the conditional and hybrid
uniformity results would require an extension to a growing sequence
of polyhedral conditioning events.  We leave this extension to
future research.

\subsection{Empirical Welfare Maximization with Observational Data\label{subsec:Empirical-Welfare-Maximization}}

Consider allocating a binary treatment based on observed covariates
$X\in\mathcal{X}$. A treatment allocation rule can be defined as
a function $\delta:\mathcal{X}\rightarrow\{0,1\}$ in a class of rules
$\cD$. Consider the utilitarian welfare of deploying $\delta$
relative to treating no one.
The optimal allocation $\delta^{*}$ satisfies
\begin{align*}
\delta^{*} & \in\arg\max_{\delta\in\cD}W(\delta).
\end{align*}
Note that $\mathbb{E}[Y(\delta(X))-Y(0)] =E\left[\delta(X)\Delta(X)\right]$, 
where $\Delta(X)\equiv \mathbb{E}[Y(1)-Y(0)|X]$. This problem is considered
in \citet{KT18} and \citet{AW21}, among others. When only
observational data for $(Y,D,X)$ are available with $D$ being endogenous,
$W(\delta)$ is only partially identified unless strong treatment
effect homogeneity is assumed. This problem has been studied in \citet{kallus2021minimax,pu2021estimating,d2021orthogonal,Bya22}, among others. Using instrumental variables, one can consider
bounds on the conditional ATE based on conditional versions of the bounds considered
in Sections \ref{subsec:Revisiting-Manski's-Bounds} and \ref{subsec:Bounds-Derived-from}
(i.e., Manski's bounds and bounds produced by linear programming). 

In particular, assume that $(Y(d),D(z))$ is independent of $Z$ given $X$.  Let $L(X)$ and $U(X)$ be conditional Manski bounds on $\Delta(X)$. Then, bounds on $W(\delta)$ can be
characterized as
\begin{align*}
L(\delta) & \equiv \mathbb{E}[\delta(X)L(X)],\qquad U(\delta) \equiv \mathbb{E}[\delta(X)U(X)].
\end{align*}
Similarly, bounds on $W(\tilde\delta)-W({\delta})=\mathbb{E}[(\tilde\delta(X)-{\delta}(X))\Delta(X)]$
can be characterized as
\begin{align}
L(\tilde\delta,{\delta}) & \equiv \mathbb{E}[(\tilde\delta(X)-{\delta}(X))L(X)],\quad U(\tilde\delta,{\delta}) \equiv \mathbb{E}[(\tilde\delta(X)-{\delta}(X))U(X)].\label{eq:EWM3}
\end{align}
Note that $L(\tilde\delta,{\delta})\neq L(\tilde\delta)-U({\delta})$
in general (and similarly for $U(\tilde\delta,{\delta})$).

Suppose $\mathcal{X}$ is finite and $\mathcal{X}=\{x_{1},...,x_{K}\}$
where $x_{k}$ can be a vector and $K$ can potentially be large. The discrete covariate space is practically relevant because what may be feasible to the PM are only policies that admit discrete inputs. For simplicity of exposition, suppose $\mathcal{X}=\{0,1,2\}$. Then
$\cD=\{\delta_{1},...,\delta_{8}\}$ where each $\delta_{j}$
corresponds to a mapping type from $\{0,1,2\}$ to $\{0,1\}$. To
verify Assumptions \ref{ass: welfare bounds} and \ref{ass: estimated ID'd set}, we proceed as follows. For given $x\in\mathcal{X}$,
by arguments analogous to those in Section \ref{subsec:Bounds-Derived-from} (and Section \ref{subsec:Revisiting-Manski's-Bounds}), bounds $L_x$ and $U_x$
on $\Delta(x)$ satisfy, for some scalars $\tilde{\ell}_{j}$ and
$\tilde{u}_{j}$ and row vectors $\ell_{j}$ and $u_{j}$,
\begin{align*}
L_x & =\max_{j\in\{1,...,J_{L}\}}\{\tilde{\ell}_{j}+\ell_{j}p_x\},\qquad U_x =\min_{j\in\{1,...,J_{U}\}}\{\tilde{u}_{j}+u_{j}p_x\},
\end{align*}
where $p(x)$ is the vector of $p(y,d|z,x)$'s across $(y,d,z)$ fixing $x$.
Then, by Jensen's inequality, for each $\delta\in\cD$,
\begin{align*}
L(\delta) & \ge\tilde{L}(\delta)\equiv\max_{j\in\{1,...,J_{L}\}}\{\tilde{\ell}_{j}\mathbb{E}[\delta(X)]+\ell_{j}\mathbb{E}[\delta(X)p(X)]\},\\
U(\delta) & \le\tilde{U}(\delta)\equiv\min_{j\in\{1,...,J_{U}\}}\{\tilde{u}_{j}\mathbb{E}[\delta(X)]+u_{j}\mathbb{E}[\delta(X)p(X)]\}.
\end{align*}
Note that $\tilde{L}(\delta)$ and $\tilde{U}(\delta)$ are non-sharp
bounds; for calculation of sharp bounds, see Section \ref{subsec:EWM with LP}.
We can verify Assumption \ref{ass: welfare bounds} with $\tilde{L}(\delta)$ and $\tilde{U}(\delta)$
by defining
\begin{align*}
p & =\left(\begin{array}{c}
\mathbb{E}[\delta_{1}(X)]\\
\mathbb{E}[\delta_{1}(X)p(X)]\\
\vdots\\
\mathbb{E}[\delta_{8}(X)]\\
\mathbb{E}[\delta_{8}(X)p(X)]
\end{array}\right)
\end{align*}
and, for $\delta=\delta_{1}$ as an example, by using $\ell_{\delta_{1},j} =(\begin{array}{ccccc}
\tilde{\ell}_{j} & \ell_{j} & 0 & \cdots & 0\end{array})$. Similarly, we can verify Assumption \ref{ass: estimated ID'd set} by estimating $p(X)$ and $\mathbb{E}[\delta(X)]$ with sample means and ${E}[\delta(X){p}(X)]$ with
$\frac{1}{n}\sum_{i=1}^{n}\delta(X_{i})\widehat{p}(X_{i}).$  Then Assumption \ref{ass: estimated ID'd set} holds for $\cD(\hp)=\widehat\cD=\{\delta\in\cD:\widehat{L}(\tilde{\delta},\delta)\le0\text{ }\forall\tilde{\delta}\neq \delta\}$ for $\widehat{L}(\tilde{\delta},\delta)$ defined the same as ${L}(\tilde{\delta},\delta)$ in \eqref{eq:EWM3} after substituting $\widehat p$ for $p$.

This framework can be generalized to settings where $W(\delta)$ is
partially identified, not necessarily due to treatment endogeneity
but because $W(\delta)$ is a non-utilitarian welfare defined as a
functional of the joint distribution of counterfactual outcomes (e.g.,
\citealp{cui2026policy}): $W(\delta)=f(F_{Y(1),Y(0)|X})$ where $f$ is
some functional and $F_{Y(1),Y(0)|X}$ is the joint distribution of
$(Y(1),Y(0))$ conditional on $X$.

\subsection{Empirical Welfare Maximization via Linear Programming\label{subsec:EWM with LP}}

Continuing from Section \ref{subsec:Empirical-Welfare-Maximization},
we show how sharp bounds on $W(\delta)$ can be computed using linear
programming. This can be done by extending the example in Section
\ref{subsec:Bounds-Derived-from} with binary $Y$. Again, let $\mathcal{X}=\{x_{1},...,x_{K}\}$.
Let $q(e|x)\equiv \mathbb{P}(\varepsilon=e|X=x)$. In analogy,
\begin{align*}
\mathbb{E}[Y(d)|X=x] & =\mathbb{P}[Y(d)=1|X=x]=\sum_{e:y(d)=1}q(e|x)\equiv A_{d}q(x),
\end{align*}
where $q(x)$ is a vector with entries $q(e|x)$ across $e$, and 
\begin{gather*}
\mathbb{P}[Y=1,D=d|Z=z,X=x] =\mathbb{P}[Y(d)=1,D(z)=d|X=x] \\
=\sum_{e:y(d)=1,d(z)=d}\mathbb{P}[\varepsilon=e|X=x]\equiv B_{d,z}q(x),
\end{gather*}
where the first equality holds by the independence assumption in the previous section. Then the
constraint for each $x\in\mathcal{X}$ becomes 
\begin{align*}
Bq(x) & =p(x),
\end{align*}
where $p(x)$ is the vector of $p(y,d|z,x)$'s across $(y,d,z)$ fixing $x$. Now we can construct a linear program for welfare:
\begin{align*}
W(\delta)=\mathbb{E}[\delta(X)\Delta(X)] & =\sum_{x_{k}\in\mathcal{X}}p(x_{k})\delta(x_{k})\Delta(x_{k})\\
 & =\sum_{x_{k}\in\mathcal{X}}p(x_{k})\delta(x_{k})(A_{1}-A_{0})q(x_{k}).
\end{align*}
Therefore $W(\delta)$ satisfies the structure of Section \ref{subsec:Bounds-Derived-from} and by analogous arguments, Assumptions \ref{ass: welfare bounds} and \ref{ass: estimated ID'd set} hold for $\cD(\hp)=\widehat\cD$.

\section{Relationship with \cite{AKM24}}\label{sec: AKM generalization}

The setting of \citet{AKM24} (AKM henceforth) has a finite set of options, denoted here by $\cD=\{d^0,\ldots,d^K\}.$
For a sample of size $n$, AKM observe $(K+1)$-dimensional estimators $\widetilde X_n$ and $\widetilde Y_n$, where $\widetilde X_n(d)$ is used to select among the options and $\widetilde Y_n(d)$ estimates the point-identified target associated with option $d$, $\widetilde\mu_{Y,n}(d)$.  Their selected option and inferential target are $\widehat d=\argmax_{d\in\cD}\widetilde X_n(d)$ and $\widetilde\mu_{Y,n}(\widehat d)$.  AKM work in Online Appendix F with the scaled estimates $(X_n,Y_n)=\sqrt{n}(\widetilde X_n,\widetilde Y_n)$ and corresponding centering vectors $\mu_{X,n}=\sqrt n\,\widetilde\mu_{X,n}$ and $\mu_{Y,n}=\sqrt n\,\widetilde\mu_{Y,n}$, which may depend upon $\mathbb P\in\mathcal{P}_n$.  Their Assumptions 6--8 deliver the uniform asymptotic validity of the conditional, projection, and hybrid procedures. To embed the AKM framework in the present paper, define
\begin{equation}
    \widehat p
    =
    \begin{pmatrix}
        \widetilde X_n\\
        \widetilde Y_n
    \end{pmatrix},
    \qquad
    p(\mathbb P)
    =
    \begin{pmatrix}
        \widetilde\mu_{X,n}(\mathbb P)\\
        \widetilde\mu_{Y,n}(\mathbb P)
    \end{pmatrix}.
    \label{eq:akm-asymptotic-p-embedding}
\end{equation}
Thus,
\begin{equation}
    \sqrt n(\widehat p-p(\mathbb P))
    =
    \begin{pmatrix}
        X_n-\mu_{X,n}(\mathbb P)\\
        Y_n-\mu_{Y,n}(\mathbb P)
    \end{pmatrix}.
    \label{eq:akm-asymptotic-scaled-embedding}
\end{equation}
For each $d\in\cD$, let $r_d^X$ and $r_d^Y$ be selector row vectors satisfying
\begin{equation}
    r_d^X\widehat p=\widetilde X_n(d),
    \qquad
    r_d^Y\widehat p=\widetilde Y_n(d).
    \label{eq:akm-selector-rows}
\end{equation}

\subsection{Step-by-Step Comparison of Assumptions}
\label{app:akm-asymptotic-assumption-mapping}

\textbf{Step 1:~AKM point identification as a special case of Assumption \ref{ass: welfare bounds}.} \\
Temporarily suppressing dependence on $\mathbb P$ for notational brevity, for option $d$, define $W(d)=\widetilde\mu_{Y,n}(d)=r_d^Yp$.
Set
\begin{equation}
    J_L=J_U=1,
    \qquad
    \widetilde\ell_{d,1}=\widetilde u_{d,1}=0,
    \qquad
    \ell_{d,1}=u_{d,1}=r_d^Y.
    \label{eq:akm-asymptotic-ass31}
\end{equation}
Then Assumption \ref{ass: welfare bounds} becomes $L(d)=U(d)=W(d)$.  The estimated lower and upper endpoints both equal $\widetilde Y_n(d)$, and the active component indices are nonrandom:~$\widehat j_L(d)=\widehat j_U(d)=1$.  Consequently, the point-identified AKM setting removes the additional max/min nonsmoothness that arises from estimating which component attains an intersection bound.

\textbf{Step 2:~the AKM winner event is a special case of Assumption \ref{ass: estimated ID'd set}.} \\
For each $d\in\cD$, let $A_d$ contain the row $r_{d'}^X-r_d^X$ for every $d'\neq d$.  Ignoring the treatment of exact ties for the moment, $\{\widehat d=d\}=\{\widetilde X_n(d)\geq\widetilde X_n(d')\text{ for every }d'\neq d\}=\{A_d\widehat p\leq0\}$.  Hence, Assumption \ref{ass: estimated ID'd set} holds with $\hcD=\{\widehat d\}$, $A^L(d,1,\gamma)=A^U(d,1,\gamma)=A_d$ and $c^L(d,1,\gamma)=c^U(d,1,\gamma)=0$, where $\widehat\gamma_L(d)$ and $\widehat\gamma_U(d)$ are vacuously set equal to fixed constants.  The conditioning events therefore contain only the realized winner; the binding-bound and additional-conditioning labels in the general framework carry no information in this specialization.

\textbf{Step 3:~Assumption \ref{ass: selection rule} differs from AKM only on the tie event.} \\
In the AKM specialization, the rows determining the winner event are the pairwise contrasts $r_{d'}^X-r_d^X$.  As shown in Step 6 below, AKM Assumption 8 implies that each such contrast has asymptotic variance bounded away from zero.  Together with AKM Assumption 6, this implies that the probability of a non-unique winner converges to zero.  Assumption \ref{ass: selection rule} is stated as exact almost-sure uniqueness in order to define the unconditional target without additional notation.  The formulations therefore differ only on an asymptotically negligible tie event, which can alternatively be resolved by a deterministic tie-breaking rule.

\textbf{Step 4:~AKM Assumption 6 is equivalent to Assumption \ref{ass: joint normality}.} \\
AKM Assumption 6 requires a uniform Gaussian approximation, in bounded-Lipschitz metric, for
\begin{equation*}
    \begin{pmatrix}
        X_n-\mu_{X,n}(\mathbb P)\\
        Y_n-\mu_{Y,n}(\mathbb P)
    \end{pmatrix}.
\end{equation*}
By \eqref{eq:akm-asymptotic-scaled-embedding}, this vector is $\sqrt n\{\widehat p-p(\mathbb P)\}$.  Therefore, AKM Assumption 6 is identical to Assumption \ref{ass: joint normality} in the special case setting of AKM.

\textbf{Step 5:~AKM Assumption 7 is equivalent to Assumption \ref{ass: variance estimation}.} \\
AKM Assumption 7 requires their covariance estimator $\widehat\Sigma_n$ to be uniformly consistent for $\Sigma(\mathbb P)$.  Under \eqref{eq:akm-asymptotic-p-embedding}, this is exactly Assumption \ref{ass: variance estimation} of the present paper.

\textbf{Step 6:~AKM Assumption 8 implies Assumption \ref{ass: variance restriction}.} \\

AKM Assumption 8 requires the asymptotic variances $\Sigma_X(d;\mathbb P)$ and $\Sigma_Y(d;\mathbb P)$ to be uniformly bounded above and away from zero for every $d\in\cD$.  It also requires
\begin{equation}
    \sqrt{\Sigma_X(d;\mathbb P)\Sigma_X(d';\mathbb P)}
    -\Sigma_X(d,d';\mathbb P)
    \label{eq:akm-positive-correlation-gap}
\end{equation}
to be uniformly bounded away from zero for every $d,d'\in\cD$ such that $d\neq d'$.  Since $\cD$ is fixed, the constants appearing in these bounds can be replaced by a common finite constant $\bar\lambda$ after increasing it if necessary.

Under the embedding in \eqref{eq:akm-asymptotic-p-embedding}, $\Sigma(\mathbb P)$ is the covariance matrix of the stacked vector $(X_n',Y_n')'$ and is therefore positive semidefinite.  Moreover, the upper bounds on the component variances in AKM Assumption 8 imply $\lambda_{\max}\{\Sigma(\mathbb P)\}\leq \operatorname{tr}\{\Sigma(\mathbb P)\} =\sum_{d\in\cD}\left\{\Sigma_X(d;\mathbb P)+\Sigma_Y(d;\mathbb P)\right\}\leq \bar\lambda$
uniformly over $\mathbb P\in\mathcal P_n$, after increasing $\bar\lambda$ to account for the fixed cardinality of $\cD$.  Hence, Part 1.~of Assumption \ref{ass: variance restriction} is implied.

Next, under \eqref{eq:akm-asymptotic-ass31}, $\ell_{d,1}=u_{d,1}=r_d^Y.$  Since $r_d^Y$ is a selector row vector, $\|r_d^Y\|^2=1$, and AKM Assumption 8 implies
\begin{align*}
    \ell_{d,1}\Sigma(\mathbb P)\ell_{d,1}'
    &=
    r_d^Y\Sigma(\mathbb P)r_d^{Y\prime}
    =
    \Sigma_Y(d;\mathbb P)
    \geq \frac{1}{\bar\lambda}
    =
    \frac{\|\ell_{d,1}\|^2}{\bar\lambda},\\
    u_{d,1}\Sigma(\mathbb P)u_{d,1}'
    &=
    r_d^Y\Sigma(\mathbb P)r_d^{Y\prime}
    =
    \Sigma_Y(d;\mathbb P)
    \geq \frac{1}{\bar\lambda}
    =
    \frac{\|u_{d,1}\|^2}{\bar\lambda}.
\end{align*}
Thus, Part 2.~of Assumption \ref{ass: variance restriction} is implied.

Finally, by Step 2, every nonzero row of $A^L(d,1,\gamma)=A^U(d,1,\gamma)=A_d$ takes the form $r_{d'}^X-r_d^X$ for some $d'\neq d$.  Since $r_d^X$ and $r_{d'}^X$ select distinct coordinates, $\left\|r_{d'}^X-r_d^X\right\|^2=2$.  Furthermore,
\begin{align*}
    &(r_{d'}^X-r_d^X)
    \Sigma(\mathbb P)
    (r_{d'}^X-r_d^X)'\\
    &\qquad=
    \Sigma_X(d';\mathbb P)
    +\Sigma_X(d;\mathbb P)
    -2\Sigma_X(d,d';\mathbb P)\\
    &\qquad=
    \left\{
        \sqrt{\Sigma_X(d';\mathbb P)}
        -
        \sqrt{\Sigma_X(d;\mathbb P)}
    \right\}^2\\
    &\qquad\quad+
    2\left\{
        \sqrt{\Sigma_X(d;\mathbb P)\Sigma_X(d';\mathbb P)}
        -
        \Sigma_X(d,d';\mathbb P)
    \right\}\\
    &\qquad\geq
    \frac{2}{\bar\lambda}
    =
    \frac{\left\|r_{d'}^X-r_d^X\right\|^2}{\bar\lambda},
\end{align*}
where the inequality follows from \eqref{eq:akm-positive-correlation-gap}.  Hence, Part 3.~of Assumption \ref{ass: variance restriction} is also implied.

\subsection{Specialization of Conditional Inference to AKM Setting}
\label{app:akm-asymptotic-conditional}

\textbf{Step C1: the endpoint statistic becomes the selected AKM inferential statistic.} \\
Under \eqref{eq:akm-asymptotic-ass31}, both estimated endpoints for option $d$ equal $\widetilde Y_n(d)$.  The lower- and upper-endpoint residual statistics of the present paper therefore coincide.  Let $\widehat b(d)=\widehat\Sigma r_d^{Y\prime}\big(r_d^Y\widehat\Sigma r_d^{Y\prime}\big)^{-1}$
and define
\begin{equation}
    \widehat Z_n(d)
    =\sqrt n\widehat p
    -\widehat b(d)\sqrt n\,r_d^Y\widehat p.
    \label{eq:akm-asymptotic-residual-full}
\end{equation}
The $X$ block of \eqref{eq:akm-asymptotic-residual-full} is
\begin{equation}
     Z_{n}(d)
    =X_n-
    \frac{\widehat\Sigma_{XY,n}(\cdot,d)}
         {\widehat\Sigma_{Y,n}(d)}Y_n(d),
    \label{eq:akm-asymptotic-residual-x}
\end{equation}
which is the feasible residual sufficient statistic used by AKM.  The remaining coordinates in \eqref{eq:akm-asymptotic-residual-full} are residualized inferential statistics for options other than $d$.  They do not enter the winner inequalities and therefore do not alter the scalar truncation interval.  Thus, conditioning on the full residual vector in the present paper gives the same truncated distribution for $Y_n(d)$ as conditioning on the AKM residual in \eqref{eq:akm-asymptotic-residual-x}.

\textbf{Step C2: the polyhedral truncation set is the AKM winner truncation set.} \\
Let $\mathcal X(d)=\{x\in\Rr^{K+1}:x(d)\geq x(d')\text{ for every }d'\neq d\}$.  Conditional on $Z_{n}(d)=z$, the event $\widehat d=d$ is equivalent to $Y_n(d)\in\mathcal Y_n(d,z)$, where
\begin{equation}
    \mathcal Y_n(d,z)
    =\left\{y:
    z+\frac{\widehat\Sigma_{XY,n}(\cdot,d)}
             {\widehat\Sigma_{Y,n}(d)}y
    \in\mathcal X(d)
    \right\}.
    \label{eq:akm-asymptotic-akm-yset}
\end{equation}
The polyhedral lemma used in the present paper expresses \eqref{eq:akm-asymptotic-akm-yset} as an interval with endpoints $\widehat{\mathcal V}^-(d,z)$ and $\widehat{\mathcal V}^+(d,z)$ (dropping the unnecessary dependence on $j$ and $\gamma$ for the present special case).  These are the same lower and upper truncation bounds denoted by AKM's $\mathcal L(d,z)$ and $\mathcal U(d,z)$, after translating notation and replacing the covariance matrix $\Sigma$ by its consistent estimator $\widehat\Sigma$.

\textbf{Step C3: the quantile inversion and equal-tailed CIs coincide.} \\
Let $F_{TN,n}(\cdot;\mu,d,z)$ be the CDF of a normal random variable with mean $\mu$ and variance $\widehat\Sigma_{Y,n}(d)$ truncated to \eqref{eq:akm-asymptotic-akm-yset}.  In the AKM point-identified specialization described above, the conditional endpoint estimators in the present paper both solve $F_{TN,n}\big(Y_n(\widehat d);\mu,\widehat d,Z_{n}(\widehat d)\big)=1-q$ in $\mu$.  This is the feasible AKM quantile-unbiased estimator $\widehat\mu_{q,n}$.  Combining the lower and upper inversions at $q=\alpha/2$ and $q=1-\alpha/2$ gives the CI $\left[\widehat\mu_{\alpha/2,n},\widehat\mu_{1-\alpha/2,n}\right]$, which is the feasible equal-tailed conditional CI in AKM.  Thus, after setting $J_L=J_U=1$ and making the auxiliary conditioning vectors constant, the conditional method of the present paper is exactly the AKM feasible conditional method.

\subsection{Specialization of Projection Inference to AKM Setting}
\label{app:akm-asymptotic-projection}

\textbf{Step P1: simultaneous endpoint inference collapses to simultaneous inference on $\mu_{Y,n}$.} \\
Under point identification, the lower and upper coefficient matrices in the present paper both reduce to the selector matrix for $\widetilde Y_n$:~$\ell_{\mathrm{mat}}=u_{\mathrm{mat}}=(r_{d^0}^Y,\ldots,r_{d^K}^Y)'$.  Hence, the lower- and upper-endpoint covariance matrices both reduce to the asymptotic covariance matrix $\Sigma_Y$ of $Y_n$.  Both projection methods therefore begin from simultaneous inference for the same vector:~$\mu_{Y,n}=\big(\mu_{Y,n}(d^0),\ldots,\mu_{Y,n}(d^K)\big)'$.

\textbf{Step P2:~the exact construction of the projection CIs differs.} \\
Treating $\widehat\Sigma_{Y,n}$ as fixed, let $\widehat c_{1-q}^+$ be the $1-q$ quantile of $\max_{d\in\cD}\xi_Y(d)/\sqrt{\widehat\Sigma_{Y,n}(d)}$ for $\xi_Y\sim\mathcal N(0,\widehat\Sigma_{Y,n})$.  If the separate lower- and upper-endpoint formula in the present paper is applied mechanically to the AKM setting with $\alpha_1=\alpha_2=\alpha/2$, the projection CI for $\mu_{y,n}(\widehat d)$ is
\begin{equation}
    \left[
    Y_n(\widehat d)-\widehat c_{1-\alpha/2}^+
    \sqrt{\widehat\Sigma_{Y,n}(\widehat d)},
    Y_n(\widehat d)+\widehat c_{1-\alpha/2}^+
    \sqrt{\widehat\Sigma_{Y,n}(\widehat d)}
    \right].
    \label{eq:akm-asymptotic-bonf-projection}
\end{equation}
On the other hand, the feasible projection CI for $\mu_{y,n}(\widehat d)$ in AKM is given by
\begin{equation}
\left[
    Y_n(\widehat d)-\widehat c_{1-\alpha}^{\mathrm{abs}}
    \sqrt{\widehat\Sigma_{Y,n}(\widehat d)},
    Y_n(\widehat d)+\widehat c_{1-\alpha}^{\mathrm{abs}}
    \sqrt{\widehat\Sigma_{Y,n}(\widehat d)}
    \right],
    \label{eq:akm-asymptotic-projection-ci}
\end{equation}
where $\widehat c_{1-\alpha}^{\mathrm{abs}}$ is the $1-\alpha$ quantile of $\max_{d\in\cD}
    |\xi_Y(d)|/\sqrt{\widehat\Sigma_{Y,n}(d)}$ when treating $\widehat\Sigma_{Y,n}$ as fixed.  Therefore, the projection \emph{principle} in the present paper specializes directly to AKM but the projection CIs do not coincide in the point-identified winners special case primarily because the endpoints of the projection CI of the present paper are designed to separately cover the lower- and upper-bounds of the identified set under partial identification while the CI of AKM is not robust to partial identification.  

\subsection{Specialization of Hybrid Inference to the AKM Setting}
\label{app:akm-asymptotic-hybrid}

In analogy with their projection counterparts, the hybrid CI of the present paper uses the same hybrid \emph{principle} as that of AKM when specialized to the point-identified winners setting.  However, they differ in their details because the hybrid CI of the present paper is designed to separately cover the lower- and upper-bounds of the identified set under partial identification while the CI of AKM is not robust to partial identification.  In particular, both hybrid CIs are formed as conditional CIs after intersecting the selection event $\{\widehat d=d\}$ with the event that $\mu_{y,n}(\widehat d)$ lies inside of a projection CI.  However, the feasible hybrid CI of AKM is based upon intersecting the selection event with the event that $\mu_{y,n}(\widehat d)$ lies in \eqref{eq:akm-asymptotic-projection-ci}, whereas the hybrid CI of the present paper intersects the selection event with the event that $\mu_{y,n}(\widehat d)$ lies in one-sided versions of \eqref{eq:akm-asymptotic-bonf-projection}.  Moreover, unlike AKM, the hybrid CI of the present paper treats the lower and upper endpoints separately in order to cover an identified set.  It adds a one-sided projection safeguard for the lower-endpoint and the opposite one-sided safeguard for the upper-endpoint experiment, and then combines the two endpoint bounds.  In the point-identified setting of AKM, this is unnecessary.

\section{Reality Check Power Comparison}\label{sec: power comp}

To our knowledge, the CIs proposed in this paper are the first with proven asymptotic validity for partially-identified parameters selected from the data.  Therefore, we have no existing inference method to compare the performance of our CIs to when the partially-identified parameter is data-dependent.  However, as discussed in Section \ref{sec: general framework} above, our inference framework covers cases for which the partially-identified parameter is chosen a priori.  For these special cases, there is a large literature on inference on partially-identified parameters or their identified sets that can be applied to form CIs.  Although these special cases are not of primary interest for this paper, in this section we compare the performance of our proposed inference methods to one of the leading inference methods in the partial identification literature as a ``reality check'' on whether our proposed methods are reasonably informative.  

In particular, we compare the power of the test implied by our hybrid CI (i.e.,~a test that rejects when the value of the parameter under the null hypothesis lies outside of the hybrid CI) to the power of the hybrid test of \cite{ARP23}, which applies to a general class of moment-inequality models.  When $d$ is chosen a priori and the parameter $p$ is equal to a vector of moments of underlying data, Assumption \ref{ass: welfare bounds} implies that $L(d)\leq W(d)\leq U(d)$ can be written as a set of (unconditional) moment inequalities, a special case of the general framework of that paper.  Of the many papers on inference for moment inequalities, we choose the test of \cite{ARP23} for comparison for two reasons:~(i) it has been shown to be quite competitive in terms of power and (ii) it is also based upon a (different) inference method that is a hybrid between conditional and projection-based inference. 

When $d$ is chosen a priori, our conditional construction and that of \citet{ARP23} share the same general polyhedral-conditioning logic.  In both cases, inference is based upon an asymptotically Gaussian scalar statistic by conditioning on a residual sufficient statistic for the remaining mean parameters and on a polyhedral optimizer event but the commonality basically ends there.  In our setting, the conditioning event identifies the affine component attaining an intersection bound and the scalar statistic is the corresponding estimate of the lower or upper endpoint.  We invert its truncated-normal distribution in the endpoint mean.  In \citet{ARP23}, the statistic is the profiled maximum violation of a system of moment inequalities.  Its dual representation is a maximum over Lagrange multipliers and their procedure conditions on the optimal dual multiplier before using the null restriction that its mean is nonpositive.  Another difference between the inference methods is that we conduct endpoint-specific inference and combine separate lower- and upper-endpoint confidence bounds whereas their test treats the moment inequalities for a candidate parameter value jointly through a profiled max statistic.  Finally, our hybrid procedure combines conditional inference with a simultaneous projection bound while their hybrid procedure combines a conditional test with a least-favorable first-stage test.

\begin{figure}
\centering
\includegraphics[scale=0.29,trim={2.5cm 7cm 1cm 7cm}]{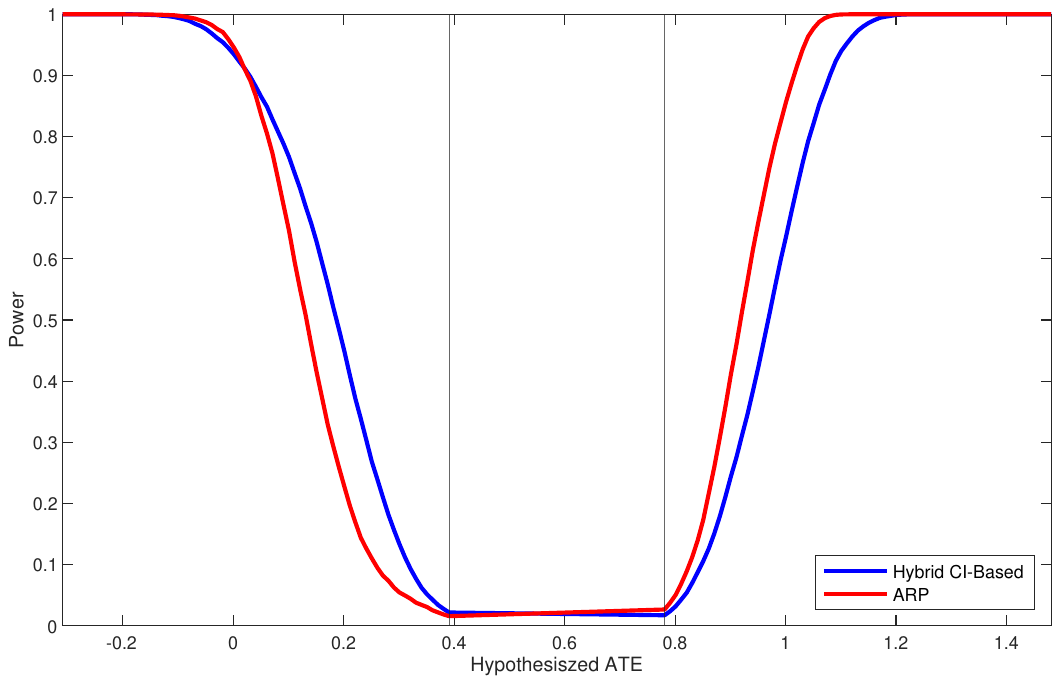}\includegraphics[scale=0.29,trim={1cm 7cm 1cm 7cm}]{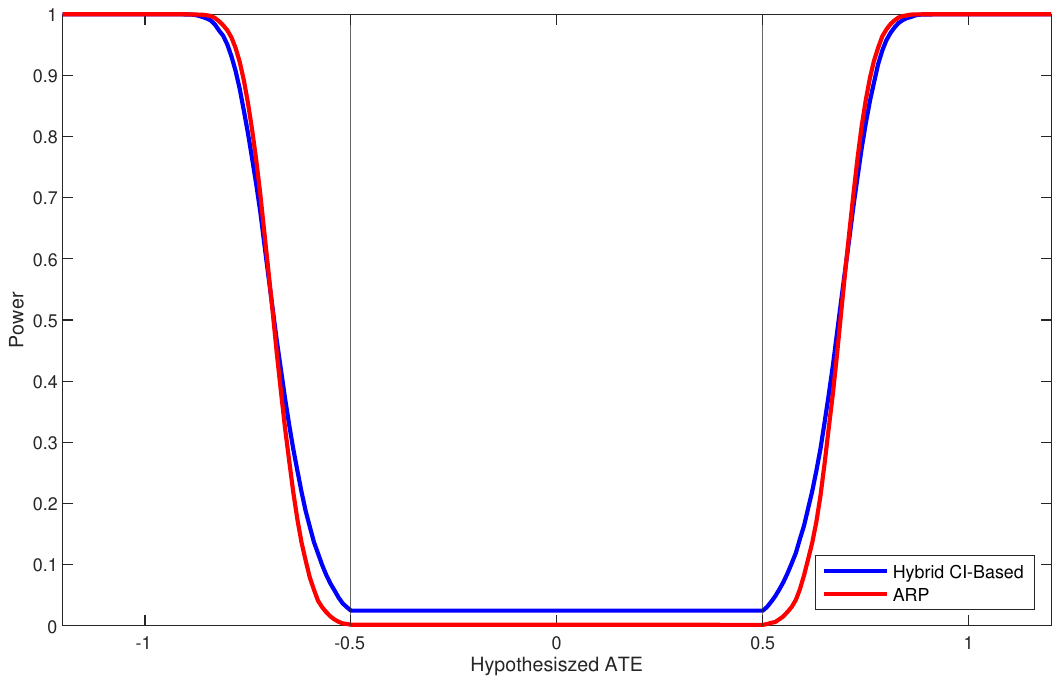}\includegraphics[scale=0.29,trim={1cm 7cm 1cm 7cm}]{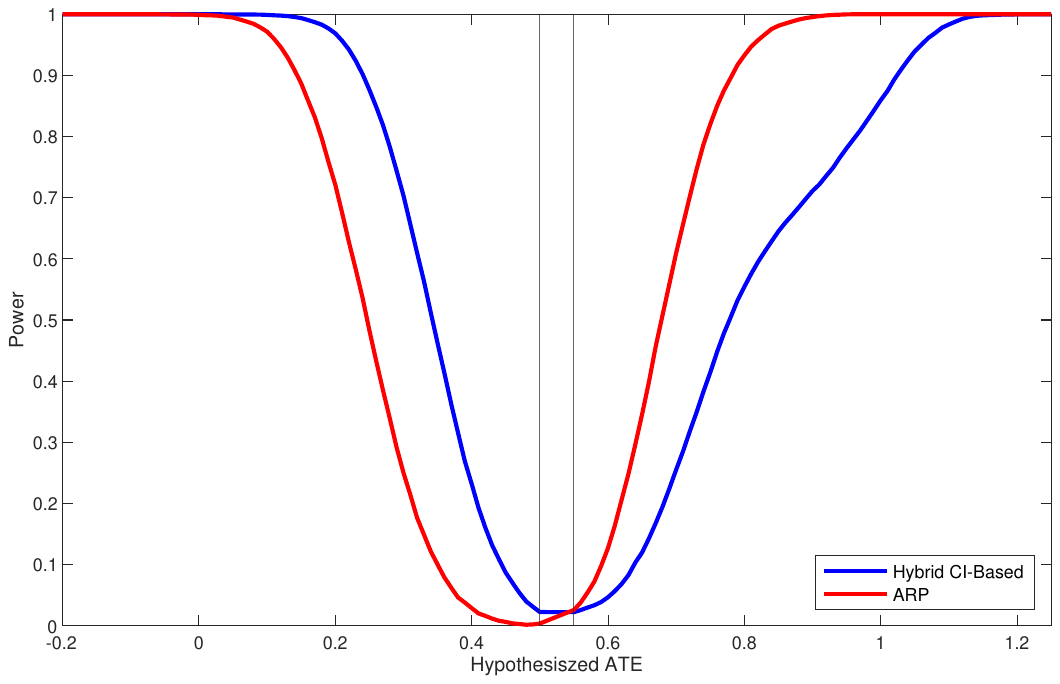}
\caption{Power curves for hybrid \cite{ARP23} test (red) and test implied by hybrid CI (blue) of ATE using bounds from \cite{BP97,BP13} for $p=(.08,.001,.001,.073,.139,.473)'$ (left), $p=(.25,.25,.25,.25,.25,.25)'$ (middle) and $p=(.01,.44,.01,.01,.01,.54)'$ (right) and $n=100$.  The vertical lines illustrate the true identified interval in each of these settings.}
\label{fig: ARP comp}
\end{figure}

We compare the power of tests on the ATE using the
Balke--Pearl bounds described in Section
\ref{sec: Monte Carlo}.  The second and third DGPs are
the same as in that section.  For the first DGP, the
power comparison uses
$p=(.08,.001,.001,.073,.139,.473)'$, a perturbation of
the exact Balke--Pearl calibration
$p=(.081,0,0,.073,.139,.473)'$ used in Section
\ref{sec: Monte Carlo}.  Figure \ref{fig: ARP comp}
plots the power curves of the hybrid \cite{ARP23}
test and the test implied by our hybrid CI, together
with the endpoints of the true identified interval
for the ATE in each of these three DGPs.

We can see that overall, the power of the test implied by our hybrid CI is quite competitive with that of \cite{ARP23}.  Interestingly, it appears that our test tends to be more powerful than that of \cite{ARP23} when the true ATE is larger than the hypothesized one, which can be seen from the hypothesized ATE lying to the left of the lower bound of the identified interval, and vice versa.  Although the main innovation of our inference procedures is really their validity in the presence of data-dependent selection, the exercise of this section is reassuring for the informativeness of the procedures we propose as they are quite competitive in the absence of selection.

\section{Additional Monte Carlo Simulations}\label{sec: additional simulation}

In analogy with Section \ref{sec: Monte Carlo}, we compute the unconditional coverage frequencies of the conditional, projection and hybrid CIs for DGPs in the dynamic treatment regime setting of the empirical application (Section \ref{sec: application}).  In particular, we consider two DGPs: DGP 1 generates $\widehat p$ from a multinomial distribution based on $p_{d_1,y_1,d_2,y_2|z_1,z_2}=0.25$ for $(d_1,y_1,d_2,y_2)\in\{(1,0,0,1),(1,1,1,1)\}$ and all $(z_1,z_2)$ and $p_{d_1,y_1,d_2,y_2|z_1,z_2}=0.0357$ for all other $(d_1,y_1,d_2,y_2)$ and all $(z_1,z_2)$; DGP 2 generates $\widehat p$ based on $p_{d_1,y_1,d_2,y_2|z_1,z_2}=0.375$ for $(d_1,y_1,d_2,y_2)\in\{(1,0,0,1),(1,1,1,1)\}$ and all $(z_1,z_2)$ and $p_{d_1,y_1,d_2,y_2|z_1,z_2}=0.0179$ for all other $(d_1,y_1,d_2,y_2)$ and all $(z_1,z_2)$.  The coverage frequencies of the CIs are reported in Table \ref{table:cov freqs app}.  Again, consistent with the asymptotic results of Theorems \ref{thm: cond'l coverage for cond'l bounds}--\ref{thm: uncond'l coverage for hybrid bounds}, the conditional, projection and hybrid CIs all have correct coverage for all DGPs and the sample size of $n=1000$.  Also, the projection CI tends to be conservative with true coverage above the nominal level of 95\%, and the conventional CI can substantially under-cover.

\begin{table}[htbp]
\caption{Unconditional Coverage Frequencies}\label{table:cov freqs app}
\centering{}%
\begin{tabular}{lccccc}
 & & \multicolumn{4}{c}{Confidence Interval}\tabularnewline
Data-Generating Process &  & Conv & Cond  & Proj & Hyb \tabularnewline
\hline
DGP 1  &  &  0.66	&0.94	&0.99	&0.94  \tabularnewline
DGP 2  &  & 0.88	&0.94	&0.99	&0.95 \tabularnewline
\hline
\end{tabular}
\vspace{0.2in}
\caption*{\footnotesize This table reports unconditional coverage frequencies for the counterfactual outcome selected by maximizing the estimated lower bound on the counterfactual outcomes of one of the eight dynamic regimes, all evaluated at the nominal coverage level of 95\%.  Coverage frequencies are reported for conventional (``Conv''), conditional (``Cond''), projection (``Proj'') and hybrid (``Hyb'') CIs for a sample size of $n=1000$.}
\end{table}

\begin{figure}
\centering
\includegraphics[scale=0.29,trim={2.5cm 7cm 1cm 7cm}]{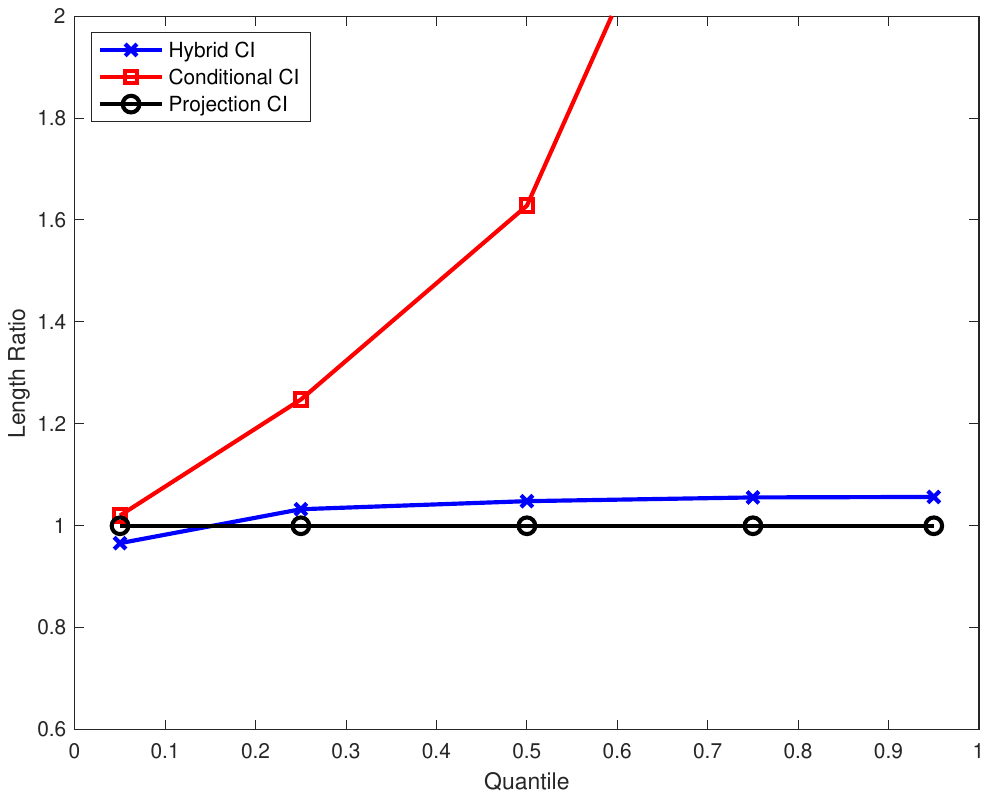}\includegraphics[scale=0.29,trim={1cm 7cm 1cm 7cm}]{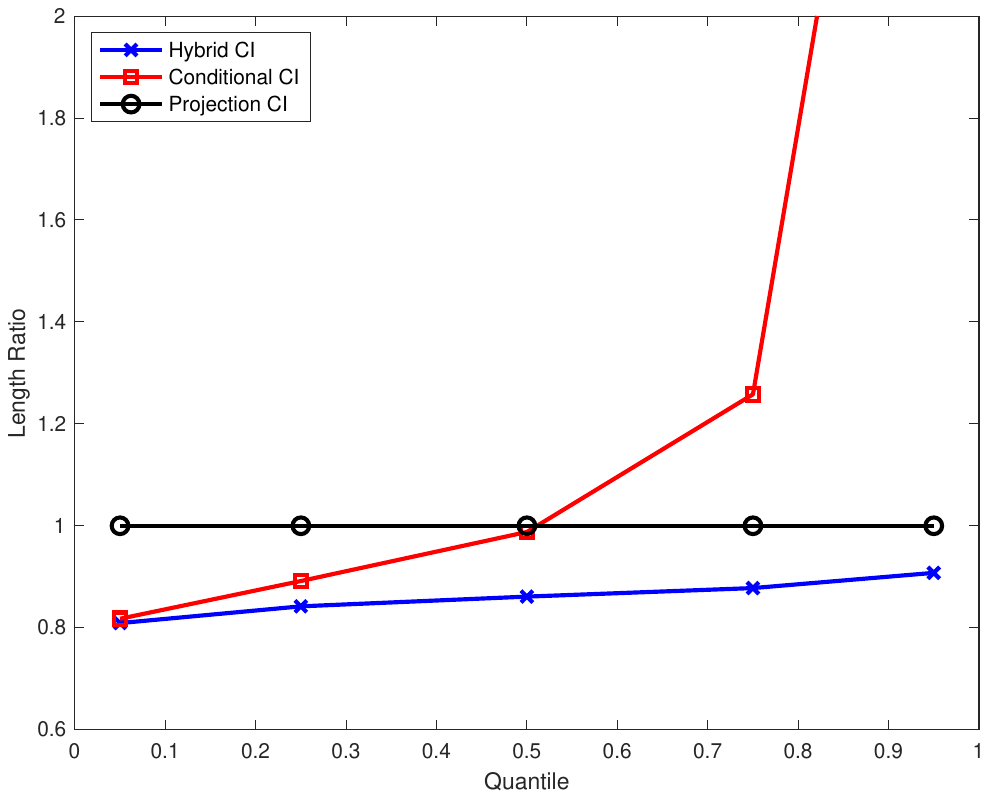}
\caption{Ratios of CI length quantiles relative to those of the projection CI for the conditional CI (red), projection CI (black) and hybrid CI (blue) of the average counterfactual outcome selected by maximizing the estimated lower bound for DGP 1 (left) and DGP 2 (right) and $n=1000$.}
\label{fig: CI length comp app}
\end{figure}

Figure \ref{fig: CI length comp app} plots the ratios of the $5^{th}$, $25^{th}$, $50^{th}$, $75^{th}$ and $95^{th}$ quantiles of the length of the conditional, projection and hybrid CIs relative to those same length quantiles of the projection CI.  The figure shows that the conditional CI has the tendency to become very long, especially at high quantile levels for certain DGPs, whereas the hybrid CI tends to perform the best overall by limiting the worst-case length performance of the conditional CI relative to the projection CI.  Relative to projection, the hybrid CI enjoys length reductions of 10-20\% for favorable DGPs while showing length increases of 5-10\% for unfavorable DGPs.

\section{Technical Appendix}\label{sec: appendix}

\textbf{Proof of Proposition~\ref{prop: weak polyhedral examples}:} In Example \ref{ex:maximin}, $d=\hd$ if and only if, for some
index $j^*_L\in\{1,\ldots,J_L\}$,
\begin{equation}
    (\ell_{d',j'}-\ell_{d,j^*_L})\hp
\leq \tilde\ell_{d,j^*_L}-\tilde\ell_{d',j'}
\qquad\text{for all }d'\in\cD,\ 
j'\in\{1,\ldots,J_L\}.
\label{eq:maximin_lower_active}
\end{equation}
Thus, the selection event is a union of $J_L$ polyhedra. An analogous expression can be derived for Example \ref{ex:hurwicz}. 

In Example \ref{ex:minimax}, $d=\hd$ if and only if, for some indices
$j^*_{a,L}\in\{1,\ldots,J_L\}$,
$j^*_{a,U}\in\{1,\ldots,J_U\}$, and $b^*_a\in\cD$
for each $a\in\cD$,
\begin{align}
(\ell_{a,j}-\ell_{a,j^*_{a,L}})\hp
&\leq \tilde\ell_{a,j^*_{a,L}}-\tilde\ell_{a,j}
\qquad\text{for all }a\in\cD,\ j=1,\ldots,J_L,
\label{eq:minimax-lower-active}\\
(u_{a,j^*_{a,U}}-u_{a,j})\hp
&\leq \tilde u_{a,j}-\tilde u_{a,j^*_{a,U}}
\qquad\text{for all }a\in\cD,\ j=1,\ldots,J_U,
\label{eq:minimax-upper-active}\\
\widehat U(b,d)
&\leq\widehat U(b^*_a,a)
\qquad\text{for all }a,b\in\cD.
\label{eq:minimax-comparisons}
\end{align}
Equations~\eqref{eq:minimax-lower-active}
and~\eqref{eq:minimax-upper-active} imply
\begin{align}
\widehat U(b,a)
&=
\begin{cases}
0, & b=a,\\
\tilde u_{b,j^*_{b,U}}
-\tilde\ell_{a,j^*_{a,L}}
+(u_{b,j^*_{b,U}}-\ell_{a,j^*_{a,L}})\hp,
& b\neq a,
\end{cases}
\qquad a,b\in\cD.
\label{eq:minimax-pairwise-affine}
\end{align}
By \eqref{eq:minimax-pairwise-affine},
\eqref{eq:minimax-comparisons} also consists of linear inequalities.
The existence of indices $b^*_a$ satisfying
\eqref{eq:minimax-comparisons} is equivalent to
$\widehat R(d)\leq\widehat R(a)$ for every $a\in\cD$, which characterizes selection. Therefore, the selection event is the union of the polyhedra defined by
\eqref{eq:minimax-lower-active}--\eqref{eq:minimax-comparisons}
over the finitely many index choices.

In Example \ref{ex:undominated}, we have that $d\in\widehat{\cD}$ if and only if
\begin{equation}
    (\ell_{d',j'}-u_{d,j})\widehat p \leq \tilde u_{d,j}-\tilde\ell_{d',j'}
    \label{eq:estimated_set_ineq}
\end{equation}
for all $d'\in\{d^0,\ldots,d^K\}$, $j\in \{1,\ldots,J_U\}$ and $j'\in\{1,\ldots,J_L\}$.  $\blacksquare$

\textbf{Proof of Proposition~\ref{prop: selection rule}:}
All event equivalences below hold a.s. We consider consistent
conditioning values; inconsistent values give the empty
polyhedron. We suppress the conditioning arguments of
$A^M,c^M$ and write $j^*_{a,M}$ for a fixed value of
$\widehat j_M(a)$, $M\in\{L,U\}$.

For Examples \ref{ex:maximin}--\ref{ex:hurwicz}, $d=\hd$ if and only if
\begin{align}
w_L\widehat L(d)+w_U\widehat U(d)
&\geq w_L\widehat L(a)+w_U\widehat U(a),
\qquad\text{for all }a\in\cD.
\label{eq: wtd combo equiv}
\end{align}
Define the endpoint-activity matrices and vectors by
\begin{align}
A_L(a,j)
&=\begin{pmatrix}
\ell_{a,1}-\ell_{a,j}\\
\vdots\\
\ell_{a,J_L}-\ell_{a,j}
\end{pmatrix},
&
c_L(a,j)
&=\begin{pmatrix}
\tilde\ell_{a,j}-\tilde\ell_{a,1}\\
\vdots\\
\tilde\ell_{a,j}-\tilde\ell_{a,J_L}
\end{pmatrix},
\label{eq:selection-lower-activity}\\
A_U(a,j)
&=\begin{pmatrix}
u_{a,j}-u_{a,1}\\
\vdots\\
u_{a,j}-u_{a,J_U}
\end{pmatrix},
&
c_U(a,j)
&=\begin{pmatrix}
\tilde u_{a,1}-\tilde u_{a,j}\\
\vdots\\
\tilde u_{a,J_U}-\tilde u_{a,j}
\end{pmatrix}.
\label{eq:selection-upper-activity}
\end{align}
Thus, $\widehat j_M(a)=j$ is characterized by
$A_M(a,j)\hp\leq c_M(a,j)$.

\begin{enumerate}
\item When $w_U=0<w_L$, the event
$\{\hd=d,\widehat j_L(d)=j^*_{d,L}\}$ is characterized by
\eqref{eq:maximin_lower_active}, with $j_L^*=j^*_{d,L}$.
Consequently,
\begin{align}
A^L
&=\begin{pmatrix}
\ell_{d^0,1}-\ell_{d,j^*_{d,L}}\\
\vdots\\
\ell_{d^0,J_L}-\ell_{d,j^*_{d,L}}\\
\vdots\\
\ell_{d^K,1}-\ell_{d,j^*_{d,L}}\\
\vdots\\
\ell_{d^K,J_L}-\ell_{d,j^*_{d,L}}
\end{pmatrix},
&
c^L
&=\begin{pmatrix}
\tilde\ell_{d,j^*_{d,L}}-\tilde\ell_{d^0,1}\\
\vdots\\
\tilde\ell_{d,j^*_{d,L}}-\tilde\ell_{d^0,J_L}\\
\vdots\\
\tilde\ell_{d,j^*_{d,L}}-\tilde\ell_{d^K,1}\\
\vdots\\
\tilde\ell_{d,j^*_{d,L}}-\tilde\ell_{d^K,J_L}
\end{pmatrix}.
\label{eq:selection-maximin-matrices}
\end{align}
For upper-endpoint inference, additionally impose
$\widehat j_U(d)=j^*_{d,U}$ using
\eqref{eq:selection-upper-activity}. Hence,
\begin{align}
A^U
&=\begin{pmatrix}A^L\\A_U(d,j^*_{d,U})\end{pmatrix},
&
c^U
&=\begin{pmatrix}c^L\\c_U(d,j^*_{d,U})\end{pmatrix}.
\label{eq:selection-maximin-upper-matrices}
\end{align}

\item When $w_L=0<w_U$, fix
$\widehat j_U(a)=j^*_{a,U}$ for every $a\in\cD$.
Substituting these indices into \eqref{eq: wtd combo equiv}
and adding their activity restrictions gives
\begin{align}
A^U
&=\begin{pmatrix}
u_{d^0,j^*_{d^0,U}}-u_{d,j^*_{d,U}}\\
\vdots\\
u_{d^K,j^*_{d^K,U}}-u_{d,j^*_{d,U}}\\
A_U(d^0,j^*_{d^0,U})\\
\vdots\\
A_U(d^K,j^*_{d^K,U})
\end{pmatrix},
&
c^U
&=\begin{pmatrix}
\tilde u_{d,j^*_{d,U}}-\tilde u_{d^0,j^*_{d^0,U}}\\
\vdots\\
\tilde u_{d,j^*_{d,U}}-\tilde u_{d^K,j^*_{d^K,U}}\\
c_U(d^0,j^*_{d^0,U})\\
\vdots\\
c_U(d^K,j^*_{d^K,U})
\end{pmatrix}.
\label{eq:selection-maximax-matrices}
\end{align}
For lower-endpoint inference, additionally impose
$\widehat j_L(d)=j^*_{d,L}$ using
\eqref{eq:selection-lower-activity}. Hence,
\begin{align}
A^L
&=\begin{pmatrix}A^U\\A_L(d,j^*_{d,L})\end{pmatrix},
&
c^L
&=\begin{pmatrix}c^U\\c_L(d,j^*_{d,L})\end{pmatrix}.
\label{eq:selection-maximax-lower-matrices}
\end{align}

\item When $w_L,w_U>0$, fix
$\widehat j_L(a)=j^*_{a,L}$ and $\widehat j_U(a)=j^*_{a,U}$
for every $a\in\cD$. Substitution into
\eqref{eq: wtd combo equiv}, together with
\eqref{eq:selection-lower-activity}--\eqref{eq:selection-upper-activity},
gives
\begin{align}
A^L=A^U
&=\begin{pmatrix}
w_L(\ell_{d^0,j^*_{d^0,L}}-\ell_{d,j^*_{d,L}})
+w_U(u_{d^0,j^*_{d^0,U}}-u_{d,j^*_{d,U}})\\
\vdots\\
w_L(\ell_{d^K,j^*_{d^K,L}}-\ell_{d,j^*_{d,L}})
+w_U(u_{d^K,j^*_{d^K,U}}-u_{d,j^*_{d,U}})\\
A_L(d^0,j^*_{d^0,L})\\
\vdots\\
A_L(d^K,j^*_{d^K,L})\\
A_U(d^0,j^*_{d^0,U})\\
\vdots\\
A_U(d^K,j^*_{d^K,U})
\end{pmatrix},
\label{eq:selection-hurwicz-matrix}\\
c^L=c^U
&=\begin{pmatrix}
w_L(\tilde\ell_{d,j^*_{d,L}}-\tilde\ell_{d^0,j^*_{d^0,L}})
+w_U(\tilde u_{d,j^*_{d,U}}-\tilde u_{d^0,j^*_{d^0,U}})\\
\vdots\\
w_L(\tilde\ell_{d,j^*_{d,L}}-\tilde\ell_{d^K,j^*_{d^K,L}})
+w_U(\tilde u_{d,j^*_{d,U}}-\tilde u_{d^K,j^*_{d^K,U}})\\
c_L(d^0,j^*_{d^0,L})\\
\vdots\\
c_L(d^K,j^*_{d^K,L})\\
c_U(d^0,j^*_{d^0,U})\\
\vdots\\
c_U(d^K,j^*_{d^K,U})
\end{pmatrix}.
\label{eq:selection-hurwicz-vector}
\end{align}

\item In Example \ref{ex:minimax}, fix the indices $\bigl(\widehat j_L(a),\widehat j_U(a),\widehat b(a)\bigr)_{a\in\cD}$ at
$(j^*_{a,L},j^*_{a,U},b^*_a)_{a\in\cD}$.
Their joint realization is characterized by
\eqref{eq:minimax-lower-active}--\eqref{eq:minimax-upper-active}
and
\begin{align}
\widehat U(b,a)
&\leq\widehat U(b^*_a,a),
\qquad\text{for all }a,b\in\cD.
\label{eq:minimax-regret-active}
\end{align}
Under \eqref{eq:minimax-regret-active},
$\widehat R(a)=\widehat U(b^*_a,a)$, so selection additionally
requires
\begin{align}
\widehat U(b^*_d,d)
&\leq\widehat U(b^*_a,a),
\qquad\text{for all }a\in\cD.
\label{eq:minimax-selected-comparisons}
\end{align}
By \eqref{eq:minimax-pairwise-affine} and
$\widehat U(a,a)=0$, all these inequalities are affine in
$\hp$. Stacking their coefficients gives
$A^L=A^U$ and $c^L=c^U$.
\end{enumerate}

Finally, Examples \ref{ex:undominated} and \ref{ex:predetermined} both condition on the endpoint realizations and therefore involve the inequalities $A_M(d,j_M^*)\hp\leq c_M(d,j_M^*)$ for the relevant
$M\in\{L,U\}$ characterized by \eqref{eq:selection-lower-activity}--\eqref{eq:selection-upper-activity}. These inequalities completely characterize $A^M$ and $c^M$, $M\in\{L,U\}$, for Example \ref{ex:predetermined} while these inequalities are concatenated with all of the inequalities in \eqref{eq:estimated_set_ineq} to produce $A^M$ and $c^M$ for Example \ref{ex:undominated}.  $\blacksquare$


\begin{lemma} \label{lem: double cond'l coverage for pseudo-bounds}
Suppose Assumptions \ref{ass: welfare bounds}, \ref{ass: estimated ID'd set} and \ref{ass: joint normality}--\ref{ass: variance restriction} hold.  Then, for any $0<\alpha<1$,
\begin{gather}
\lim_{n\rightarrow\infty}\sup_{\mathbb{P}\in\mathcal{P}_n}\left|\mathbb{P}\left(\left.\sqrt{n}(\tilde\ell_{d,\widehat j_L(d)}+\ell_{d,\widehat j_L(d)}p)\leq\widehat L(d)_{\alpha}^C\right|d\in\cD(\hp),\widehat j_L(d)=j_L^*,\widehat \gamma_L(d)=\gamma_L^*\right)-\alpha\right| \notag\\
\cdot\mathbb{P}\left(d\in\cD(\hp),\widehat j_L(d)=j_L^*,\widehat \gamma_L(d)=\gamma_L^*\right)=0, \label{eq: lb unif double cond cov}\\
\lim_{n\rightarrow\infty}\sup_{\mathbb{P}\in\mathcal{P}_n}\left|\mathbb{P}\left(\left.\sqrt{n}(\tilde u_{d,\widehat j_U(d)}+u_{d,\widehat j_U(d)}p)\leq\widehat U(d)_{\alpha}^C\right|d\in\cD(\hp),\widehat j_U(d)=j_U^*,\widehat \gamma_U(d)=\gamma_U^*\right)-\alpha\right| \notag \\
\cdot\mathbb{P}\left(d\in\cD(\hp),\widehat j_U(d)=j_U^*,\widehat \gamma_U(d)=\gamma_U^*\right)=0, \label{eq: ub unif double cond cov}
\end{gather}
for all $d\in\{d^0,\ldots,d^K\}$, $j_L^*\in \{1,\ldots,J_L\}$, $j_U^*\in \{1,\ldots,J_U\}$, $\gamma_L^*$ in the support of $\widehat\gamma_L(d)$ and $\gamma_U^*$ in the support of $\widehat\gamma_U(d)$.
\end{lemma}

\textbf{Proof:} The proof of \eqref{eq: ub unif double cond cov} is nearly identical to the proof of \eqref{eq: lb unif double cond cov} so that we only show the latter.  Lemma A.1 of \cite{LSST16} implies that $F_{TN}(t;\mu,\sigma^2,v^-,v^+)$ is strictly decreasing in $\mu$ so that $\sqrt{n}(\tilde\ell_{d,\widehat j_L(d)}+\ell_{d,\widehat j_L(d)}p)\leq\widehat L(d)_{\alpha}^C$ is equivalent to
\begin{gather*}
F_{TN}\left(\sqrt{n}(\tilde\ell_{d,\widehat j_L(d)}+\ell_{d,\widehat j_L(d)}\widehat p);\sqrt{n}(\tilde\ell_{d,\widehat j_L(d)}+\ell_{d,\widehat j_L(d)}p),\ell_{d,\widehat j_L(d)}\widehat\Sigma\ell_{d,\widehat j_L(d)}^{\prime}\right|\widehat{\mathcal{V}}_{L}^-\left(\widehat{\mathcal{Z}}_{L}(d,\widehat j_L(d))\right), \\
\left.\widehat{\mathcal{V}}_{L}^+\left(\widehat{\mathcal{Z}}_{L}(d,\widehat j_L(d))\right)\right)\geq 1-\alpha,
\end{gather*}
where we use $\widehat{\mathcal{V}}_{L}^-\left(\widehat{\mathcal{Z}}_{L}(d,\widehat j_L(d))\right)$ and $\widehat{\mathcal{V}}_{L}^+\left(\widehat{\mathcal{Z}}_{L}(d,\widehat j_L(d))\right)$ as shorthand for 
\[\widehat{\mathcal{V}}_{L}^-\left(\widehat{\mathcal{Z}}_{L}(d,\widehat j_L(d)),d,\widehat j_L(d),\widehat \gamma_L(d)\right)\] 
and 
\[\widehat{\mathcal{V}}_{L}^+\left(\widehat{\mathcal{Z}}_{L}(d,\widehat j_L(d)),d,\widehat j_L(d),\widehat \gamma_L(d)\right).\] 
In addition, Lemma 2 of \cite{McC24} implies
\begin{gather*}
F_{TN}\left(\sqrt{n}(\tilde\ell_{d,\widehat j_L(d)}+\ell_{d,\widehat j_L(d)}\widehat p);\sqrt{n}(\tilde\ell_{d,\widehat j_L(d)}+\ell_{d,\widehat j_L(d)}p),\ell_{d,\widehat j_L(d)}\widehat\Sigma\ell_{d,\widehat j_L(d)}^{\prime}\right|\widehat{\mathcal{V}}_{L}^-\left(\widehat{\mathcal{Z}}_{L}(d,\widehat j_L(d))\right), \\
\left.\widehat{\mathcal{V}}_{L}^+\left(\widehat{\mathcal{Z}}_{L}(d,\widehat j_L(d))\right)\right) \\
=F_{TN}\left(\ell_{d,\widehat j_L(d)}\sqrt{n}(\widehat p-p);0,\ell_{d,\widehat j_L(d)}\widehat\Sigma\ell_{d,\widehat j_L(d)}^{\prime}\right|\left.\widehat{\mathcal{V}}_{L}^-\left(\widehat{\mathcal{Z}}_{L}^*(d,\widehat j_L(d))\right),\widehat{\mathcal{V}}_{L}^+\left(\widehat{\mathcal{Z}}_{L}^*(d,\widehat j_L(d))\right)\right), 
\end{gather*}
where $\widehat{\mathcal{Z}}_{L}^*(d,j)=\sqrt{n}\widehat p-\widehat b_{L}(d,j)\ell_{d,j}\sqrt{n}(\widehat p-p)$.  Therefore, $\sqrt{n}(\tilde\ell_{d,\widehat j_L(d)}+\ell_{d,\widehat j_L(d)}p)\leq\widehat L(d)_{\alpha}^C$ is equivalent to
\begin{equation}
F_{TN}\left(\ell_{d,\widehat j_L(d)}\sqrt{n}(\widehat p-p);0,\ell_{d,\widehat j_L(d)}\widehat\Sigma\ell_{d,\widehat j_L(d)}^{\prime}\right|\left.\widehat{\mathcal{V}}_{L}^-\left(\widehat{\mathcal{Z}}_{L}^*(d,\widehat j_L(d))\right),\widehat{\mathcal{V}}_{L}^+\left(\widehat{\mathcal{Z}}_{L}^*(d,\widehat j_L(d))\right)\right)\geq 1-\alpha.\label{eq: est ineq equiv}
\end{equation}

Note that the set of matrices satisfying Assumption \ref{ass: variance restriction} is compact because it is a closed subset of the finite-dimensional set of positive semidefinite matrices whose largest eigenvalue is at most $\bar\lambda$.  Therefore, under Assumptions \ref{ass: welfare bounds}, \ref{ass: joint normality} and \ref{ass: variance restriction}, a slight modification of Lemma 5 of \cite{AKM24} implies that to prove \eqref{eq: lb unif double cond cov}, it suffices to show that for all subsequences $\{n_s\}\subset \{n\}$, $\{\mathbb{P}_{n_s}\}\subset \times_{n=1}^{\infty}\mathcal{P}_n$ satisfying
\begin{enumerate}
\item $\Sigma(\mathbb{P}_{n_s})\rightarrow \Sigma^*$, where $\Sigma^*$ satisfies Assumption \ref{ass: variance restriction};
\item $\mathbb{P}_{n_s}\left(d\in\cD(\hp),\widehat j_L(d)=j_L^*,\widehat \gamma_L(d)=\gamma_L^*\right)\rightarrow q^*\in(0,1]$;
\item $\sqrt{n_s}\left\{c^L(d,j_L^*,\gamma_L^*)
-A^L(d,j_L^*,\gamma_L^*)p(\mathbb P_{n_s})\right\}
\to \omega(d,j_L^*,\gamma_L^*)
\in(-\infty,\infty]^{\operatorname{rows}(A^L(d,j_L^*,\gamma_L^*))}$ component-wise;
	\end{enumerate}
we have
\begin{equation*}
\lim_{s\rightarrow\infty}\mathbb{P}_{n_s}\left(\left.\sqrt{n_s}(\tilde\ell_{d,\widehat j_L(d)}+\ell_{d,\widehat j_L(d)}p(\mathbb{P}_{n_s}))\leq\widehat L(d)_{\alpha}^C\right|d\in\cD(\hp),\widehat j_L(d)=j_L^*,\widehat \gamma_L(d)=\gamma_L^*\right)=\alpha 
\end{equation*}
for all $d\in\{d^0,\ldots,d^K\}$, $j_L^*\in \{1,\ldots,J_L\}$ and $\gamma_L^*$ in the support of $\widehat \gamma_L(d)$.  A  subsequence satisfying condition 3.~always exists by compactness of $[-\infty,\infty]$ applied componentwise.  No component of $\omega(d,j_L^*,\gamma_L^*)$ can equal $-\infty$.  Indeed, if its $k$th component were $-\infty$, then Assumption \ref{ass: joint normality} would imply
\begin{align*}
&\mathbb P_{n_s}\!\left(
A_k^L(d,j_L^*,\gamma_L^*)\sqrt{n_s}(\widehat p-p(\mathbb P_{n_s}))
\leq
\sqrt{n_s}\{c_k^L(d,j_L^*,\gamma_L^*)
-A_k^L(d,j_L^*,\gamma_L^*)p(\mathbb P_{n_s})\}
\right)\to0,
\end{align*}
where $A_k^L(d,j_L^*,\gamma_L^*)$ denotes the $k$th row of $A^L(d,j_L^*,\gamma_L^*)$.  Since the conditioning event is a subset of the event in the preceding display, this would contradict condition 2.

Let $\{\mathbb{P}_{n_s}\}$ be a sequence satisfying conditions 1.--3.  Under $\{\mathbb{P}_{n_s}\}$, $(\sqrt{n_s}(\widehat p-p(\mathbb{P}_{n_s})),\widehat{\Sigma})\overset{d}\longrightarrow (\xi^*,{\Sigma}^*)$ by Assumptions \ref{ass: joint normality} and \ref{ass: variance estimation}, where $\xi^*\sim\mathcal{N}(0,\Sigma^*)$.  Thus, under Assumptions \ref{ass: welfare bounds}, \ref{ass: estimated ID'd set} and \ref{ass: joint normality}--\ref{ass: variance restriction}, similar arguments to those used in the proof of Lemma 8 in \cite{AKM24} show that for any $d\in\{d^0,\ldots,d^K\}$, $j_L^*\in \{1,\ldots,J_L\}$ and $\gamma_L^*$ in the support of $\widehat\gamma_L(d)$, $(\widehat{\mathcal{V}}_{L}^-(\widehat{\mathcal{Z}}_{L}^*(d,j_L^*)),\widehat{\mathcal{V}}_{L}^+(\widehat{\mathcal{Z}}_{L}^*(d,j_L^*)))\overset{d}\longrightarrow (\mathcal{V}_{L}^{-,*}(d,j_L^*,\gamma_L^*),\mathcal{V}_{L}^{+,*}(d,j_L^*,\gamma_L^*))$ under $\{\mathbb{P}_{n_s}\}$, where
\begin{gather*}
\mathcal{V}_{L}^{-,*}(d,j_L^*,\gamma_L^*)=\max_{k:(A^L(d,j_L^*,\gamma_L^*) b_{L}(d,j_L^*))_k<0}\frac{(\omega(d,j_L^*,\gamma_L^*))_k-(A^L(d,j_L^*,\gamma_L^*)(I-b_L(d,j_L^*)\ell_{d,j_L^*})\xi^*)_k}{(A^L(d,j_L^*,\gamma_L^*) b_{L}(d,j_L^*))_k}, \\
\mathcal{V}_{L}^{+,*}(d,j_L^*,\gamma_L^*)=\min_{k:(A^L(d,j_L^*,\gamma_L^*) b_{L}(d,j_L^*))_k>0}\frac{(\omega(d,j_L^*,\gamma_L^*))_k-(A^L(d,j_L^*,\gamma_L^*)(I-b_L(d,j_L^*)\ell_{d,j_L^*})\xi^*)_k}{(A^L(d,j_L^*,\gamma_L^*) b_{L}(d,j_L^*))_k},
\end{gather*}
with $b_{L}(d,j)=\Sigma^*\ell_{d,j}^{\prime}\left(\ell_{d,j}\Sigma^*\ell_{d,j}^{\prime}\right)^{-1}$.  This convergence is joint with that of $(\sqrt{n_s}(\widehat p-p(\mathbb{P}_{n_s})),\widehat{\Sigma})$ so that under $\{\mathbb{P}_{n_s}\}$,
\begin{gather}
\left(\sqrt{n_s}\ell_{d,j_L^*}(\widehat p-p(\mathbb{P}_{n_s})),\widehat{\Sigma},\widehat{\mathcal{V}}_{L}^-\left(\widehat{\mathcal{Z}}_{L}^*(d,j_L^*)\right),\widehat{\mathcal{V}}_{L}^+\left(\widehat{\mathcal{Z}}_{L}^*(d,j_L^*)\right)\right) \notag \\
\overset{d}\longrightarrow \left(\ell_{d,j_L^*}\xi^*,\Sigma^*,\mathcal{V}_{L}^{-,*}(d,j_L^*,\gamma_L^*),\mathcal{V}_{L}^{+,*}(d,j_L^*,\gamma_L^*)\right)\label{eq: joint conv}
\end{gather}
for all $d\in\{d^0,\ldots,d^K\}$, $j_L^*\in \{1,\ldots,J_L\}$ and $\gamma_L^*$ in the support of $\widehat\gamma_L(d)$.

Using \eqref{eq: joint conv} and the equivalence in \eqref{eq: est ineq equiv}, the remaining arguments to prove \eqref{eq: lb unif double cond cov} are nearly identical to those used in the proof of Proposition 1 of \cite{McC24} and therefore omitted for brevity.  $\blacksquare$


\begin{lemma} \label{lem: single cond'l coverage for pseudo-bounds}
Suppose Assumptions \ref{ass: welfare bounds}, \ref{ass: estimated ID'd set} and \ref{ass: joint normality}--\ref{ass: variance restriction} hold.  Then, for any $0<\alpha<1$,
\begin{gather*}
\lim_{n\rightarrow\infty}\sup_{\mathbb{P}\in\mathcal{P}_n}\left|\mathbb{P}\left(\left.\sqrt{n}(\tilde\ell_{d,\widehat j_L(d)}+\ell_{d,\widehat j_L(d)}p)\leq\widehat L(d)_{\alpha}^C\right|d\in\cD(\hp)\right)-\alpha\right|\cdot\mathbb{P}\left(d\in\cD(\hp)\right)=0, \\
\lim_{n\rightarrow\infty}\sup_{\mathbb{P}\in\mathcal{P}_n}\left|\mathbb{P}\left(\left.\sqrt{n}(\tilde u_{d,\widehat j_U(d)}+u_{d,\widehat j_U(d)}p)\leq\widehat U(d)_{\alpha}^C\right|d\in\cD(\hp)\right)-\alpha\right|\cdot\mathbb{P}\left(d\in\cD(\hp)\right)=0, 
\end{gather*}
for all $d\in\{d^0,\ldots,d^K\}$.
\end{lemma}

\textbf{Proof:} The results of this lemma follow from Lemma \ref{lem: double cond'l coverage for pseudo-bounds} since, e.g.,
\begin{gather*}
\lim_{n\rightarrow\infty}\sup_{\mathbb{P}\in\mathcal{P}_n}\left|\mathbb{P}\left(\left.\sqrt{n}(\tilde\ell_{d,\widehat j_L(d)}+\ell_{d,\widehat j_L(d)}p)\leq\widehat L(d)_{\alpha}^C\right|d\in\cD(\hp)\right)-\alpha\right|\cdot\mathbb{P}\left(d\in\cD(\hp)\right) \\
=\lim_{n\rightarrow\infty}\sup_{\mathbb{P}\in\mathcal{P}_n}\left|\mathbb{P}\left(\sqrt{n}(\tilde\ell_{d,\widehat j_L(d)}+\ell_{d,\widehat j_L(d)}p)\leq\widehat L(d)_{\alpha}^C,d\in\cD(\hp)\right)-\alpha\cdot\mathbb{P}\left(d\in\cD(\hp)\right)\right| \\
=\lim_{n\rightarrow\infty}\sup_{\mathbb{P}\in\mathcal{P}_n}\left|\sum_{j_L^*=1}^{J_L}\sum_{\gamma_L^*}\left[\mathbb{P}\left(\sqrt{n}(\tilde\ell_{d,\widehat j_L(d)}+\ell_{d,\widehat j_L(d)}p)\leq\widehat L(d)_{\alpha}^C,d\in\cD(\hp),\widehat j_L(d)=j_L^*,\widehat \gamma_L(d)=\gamma_L^*\right) \right.\right. \\
\left.\left.-\alpha\cdot\mathbb{P}\left(d\in\cD(\hp),\widehat j_L(d)=j_L^*,\widehat \gamma_L(d)=\gamma_L^*\right)\right]\right| \\
=\lim_{n\rightarrow\infty}\sup_{\mathbb{P}\in\mathcal{P}_n}\left|\sum_{j_L^*=1}^{J_L}\sum_{\gamma_L^*}\left[\mathbb{P}\left(\left.\sqrt{n}(\tilde\ell_{d,\widehat j_L(d)}+\ell_{d,\widehat j_L(d)}p)\leq\widehat L(d)_{\alpha}^C\right|d\in\cD(\hp),\widehat j_L(d)=j_L^*,\widehat \gamma_L(d)=\gamma_L^*\right)-\alpha\right]\right. \\
\left.\cdot\mathbb{P}\left(d\in\cD(\hp),\widehat j_L(d)=j_L^*,\widehat \gamma_L(d)=\gamma_L^*\right)\right| \\
\leq \lim_{n\rightarrow\infty}\sup_{\mathbb{P}\in\mathcal{P}_n}\sum_{j_L^*=1}^{J_L}\sum_{\gamma_L^*}\left|\mathbb{P}\left(\left.\sqrt{n}(\tilde\ell_{d,\widehat j_L(d)}+\ell_{d,\widehat j_L(d)}p)\leq\widehat L(d)_{\alpha}^C\right|d\in\cD(\hp),\widehat j_L(d)=j_L^*,\widehat \gamma_L(d)=\gamma_L^*\right)-\alpha\right| \\
\cdot\mathbb{P}\left(d\in\cD(\hp),\widehat j_L(d)=j_L^*,\widehat \gamma_L(d)=\gamma_L^*\right) \\
\leq \sum_{j_L^*=1}^{J_L}\sum_{\gamma_L^*}\lim_{n\rightarrow\infty}\sup_{\mathbb{P}\in\mathcal{P}_n}\left|\mathbb{P}\left(\left.\sqrt{n}(\tilde\ell_{d,\widehat j_L(d)}+\ell_{d,\widehat j_L(d)}p)\leq\widehat L(d)_{\alpha}^C\right|d\in\cD(\hp),\widehat j_L(d)=j_L^*,\widehat \gamma_L(d)=\gamma_L^*\right)-\alpha\right| \\
\cdot\mathbb{P}\left(d\in\cD(\hp),\widehat j_L(d)=j_L^*,\widehat \gamma_L(d)=\gamma_L^*\right)=0,  
\end{gather*}
where the inner sums $\sum_{\gamma_L^*}$ are over the elements of the support of $\widehat\gamma_L(d)$. $\blacksquare$

\textbf{Proof of Theorem \ref{thm: cond'l coverage for cond'l bounds}:} The result of this theorem follows from Lemma \ref{lem: single cond'l coverage for pseudo-bounds} since
\begin{align*}
&\liminf_{n\rightarrow\infty}\inf_{\mathbb{P}\in\mathcal{P}_n}\left[\mathbb{P}\left(\left.\sqrt{n}[L(d),U(d)]\subseteq\left(\widehat L(d)_{\alpha_1}^C,\widehat U(d)_{1-\alpha_2}^C\right)\right|d\in\cD(\hp)\right)-(1-\alpha_1-\alpha_2)\right]\cdot\mathbb{P}\left(d\in\cD(\hp)\right) \\
&\geq \liminf_{n\rightarrow\infty}\inf_{\mathbb{P}\in\mathcal{P}_n}\left[1-\mathbb{P}\left(\left.\sqrt{n}L(d)\leq\widehat L(d)_{\alpha_1}^C\right|d\in\cD(\hp)\right)-\mathbb{P}\left(\left.\sqrt{n}U(d)\geq\widehat U(d)_{1-\alpha_2}^C\right|d\in\cD(\hp)\right)\right. \\
&\qquad\qquad\qquad-\left.(1-\alpha_1-\alpha_2)\right]\cdot\mathbb{P}\left(d\in\cD(\hp)\right) \\
&\geq \liminf_{n\rightarrow\infty}\inf_{\mathbb{P}\in\mathcal{P}_n}\left[1-\mathbb{P}\left(\left.\sqrt{n}(\tilde\ell_{d,\widehat j_L(d)}+\ell_{d,\widehat j_L(d)}p)\leq\widehat L(d)_{\alpha_1}^C\right|d\in\cD(\hp)\right)\right. \\
&\qquad\qquad\qquad-\left.\mathbb{P}\left(\left.\sqrt{n}(\tilde u_{d,\widehat j_U(d)}+u_{d,\widehat j_U(d)}p)\geq\widehat U(d)_{1-\alpha_2}^C\right|d\in\cD(\hp)\right)-(1-\alpha_1-\alpha_2)\right]\cdot\mathbb{P}\left(d\in\cD(\hp)\right) \\
&=\liminf_{n\rightarrow\infty}\inf_{\mathbb{P}\in\mathcal{P}_n}\left[\alpha_1-\mathbb{P}\left(\left.\sqrt{n}(\tilde\ell_{d,\widehat j_L(d)}+\ell_{d,\widehat j_L(d)}p)\leq\widehat L(d)_{\alpha_1}^C\right|d\in\cD(\hp)\right)\right. \\
&\qquad\qquad\qquad+\left.\mathbb{P}\left(\left.\sqrt{n}(\tilde u_{d,\widehat j_U(d)}+u_{d,\widehat j_U(d)}p)\leq\widehat U(d)_{1-\alpha_2}^C\right|d\in\cD(\hp)\right)-(1-\alpha_2)\right]\cdot\mathbb{P}\left(d\in\cD(\hp)\right) \\
&\geq \liminf_{n\rightarrow\infty}\inf_{\mathbb{P}\in\mathcal{P}_n}\left[\alpha_1-\mathbb{P}\left(\left.\sqrt{n}(\tilde\ell_{d,\widehat j_L(d)}+\ell_{d,\widehat j_L(d)}p)\leq\widehat L(d)_{\alpha_1}^C\right|d\in\cD(\hp)\right)\right]\cdot\mathbb{P}\left(d\in\cD(\hp)\right) \\
&\quad + \liminf_{n\rightarrow\infty}\inf_{\mathbb{P}\in\mathcal{P}_n}\left[\mathbb{P}\left(\left.\sqrt{n}(\tilde u_{d,\widehat j_U(d)}+u_{d,\widehat j_U(d)}p)\leq\widehat U(d)_{1-\alpha_2}^C\right|d\in\cD(\hp)\right)-(1-\alpha_2)\right]\cdot\mathbb{P}\left(d\in\cD(\hp)\right)=0,
\end{align*}
where the second inequality follows from the facts that $L(d)\geq \tilde\ell_{d,\widehat j_L(d)}+\ell_{d,\widehat j_L(d)}p$ and $U(d)\leq \tilde u_{d,\widehat j_U(d)}+u_{d,\widehat j_U(d)}p$ almost surely and the final equality follows from Lemma \ref{lem: single cond'l coverage for pseudo-bounds}. $\blacksquare$

\textbf{Proof of Theorem \ref{thm: uncond'l coverage for proj bounds}:} We start by showing
\begin{equation}
\liminf_{n\rightarrow\infty}\inf_{\mathbb{P}\in\mathcal{P}_n}\mathbb{P}\left(\sqrt{n}(\tilde\ell_{\widehat d,\widehat j_L(\widehat d)}+\ell_{\widehat d,\widehat j_L(\widehat d)}p)\geq\widehat L(\widehat d)_{\alpha}^P\right)\geq 1-\alpha. \label{eq: unc lb cov}
\end{equation}
Consider arbitrary subsequences $\{n_s\}\subset\{n\}$ and $\{\mathbb{P}_{n_s}\}\subset \times_{n=1}^{\infty}\mathcal{P}_n$ along which the left-hand side of \eqref{eq: unc lb cov} attains its liminf.  The set of matrices satisfying Parts 1.~and 2.~of Assumption \ref{ass: variance restriction} is compact because it is a closed subset of the finite-dimensional set of positive semidefinite matrices whose eigenvalue is at most $\bar\lambda$ so that Parts 1.~and 2.~of Assumption \ref{ass: variance restriction} imply that there always exists a further subsequence $\{n_{s_p}\}\subset\{n_s\}$ such that $\Sigma(\mathbb P_{n_{s_p}})\to\Sigma^*$, where $\Sigma^*$ satisfies Parts 1.~and 2.~of Assumption \ref{ass: variance restriction}.  Therefore, by the same argument as in the proof of Lemma \ref{lem: double cond'l coverage for pseudo-bounds}, to prove \eqref{eq: unc lb cov}, it suffices to show
\begin{equation}
\lim_{s\rightarrow\infty}\mathbb{P}_{n_s}\left(\sqrt{n_s}(\tilde\ell_{\widehat d,\widehat j_L(\widehat d)}+\ell_{\widehat d,\widehat j_L(\widehat d)}p(\mathbb{P}_{n_s}))\geq\widehat L(\widehat d)_{\alpha}^P\right)\geq 1-\alpha \label{eq: unc lb subseq}
\end{equation}
for any subsequences $\{n_s\}\subset\{n\}$ and $\{\mathbb{P}_{n_s}\}\subset \times_{n=1}^{\infty}\mathcal{P}_n$ along which the limit exists and $\Sigma(\mathbb P_{n_{s}})\to\Sigma^*$.  Consider any such subsequences.  Assumptions \ref{ass: joint normality} and \ref{ass: variance estimation} imply
\begin{equation}
\left(\ell^{mat}\sqrt{n_s}(\widehat p-p(\mathbb P_{n_s})),
\widehat\Sigma_L\right)
\overset{d}{\longrightarrow}
(\xi_L^*,\Sigma_L^*),
\qquad
\xi_L^*\sim\mathcal N(0,\Sigma_L^*),
\quad
\Sigma_L^*=\ell^{mat}\Sigma^*\ell^{mat\prime},
\label{eq: projection joint convergence revised}
\end{equation}
where every diagonal element corresponding to $(d,j)$ satisfies $(\Sigma_L^*)_{dJ_L+j,dJ_L+j}=\ell_{d,j}\Sigma^*\ell_{d,j}'
\geq\|\ell_{d,j}\|^2/\bar\lambda>0$
by Part 2.~of Assumption \ref{ass: variance restriction}.

We next verify that
\begin{equation}
\widehat c_{1-\alpha,L}\overset{p}{\longrightarrow}c_{1-\alpha,L}^*
\label{eq: projection cv convergence revised}
\end{equation}
along any $\{\mathbb{P}_{n_s}\}$ described above for $c_{1-\alpha,L}^*$ the $1-\alpha$ quantile of $\max_{i\in\{1,\ldots,(K+1)J_L\}}\frac{\xi_{L,i}^*}{\sqrt{(\Sigma_L^*)_{ii}}}$.  Let $\Sigma_{L,s}=\ell^{mat}\Sigma(\mathbb P_{n_s})\ell^{mat\prime}$ and $Z\sim\mathcal N(0,I_{(K+1)J_L})$.  Continuity of the positive semidefinite square root provides $\Sigma_{L,s}^{1/2}\to(\Sigma_L^*)^{1/2}$, and hence
\[
\max_i\frac{(\Sigma_{L,s}^{1/2}Z)_i}{\sqrt{(\Sigma_{L,s})_{ii}}}
\longrightarrow
\max_i\frac{((\Sigma_L^*)^{1/2}Z)_i}{\sqrt{(\Sigma_L^*)_{ii}}}
\qquad\text{almost surely}.
\]
The limiting random variable has a strictly increasing cdf that is continuous in $\Sigma_L^*$ so that $\widehat\Sigma_L\overset{p}{\to}\Sigma_L^*$, established in \eqref{eq: projection joint convergence revised} implies \eqref{eq: projection cv convergence revised}.

Since $\sqrt{n_s}(\tilde\ell_{\widehat d,\widehat j_L(\widehat d)}+\ell_{\widehat d,\widehat j_L(\widehat d)}p)\geq\widehat L(\widehat d)_{\alpha}^P$ is equivalent to $\ell_{\widehat d,\widehat j_L(\widehat d)}\sqrt{n_s}(\widehat p-p)\leq \widehat{c}_{1-\alpha,L}\sqrt{\widehat{\Sigma}_{L,\widehat{d}J_L+\widehat{j}_L(\widehat d)}}$, the left-hand side of \eqref{eq: unc lb subseq} is equal to
\begin{gather*}
\lim_{s\rightarrow\infty}\mathbb{P}_{n_s}\left(\ell_{\widehat d,\widehat j_L(\widehat d)}\sqrt{n_s}(\widehat p-p(\mathbb{P}_{n_s}))\leq \widehat{c}_{1-\alpha,L}\sqrt{\widehat{\Sigma}_{L,\widehat{d}J_L+\widehat{j}_L(\widehat d)}}\right) \\
\geq \lim_{s\rightarrow\infty}\mathbb{P}_{n_s}\left(\ell^{mat}\sqrt{n_s}(\widehat p-p(\mathbb{P}_{n_s}))\leq \widehat{c}_{1-\alpha,L}\sqrt{\diag\left(\widehat{\Sigma}_{L}\right)}\right) \\
=\mathbb{P}\left(\xi_L^*\leq {c}_{1-\alpha,L}^*\sqrt{\diag\left({\Sigma}_{L}^*\right)}\right)=\mathbb{P}\left(\max_{i\in\{1,\ldots,(K+1)J_L\}}\frac{\xi_{L,i}^*}{\sqrt{{\Sigma}_{L,i}^*}}\leq {c}_{1-\alpha,L}^*\right)=1-\alpha,
\end{gather*}
where all inequalities are taken element-wise across vectors, the inequality follows from the fact that $\ell_{\widehat d,\widehat j_L(\widehat d)}\sqrt{n_s}(\widehat p-p(\mathbb{P}_{n_s}))$ is a (random) element of $\ell^{mat}\sqrt{n_s}(\widehat p-p(\mathbb{P}_{n_s}))$ and the first equality follows from \eqref{eq: projection joint convergence revised} and \eqref{eq: projection cv convergence revised}.  By way of \eqref{eq: unc lb subseq}, we have thus proved \eqref{eq: unc lb cov}.  In addition,
\begin{equation*}
\liminf_{n\rightarrow\infty}\inf_{\mathbb{P}\in\mathcal{P}_n}\mathbb{P}\left(\sqrt{n}(\tilde u_{\widehat d,\widehat j_U(\widehat d)}+u_{\widehat d,\widehat j_U(\widehat d)}p)\leq\widehat U(\widehat d)_{1-\alpha}^P\right)\geq 1-\alpha. 
\end{equation*}
follows by nearly identical arguments.  The statement of the theorem then follows by nearly identical arguments to those used in the proof of Theorem \ref{thm: cond'l coverage for cond'l bounds}. $\blacksquare$


\begin{lemma}
\label{lem: hybrid cond cov}
Suppose Assumptions \ref{ass: welfare bounds}--\ref{ass: variance restriction} hold.  Then, for any $0<\beta<\alpha<1$,
\begin{gather}
\lim_{n\rightarrow\infty}\sup_{\mathbb{P}\in\mathcal{P}_n}\left|\mathbb{P}\left(\left.\sqrt{n}(\tilde\ell_{\widehat d,\widehat j_L(\widehat d)}+\ell_{\widehat d,\widehat j_L(\widehat d)}p)\leq\widehat L(\widehat d)_{\alpha}^H\right|\widehat d=d^*,\sqrt{n}(\tilde\ell_{\widehat d,\widehat j_L(\widehat d)}+\ell_{\widehat d,\widehat j_L(\widehat d)}p)\geq\widehat L(\widehat d)_{\beta}^P\right)-\frac{\alpha-\beta}{1-\beta}\right| \notag\\
\cdot\mathbb{P}\left(\widehat d=d^*,\sqrt{n}(\tilde\ell_{\widehat d,\widehat j_L(\widehat d)}+\ell_{\widehat d,\widehat j_L(\widehat d)}p)\geq\widehat L(\widehat d)_{\beta}^P\right)=0, \label{eq: hybrid cond cov LB} \\
\lim_{n\rightarrow\infty}\sup_{\mathbb{P}\in\mathcal{P}_n}\left|\mathbb{P}\left(\left.\sqrt{n}(\tilde u_{\widehat d,\widehat j_U(\widehat d)}+u_{\widehat d,\widehat j_U(\widehat d)}p)\leq\widehat U(\widehat d)_{\alpha}^H\right|\widehat d=d^*,\sqrt{n}(\tilde u_{\widehat d,\widehat j_U(\widehat d)}+u_{\widehat d,\widehat j_U(\widehat d)}p)\leq\widehat U(\widehat d)_{1-\beta}^P\right)-\frac{\alpha-\beta}{1-\beta}\right| \notag\\
\cdot\mathbb{P}(\widehat d=d^*,\sqrt{n}(\tilde u_{\widehat d,\widehat j_U(\widehat d)}+u_{\widehat d,\widehat j_U(\widehat d)}p)\leq\widehat U(\widehat d)_{1-\beta}^P)=0, \label{eq: hybrid cond cov UB}
\end{gather}
for all $d^*\in\{d^0,\ldots,d^K\}$.
\end{lemma}

\textbf{Proof:} The proof of \eqref{eq: hybrid cond cov UB} is nearly identical to the proof of \eqref{eq: hybrid cond cov LB} so that we only show the latter.  Upon noting that $F_{TN}(t;\mu,\sigma^2,\widehat{\mathcal{V}}_L^-(z,d,j,\gamma),\widehat{\mathcal{V}}_L^{+,H}(z,d,j,\gamma,\mu))$ is strictly decreasing in $\mu$ by the same argument used in the proof of Proposition 5 of \cite{AKM24} and replacing condition 2.~in the proof of Lemma \ref{lem: double cond'l coverage for pseudo-bounds} with
\begin{enumerate}
\item[2'.] $\mathbb{P}_{n_s}\left(\widehat d=d^*,\widehat j_L(\widehat d)=j_L^*,\widehat \gamma_L(\widehat d)=\gamma_L^*,\sqrt{n}(\tilde\ell_{\widehat d,\widehat j_L(\widehat d)}+\ell_{\widehat d,\widehat j_L(\widehat d)}p)\geq\widehat L(\widehat d)_{\beta}^P\right)\rightarrow q^*\in(0,1]$, 
\end{enumerate}
completely analogous arguments to those used to prove \eqref{eq: lb unif double cond cov} in Lemma \ref{lem: double cond'l coverage for pseudo-bounds}, along with the proof of \eqref{eq: projection cv convergence revised} in the proof of Theorem \ref{thm: uncond'l coverage for proj bounds} above imply
\begin{gather*}
\lim_{n\rightarrow\infty}\sup_{\mathbb{P}\in\mathcal{P}_n}\left|\mathbb{P}\left(\left.\sqrt{n}(\tilde\ell_{\widehat d,\widehat j_L(\widehat d)}+\ell_{\widehat d,\widehat j_L(\widehat d)}p)\leq\widehat L(\widehat d)_{\alpha}^H\right|\widehat d=d^*,\widehat j_L(\widehat d)=j_L^*,\widehat \gamma_L(\widehat d)=\gamma_L^*,\sqrt{n}(\tilde\ell_{\widehat d,\widehat j_L(\widehat d)}+\ell_{\widehat d,\widehat j_L(\widehat d)}p)\geq\widehat L(\widehat d)_{\beta}^P\right)\right. \notag\\
\left.-\frac{\alpha-\beta}{1-\beta}\right|\cdot\mathbb{P}\left(\widehat d=d^*,\widehat j_L(\widehat d)=j_L^*,\widehat \gamma_L(\widehat d)=\gamma_L^*,\sqrt{n}(\tilde\ell_{\widehat d,\widehat j_L(\widehat d)}+\ell_{\widehat d,\widehat j_L(\widehat d)}p)\geq\widehat L(\widehat d)_{\beta}^P\right)=0
\end{gather*}
for all $d^*\in\{d^0,\ldots,d^K\}$, $j_L^*\in \{1,\ldots,J_L\}$ and $\gamma_L^*$ in the support of $\widehat\gamma_L(d^*)$.  Then, the same argument as in the proof of Lemma \ref{lem: single cond'l coverage for pseudo-bounds} implies \eqref{eq: hybrid cond cov LB}.
$\blacksquare$

\begin{lemma}
\label{lem: hybrid uncond cov}
Suppose Assumptions \ref{ass: welfare bounds}--\ref{ass: variance restriction}  hold.  Then, for any $0<\alpha<1$,
\begin{gather}
\liminf_{n\rightarrow\infty}\inf_{\mathbb{P}\in\mathcal{P}_n}\mathbb{P}\left(\sqrt{n}(\tilde\ell_{\widehat d,\widehat j_L(\widehat d)}+\ell_{\widehat d,\widehat j_L(\widehat d)}p)>\widehat L(\widehat d)_{\alpha}^H\right)\geq 1-\alpha, \label{eq: hybrid uncond cov LB} \\
\liminf_{n\rightarrow\infty}\inf_{\mathbb{P}\in\mathcal{P}_n}\mathbb{P}\left(\sqrt{n}(\tilde u_{\widehat d,\widehat j_U(\widehat d)}+u_{\widehat d,\widehat j_U(\widehat d)}p)<\widehat U(\widehat d)_{1-\alpha}^H\right)\geq 1-\alpha. \label{eq: hybrid uncond cov UB}
\end{gather}
\end{lemma}

\textbf{Proof:} The proof of \eqref{eq: hybrid uncond cov UB} is nearly identical to the proof of \eqref{eq: hybrid uncond cov LB} so that we only show the latter.  Lemma 6 of \cite{AKM24} and Lemma \ref{lem: hybrid cond cov} imply
\begin{gather*}
\liminf_{n\rightarrow\infty}\inf_{\mathbb{P}\in\mathcal{P}_n}\mathbb{P}\left(\sqrt{n}(\tilde\ell_{\widehat d,\widehat j_L(\widehat d)}+\ell_{\widehat d,\widehat j_L(\widehat d)}p)>\widehat L(\widehat d)_{\alpha}^H\right) \\
\geq \frac{1-\alpha}{1-\beta}\liminf_{n\rightarrow\infty}\inf_{\mathbb{P}\in\mathcal{P}_n}\sum_{d^*\in\{d^0,\ldots,d^K\}}\mathbb{P}\left(\widehat d=d^*,\sqrt{n}(\tilde\ell_{\widehat d,\widehat j_L(\widehat d)}+\ell_{\widehat d,\widehat j_L(\widehat d)}p)\geq\widehat L(\widehat d)_{\beta}^P\right) \\
=\frac{1-\alpha}{1-\beta}\liminf_{n\rightarrow\infty}\inf_{\mathbb{P}\in\mathcal{P}_n}\mathbb{P}\left(\sqrt{n}(\tilde\ell_{\widehat d,\widehat j_L(\widehat d)}+\ell_{\widehat d,\widehat j_L(\widehat d)}p)\geq\widehat L(\widehat d)_{\beta}^P\right)\geq 1-\alpha,
\end{gather*}
where the final inequality follows from \eqref{eq: unc lb cov} in the proof of Theorem \ref{thm: uncond'l coverage for proj bounds}. $\blacksquare$

\textbf{Proof of Theorem \ref{thm: uncond'l coverage for hybrid bounds}:} Using Lemma \ref{lem: hybrid uncond cov} in the place of Lemma \ref{lem: single cond'l coverage for pseudo-bounds}, the proof is nearly identical to the proof of Theorem \ref{thm: cond'l coverage for cond'l bounds}. $\blacksquare$

\section{Degenerate Affine Endpoint Components}
\label{sec: degenerate affine endpoint components}

This appendix permits an affine component of an identified-set endpoint to have zero asymptotic variance when that component is uniformly separated from the corresponding population endpoint.  Parts 1.~and 3.~of Assumption \ref{ass: variance restriction} are maintained.  Part 2.~may be replaced by the following condition: for every $d\in\{d^0,\ldots,d^K\}$ and every $\mathbb P\in\mathcal P_n$,
\begin{align}
&\ell_{d,j}\Sigma(\mathbb P)\ell_{d,j}'
\geq \|\ell_{d,j}\|^2/\bar\lambda
\quad\text{or}\quad
\left\{
\begin{array}{l}
\ell_{d,j}\Sigma(\mathbb P)\ell_{d,j}'=0,\\[1mm]
L(d)-\{\tilde\ell_{d,j}+\ell_{d,j}p(\mathbb P)\}
\geq 1/\bar\lambda,
\end{array}
\right.
\label{eq: lower zero or separated}\\
&u_{d,j}\Sigma(\mathbb P)u_{d,j}'
\geq \|u_{d,j}\|^2/\bar\lambda
\quad\text{or}\quad
\left\{
\begin{array}{l}
u_{d,j}\Sigma(\mathbb P)u_{d,j}'=0,\\[1mm]
\{\tilde u_{d,j}+u_{d,j}p(\mathbb P)\}-U(d)
\geq 1/\bar\lambda.
\end{array}
\right.
\label{eq: upper zero or separated}
\end{align}
Call a lower (upper) component ``degenerate'' when the second alternative in \eqref{eq: lower zero or separated} (\eqref{eq: upper zero or separated}) holds and call it ``regular'' otherwise.  Because Assumption \ref{ass: variance estimation} holds uniformly, there exists a deterministic sequence $\tau_n\downarrow0$ such that
\begin{equation}
\sup_{\mathbb P\in\mathcal P_n}
\mathbb P\!\left(\|\widehat\Sigma-\Sigma(\mathbb P)\|>\tau_n\right)
\longrightarrow0.
\label{eq: covariance rate degenerate}
\end{equation}
In the definitions of the conditional, projection, and hybrid bounds, classify a lower component as degenerate when $\ell_{d,j}\widehat\Sigma\ell_{d,j}'\leq \tau_n\|\ell_{d,j}\|^2$ and regular otherwise and classify an upper component analogously.  Omit components classified as degenerate when computing the studentized maxima defining the projection critical values.  If the selected lower component is classified as degenerate, set
\begin{equation}
\widehat L(d)_\alpha^C
=
\widehat L(d)_\alpha^P
=
\widehat L(d)_\alpha^H
=
\sqrt n\{\tilde\ell_{d,\widehat j_L(d)}+
\ell_{d,\widehat j_L(d)}\widehat p\}.
\label{eq: degenerate lower convention}
\end{equation}
If the selected upper component is classified as degenerate, set
\begin{equation}
\widehat U(d)_\alpha^C
=
\widehat U(d)_\alpha^P
=
\widehat U(d)_\alpha^H
=
\sqrt n\{\tilde u_{d,\widehat j_U(d)}+
 u_{d,\widehat j_U(d)}\widehat p\}.
\label{eq: degenerate upper convention}
\end{equation}
Otherwise, retain the definitions in Section \ref{sec: CIs}. 

\begin{theorem}
\label{thm: degenerate endpoint extension}
Suppose Assumptions \ref{ass: welfare bounds}--\ref{ass: variance estimation}, Parts 1 and 3 of Assumption \ref{ass: variance restriction}, and \eqref{eq: lower zero or separated}--\eqref{eq: upper zero or separated} hold.  Then Theorems \ref{thm: cond'l coverage for cond'l bounds}, \ref{thm: uncond'l coverage for proj bounds}, and \ref{thm: uncond'l coverage for hybrid bounds} continue to hold for the bounds defined above.
\end{theorem}

\textbf{Proof:}
We first show that a degenerate component is selected with probability converging to zero uniformly over $\mathbb P\in\mathcal P_n$.  Suppose that the second alternative in \eqref{eq: lower zero or separated} holds for $(d,j)$, and choose $j_L^0(d)\in\argmax_{k\in\{1,\ldots,J_L\}}
\{\tilde\ell_{d,k}+\ell_{d,k}p(\mathbb P)\}$.  If $\widehat j_L(d)=j$, then
\[0\leq\{\tilde\ell_{d,j}+\ell_{d,j}\widehat p\}-\{\tilde\ell_{d,j_L^0(d)}+\ell_{d,j_L^0(d)}\widehat p\}\leq-1/\bar\lambda+\{\ell_{d,j}-\ell_{d,j_L^0(d)}\}\{\widehat p-p(\mathbb P)\},\]
so that
\begin{equation}
\mathbb P(\widehat j_L(d)=j)
\leq
\mathbb P\!\left(
\left\|\sqrt n\{\widehat p-p(\mathbb P)\}\right\|
\geq
\frac{\sqrt n}{\bar\lambda
\|\ell_{d,j}-\ell_{d,j_L^0(d)}\|}
\right).
\label{eq: degenerate lower selection probability}
\end{equation}
Assumption \ref{ass: joint normality} and Part 1 of
Assumption \ref{ass: variance restriction} imply uniform
tightness of $\sqrt n(\widehat p-p(\mathbb P))$.  Since the
vectors $\ell_{d,j}$ are fixed and finite in number, the
right-hand side of
\eqref{eq: degenerate lower selection probability}
converges to zero uniformly over the distributions for
which the component $(d,j)$ is degenerate.  Consequently,
for every $d\in\cD$ and $j\in\{1,\ldots,J_L\}$,
\[
    \sup_{\mathbb P\in\mathcal P_n}
    \mathbb P\left(
        \widehat j_L(d)=j,\,
        \ell_{d,j}\Sigma(\mathbb P)\ell_{d,j}'=0
    \right)
    \longrightarrow0.
\]
The analogous argument for the upper endpoint gives, for
every $d\in\cD$ and $j\in\{1,\ldots,J_U\}$,
\begin{equation}
    \sup_{\mathbb P\in\mathcal P_n}
    \mathbb P\left(
        \widehat j_U(d)=j,\,
        u_{d,j}\Sigma(\mathbb P)u_{d,j}'=0
    \right)
    \longrightarrow0.
    \label{eq: degenerate upper selection probability}
\end{equation}

Next, using the operator norm for matrices, on the event
$\{\|\widehat\Sigma-\Sigma(\mathbb P)\|\leq\tau_n\}$,
\[
    \left|
    \ell_{d,j}
    \{\widehat\Sigma-\Sigma(\mathbb P)\}
    \ell_{d,j}'
    \right|
    \leq
    \tau_n\|\ell_{d,j}\|^2
\]
for every $d\in\cD$ and $j\in\{1,\ldots,J_L\}$.
If the component $(d,j)$ is degenerate, this implies
\[
    \ell_{d,j}\widehat\Sigma\ell_{d,j}'
    \leq
    \tau_n\|\ell_{d,j}\|^2.
\]
If it is regular, then, for all sufficiently large $n$,
\[
    \ell_{d,j}\widehat\Sigma\ell_{d,j}'
    \geq
    \|\ell_{d,j}\|^2(1/\bar\lambda-\tau_n)
    >
    \tau_n\|\ell_{d,j}\|^2,
\]
where the strict inequality follows from
$\tau_n<1/(2\bar\lambda)$.
The same argument applies to the upper components after
replacing $\ell_{d,j}$ by $u_{d,j}$.  Thus, for all
sufficiently large $n$,
\begin{align*}
    &\inf_{\mathbb P\in\mathcal P_n}
    \mathbb P\left(
        \text{all endpoint components are classified correctly}
    \right)\\
    &\qquad\geq
    1-
    \sup_{\mathbb P\in\mathcal P_n}
    \mathbb P\left(
        \|\widehat\Sigma-\Sigma(\mathbb P)\|>\tau_n
    \right)
    \longrightarrow1
\end{align*}
by \eqref{eq: covariance rate degenerate}.

Consider Lemma
\ref{lem: double cond'l coverage for pseudo-bounds}.  
If the component indexed by $j_L^*$ is degenerate, then the
absolute conditional-coverage error in
\eqref{eq: lb unif double cond cov} is at most one and hence
\begin{align*}
&\left|
\mathbb P\left(
\left.
\sqrt n\{\tilde\ell_{d,\widehat j_L(d)}
+\ell_{d,\widehat j_L(d)}p(\mathbb P)\}
\leq \widehat L(d)_\alpha^C
\right|
d\in\cD(\hp),
      \widehat j_L(d)=j_L^*,
      \widehat\gamma_L(d)=\gamma_L^*
\right)-\alpha
\right|
\mathbb P(d\in\cD(\hp),
      \widehat j_L(d)=j_L^*,
      \widehat\gamma_L(d)=\gamma_L^*)\\
&\qquad\leq
\mathbb P(d\in\cD(\hp),
      \widehat j_L(d)=j_L^*,
      \widehat\gamma_L(d)=\gamma_L^*)
\leq
\mathbb P(\widehat j_L(d)=j_L^*).
\end{align*}
The last probability converges to zero uniformly by
\eqref{eq: degenerate lower selection probability}.  On the other hand, if the component indexed by $j_L^*$ is regular, then by the arguments above, it will be classified as regular with probability converging to one uniformly so that the proof of the lemma goes through essentially unchanged.  The proof of
\eqref{eq: ub unif double cond cov} follows identically from
\eqref{eq: degenerate upper selection probability}.  Thus Lemmas
\ref{lem: double cond'l coverage for pseudo-bounds} and
\ref{lem: single cond'l coverage for pseudo-bounds}, and therefore
Theorem \ref{thm: cond'l coverage for cond'l bounds}, remain valid.

Let $\mathcal E_n$ denote the event that all endpoint
components are classified correctly and that the components
indexed by $\widehat j_L(\widehat d)$ and $\widehat j_U(\widehat d)$ are
regular.  For all sufficiently large $n$, the preceding
classification argument and the union bound imply
\begin{align}
    \sup_{\mathbb P\in\mathcal P_n}
    \mathbb P(\mathcal E_n^c)
    &\leq
    \sup_{\mathbb P\in\mathcal P_n}
    \mathbb P\left(
        \|\widehat\Sigma-\Sigma(\mathbb P)\|>\tau_n
    \right)
    \notag\\
    &\quad+
    \sum_{d\in\cD}\sum_{j=1}^{J_L}
    \sup_{\mathbb P\in\mathcal P_n}
    \mathbb P\left(
        \widehat j_L(d)=j,\,
        \ell_{d,j}\Sigma(\mathbb P)\ell_{d,j}'=0
    \right)
    \notag\\
    &\quad+
    \sum_{d\in\cD}\sum_{j=1}^{J_U}
    \sup_{\mathbb P\in\mathcal P_n}
    \mathbb P\left(
        \widehat j_U(d)=j,\,
        u_{d,j}\Sigma(\mathbb P)u_{d,j}'=0
    \right)
    \longrightarrow0,
    \label{eq: regular selected event}
\end{align}
where convergence follows from
\eqref{eq: covariance rate degenerate}, the preceding
bounds on the probabilities of selecting degenerate
components, and finiteness of $K$, $J_L$, and $J_U$.
On $\mathcal E_n$, lower projection noncoverage implies
\begin{align*}
\max_{d\in\{d^0,\ldots,d^K\}}
\max_{j:\,\ell_{d,j}\Sigma(\mathbb P)\ell_{d,j}'>0}
\frac{\ell_{d,j}\sqrt n\{\widehat p-p(\mathbb P)\}}
{\sqrt{\ell_{d,j}\widehat\Sigma\ell_{d,j}'}}
>\widehat c_{1-\alpha,L},
\end{align*}
and upper projection noncoverage implies the analogous event with $u_{d,j}$ and $\widehat c_{1-\alpha,U}$.  The simultaneous Gaussian-approximation argument in the proof of Theorem \ref{thm: uncond'l coverage for proj bounds}, applied to the finite subvectors appearing in these maxima, therefore gives
\begin{align*}
\limsup_{n\to\infty}\sup_{\mathbb P\in\mathcal P_n}
\mathbb P\left(\sqrt nL(\widehat d)<\widehat L(\widehat d)_\alpha^P,\mathcal E_n\right)&\leq\alpha,\\
\limsup_{n\to\infty}\sup_{\mathbb P\in\mathcal P_n}
\mathbb P\left(\sqrt nU(\widehat d)>\widehat U(\widehat d)_{1-\alpha}^P,\mathcal E_n\right)&\leq\alpha.
\end{align*}
Combining these inequalities with \eqref{eq: regular selected event} and the Bonferroni inequality proves Theorem \ref{thm: uncond'l coverage for proj bounds}.

On $\mathcal E_n$, the modified hybrid bounds coincide with the original hybrid bounds for regular selected components.  Hence, for each endpoint, the probability of hybrid noncoverage is bounded above by the corresponding noncoverage probability in Lemmas \ref{lem: hybrid cond cov}--\ref{lem: hybrid uncond cov} plus $\mathbb P(\mathcal E_n^c)$.  Equation \eqref{eq: regular selected event} and those lemmas therefore imply Theorem \ref{thm: uncond'l coverage for hybrid bounds}. \hfill$\blacksquare$

\end{appendix}

\newpage

\bibliographystyle{apalike}
\bibliography{sel_from_set}

\end{document}